\documentclass{elsart}
\usepackage{epsf}
\begin{document}
\newcommand{\Section}[1]{\setcounter{equation}{0}\section{#1}}
\renewcommand{\theequation}{\thesection.\arabic{equation}}
\begin{frontmatter}
\title{Nuclear isotope shifts within \\
the local energy-density functional approach}
\author[moscow]{S.A. Fayans\thanksref{ELM}\thanksref{DFG}},
\author[moscow]{S.V. Tolokonnikov\thanksref{DFG}},
\author[obninsk]{E.L. Trykov},
\author[hannover]{D. Zawischa}
\address[moscow]{Russian Research Centre -- Kurchatov Institute,
123182 Moscow, Russian Federation}
\address[obninsk]{Institute of Physics and Power Engineering,
249020 Obninsk, Kaluga district, Russian Federation}
\address[hannover]{Institut f\"ur Theoretische Physik, 
Universit\"at Hannover, D-30060 Hannover, Germany}
\thanks[ELM]{E-mail address: fayans@mbslab.kiae.ru}
\thanks[DFG]{Supported by the Deutsche Forschungsgemeinschaft}
\begin{abstract}
The foundation of the local energy-density functional method to
describe the nuclear ground-state properties is given. The method
is used to investigate differential observables such as the odd-even
mass differences and odd-even effects in charge radii. For a few
isotope chains of spherical nuclei, the calculations are performed
with an exact treatment of the Gor'kov equations in the
coordinate-space representation. A zero-range cutoff
density-dependent pairing interaction with a density-gradient term
is used. The evolution of charge radii and nucleon separation
energies is reproduced reasonably well including kinks at magic
neutron numbers and sizes of staggering. It is shown that the
density-dependent pairing may also induce sizeable staggering and
kinks in the evolution of the mean energies of multipole excitations. 
The results are compared with the conventional mean field Skyrme--HFB 
and relativistic Hartree--BCS calculations. With the formulated 
approach, an extrapolation from the pairing properties of finite 
nuclei to pairing in infinite matter is considered, and the dilute 
limit near the critical point, at which the regime changes from weak 
to strong pairing, is discussed. 



\end{abstract}
\end{frontmatter}

\Section{Introduction}
\label{Introd}
For a long time, the fine structure of the isotopic dependence,
i.e. dependence on the neutron number, of nuclear charge radii
could not be explained satisfactorily. Because of the apparent lack
of relevant experimental information, the effective
particle--particle (pp) force which leads to the observed pairing
properties of nuclei, has mostly been assumed in a very simple form,
just sufficient to produce reasonable gap parameters $\bar \Delta$
extracted from the observed odd-even staggering in the nucleon
separation energies. Even more sophisticated effective interactions,
e.g. the Gogny force~\cite{Gogny73,Gogny75,DeGo80}, do not include 
a density dependence in the pp channel.

It then has been demonstrated in Hartree-Fock-Bogolyubov (HFB) type
calculations that three- or four-body forces are essential to
reproduce the experimental data on isotope shifts of nuclear charge 
radii~\cite{ZRS87,ReZa88,ReZa89}, and indeed that these isotope
shifts give indirect experimental information on the effective
in-medium many-body force, or what is equivalent, on the density
dependence of the effective interaction, particularly in the pp
channel.

It should be emphasized that a better knowledge of the density
dependence of the nuclear pairing force would give the possibility
to predict the pairing gap as a function of nuclear matter density
which is, in particular, of great importance for understanding the
pairing phenomena in neutron stars. It is well known that at present
the pairing gap can not be obtained with sufficient accuracy from
nuclear matter calculations based on bare NN interaction. The 
empirical information gained from the studies of real nuclei, 
specifically from the combined analyses of the nucleon separation 
energies and isotope shifts in charge radii, seems to be
indispensable in this respect. A more general remark is that, as one
may notice, a ``good'' microscopic theory, which could supply an
effective interaction for describing the nuclear ground states and
low-energy nuclear structure on a satisfactory level, is still
lacking. The presently most successful {\em simultaneous\/}
description of the {\em bulk\/} nuclear properties, such as binding
energies and radii, throughout the periodic chart is achieved with
phenomenological density-dependent interaction such as the Skyrme
force~\cite{Skyrme56}. This can be traced to
the fact that some major effects produced by the density dependence
of the effective interaction in the particle--hole (ph) channel,
which is the second variational derivative of the corresponding
energy-density functional with respect to normal density, are now
established fairly well\footnote{One may still notice that a
universal and unique parametrization of the ph force has not yet
been found. In particular, considerable effort is still continuing
to optimize the HF part of the Skyrme-type
functional~\cite{Chab97,Brown98}.}. On the same ground, one expects
that {\em simultaneous\/} description of the {\em differential\/}
observables, such as odd-even mass differences and the odd-even
effects in radii, would shed light on the density dependence of the
effective interaction in the pp channel.

In~\cite{ZRS87,ReZa88,ReZa89} an effective interaction has been
chosen which consisted of two- and three- (or four-) body parts,
with parameters adjusted independently of those of the 
single-particle potential used for the shell model description of 
the reference nucleus. Only differences with respect of this 
reference nucleus have been calculated in a self-consistent way.

This procedure corresponds to the philosophy of the Landau-Migdal
theory of finite Fermi systems~\cite{Migdal}, where there is no
simple connection between the single particle well parameters and
the effective interaction of quasiparticles close to the Fermi
surface. However, there are consistency 
requirements~\cite{SWW77,MB69,FK73,NW74,KhSa82,Migdal2} and since
the parameters have been fitted to a large number of data, a quite
reliable set of force parameters is available~\cite{Migdal2}. The
interaction of Refs.~\cite{ReZa88,ReZa89} has been restricted by 
the requirement that it should reduce to the Migdal force in the 
ph channel, the three-body part leading to the density dependence.

In the meantime, since Migdal had formulated his theory of finite
Fermi systems (FFS), there has been much progress in Hartree-Fock
calculations~\cite{Neg70,NeV72,VAB,DeGo80}, as well as in the
self-consistent version of the FFS
theory~\cite{KhSa82,STF88,Migdal2}, demonstrating that initially
Migdal may have been too pessimistic concerning the possibility to
use the same interaction for ground state properties and for
low-lying excited states.

In deriving the HF or HFB equations, the starting point is the
minimization of the expectation value of the energy, expressed by
an effective interaction. The latter is considered as a substitute
for the $G$-matrix derived from the true NN interaction, appearing
in some kind of ``effective Hamiltonian''. Therefore, ideally, the
same interaction should be used in the pp (hh) and the 
particle-hole (ph) channel. The energy is then obtained from the 
interaction and the state vector which is assumed to be a Slater 
determinant.

The self-consistent FFS theory or energy density functional (EDF)
method starts from relations between the total energy, expressed as
a functional of the density, the single particle potential, the
effective interaction, and the quasiparticle density: Instead of
parameterizing the interaction and deriving the energy from it in a
certain approximation, an ansatz for the energy functional is
assumed from which the other quantities can be derived. The
interaction in the pp (hh) channel is obtained from the energy
functional by a different procedure than that in the ph channel, and
therefore these two interactions are different, which is of great
practical importance. This feature is shared with Migdal's theory of
finite Fermi systems. 

Otherwise, from a technical point of view, there is little difference
between the EDF method and the HF or, with pairing, HFB method.
The latter correspond to special choices of the energy functional.
On the other hand in the EDF method, to compute the density, a
Slater determinant or HFB-type state vector is employed.
Thus the differences which exist in principle, due to the necessary
approximations are of little practical importance.

In a recent paper~\cite{FTZ94} the EDF method has been applied to
magic nuclei $^{40,48}$Ca and $^{208}$Pb, investigating especially
the influence of the spin--orbit interaction on collective states.
Besides demonstrating the importance of including the two-body
spin orbit force for the consistency of the method in describing
the excited states, with the chosen set of parameters good results
on the ground state properties have been obtained.

Here an analogous EDF approach is applied to open-shell nuclei with
special attention to the pairing part of the functional. First
results obtained mainly in diagonal-pairing approximation (i.e. with
a HF--BCS formalism) have already been given in a few short
publications~\cite{FTTZ94,KTFZ95,FZ96} (the paper~\cite{KTFZ95}
gives the first application of the EDF approach to deformed nuclei
and contains the results of a HFB-type calculation for dysprosium
isotopes). In the present paper we give the results of more 
elaborate calculations based on the coordinate-space technique
developed in~\cite{BSTF87} which allows us, for a local pairing
field $\Delta(\vec r)$, to solve the Gor'kov equations in spherical
finite systems without any approximations~\cite{FTTZ98}. The latter
approach, even in the case of contact pairing force, corresponds, to
a good approximation as will be shown in the present paper, to a
full treatment of the HFB problem by applying the general
variational principle to the EDF with a fixed energy cutoff 
$\epsilon_{\rm c} > \epsilon_{\rm F}$ ($\epsilon_{\rm F}$ is the
Fermi energy). It is known from nuclear matter calculations that
$s$-wave pairing vanishes at or slightly above the equilibrium
density $\approx 0.16$~fm$^{-3}$; and so, in our local EDF approach,
the cutoff is chosen to be larger than the corresponding Fermi
energy $\epsilon_{0\rm F}\approx 37$~MeV. In actual calculations
we shall use $\epsilon_{\rm c} = 40$~MeV.

In Section~\ref{Method} we give some theoretical foundation of the
local EDF method for superfluid nuclei. In Section~\ref{Enorm} the
normal part of the energy-density functional with the parameters
used in the calculations is described. In Section~\ref{spair} the
pairing part of the functional is presented. There, a possible
extrapolation to uniform nuclear matter and the dilute limit with
weak and strong pairing is also considered. Section~\ref{disres} is
devoted to the analysis of specific numerical results for a few
isotopic chains of spherical nuclei and to the comparison of these 
results with experimental observations and other theoretical
approaches. The influence of the density-dependent pairing on the
evolution of the mean energies of multipole excitations is also
studied using the self-consistent sum rule approach. In 
Section~\ref{rpacorr}, the contribution of the ground
state (``phonon'') correlations to the charge radii is analyzed.
In Section~\ref{sumconc} the conclusions are summarized. Some 
important theoretical and technical aspects are presented in the 
Appendices A, B, and C. Appendix A contains a thorough formulation 
of the generalized variational principle for the systems with pairing
correlations which could be described by a local cutoff EDF. In 
Appendix B, the expression for the pairing energy is derived by using
the Green's function formalism. In Appendix C, we give a detailed 
description of the coordinate-space technique which we apply here to 
spherical nuclei. 

\Section{Method of the local energy functional with pairing}
\label{Method}
In this section we give some foundation of the local-energy-density
functional method which we apply here. Let us first briefly mention
a few approaches used to calculate the nuclear ground state energy
which is the expectation value of the Hamiltonian $\hat H$ (with a 
``bare'' NN-interaction) over the exact ground state vector
$\vert\Phi\rangle$:
\begin{equation}
E = \langle \Phi\vert{\hat H}\vert\Phi\rangle\,.
\end{equation}
In general, the vector $\vert\Phi\rangle$ is a functional of the
particle density $\varrho^{\rm p}$~\cite{HKohn} given by
\begin{equation}
\varrho^{\rm p}(\vec r_1) = {\rm Tr}_{s t}
\langle \Phi\vert\Psi^\dagger(1)\Psi(1)\vert\Phi\rangle\,,
\end{equation}
where $s$ and $t$ are the spin and isospin indices and
$\Psi^\dagger(1)$ and $\Psi(1)$ are the particle creation and
annihilation operators, respectively,
$(1)=(\{\vec r,s_z,\tau_3\}_1)$. Therefore, the energy is a
functional of $\varrho^{\rm p}$:
\[ E = E[\varrho^{\rm p}]\,. \]
It is a difficult problem to construct such a functional and
calculate the energy for many body systems. Formally, one can
introduce single-particle degrees of freedom and, assuming that
there is no pairing condensate, extract the single-particle motion
on the HF level by writing\footnote{Of course, in the nuclear case,
this expression with bare $\hat H$ is meaningless since
$\langle{\rm HF}\vert{\hat H}\vert{\rm HF}\rangle$
becomes infinite due to the repulsive core. It is understood then
that, for the first term in~(\ref{EHF}), the $\hat H$ is taken with
some kind of microscopically derived effective interaction, for
example with the Brueckner $G$-matrix (see, e.g., the discussion
in~\cite{RiSc80}). Because of inevitable approximations, such a
replacement would never lead to the vanishing of the second term,
$E_{\rm corr}$.}
\begin{equation}
E[\hat \rho] = \langle{\rm HF}\vert{\hat H}\vert{\rm HF}\rangle
+ E_{\rm corr}[\hat \rho]\,,
\label{EHF}
\end{equation}
where
$\vert{\rm HF}\rangle$ is the HF vacuum --- a Slater determinant of
single-particle wave functions $\phi_i$. The latter are defined
in a self-consistent manner by the equation
$\hat h \phi_i = \epsilon_i \phi_i$ in which the single-particle
Hamiltonian $\hat h$ is given by the variational derivative of the
energy functional with respect to the single-particle density matrix
$\hat \rho$: $\hat h = \delta E/\delta \hat \rho$. The second term,
$E_{\rm corr}$, is the dynamical correlation energy (short range
correlations, RPA correlations, many-particle correlations, parquet
diagrams, etc.). The density matrix $\hat \rho$ may be associated
with the Landau-Migdal quasiparticle density matrix defined as
\[\hat \rho (1,2) = \langle {\rm HF}\vert\psi^\dagger(2)
\psi(1)\vert{\rm HF}\rangle\,, \]
with $\psi^\dagger$  ($\psi$) the quasiparticle creation
(annihilation) operators. It is in turn a functional of the particle
density $\varrho^{\rm p}$:
\[\hat \rho = \hat \rho[\varrho^{\rm p}]\,. \]
In homogeneous infinite nuclear matter, because of equality of the
particle and quasiparticle numbers, the following relation between
the two densities holds:
\begin{equation}
\varrho^{\rm p} = \rho = {\rm Tr}_{st}\hat\rho(1,1)\,.
\end{equation}
In an inhomogeneous system, the local difference between the
densities could be attributed to the quasiparticle form
factor~\cite{KhSa82}. Using the effective radius approximation one
may write
\begin{equation}
\varrho^{\rm p}(\vec r)=(1+\textstyle{\frac 1 6} 
R^2_{\rm q}{\vec \nabla}^2)\rho(\vec r)\,,
\end{equation}
where $R_{\rm q}$ is the effective quasiparticle radius.

Hence the mean square nuclear radius for the particles may differ
from that for the quasiparticles:
\begin{equation}
\langle r^2\rangle^{\rm p} = \langle r^2\rangle + R^2_{\rm q}\,.
\end{equation}
Some additional effects may be related with a possible
change of the internal structure of nucleons in nuclear medium.
Such effects are not included explicitly in the ordinary approaches,
based either on a ``bare" Hamiltonian $H$ or on some kind of
effective interaction, in which the nucleons or quasiparticles are
considered as point-like objects\footnote{To get the charge density
distribution, and the charge radius of a nucleus, the point nucleon
densities are usually folded with the free nucleon charge form
factors. These may also change in the nuclear medium.}. Within the
EDF approach, however, tracing back to the Hohenberg--Kohn 
density-functional theory~\cite{HKohn}, if the functional were
known, its minimization would determine both the ground-state energy
and the correct particle density $\varrho^{\rm p}(\vec r)$. Anyway,
for the differences between mean square radii of nuclei, which is of
our main interest here, one has
\begin{equation}
\delta\langle r^2\rangle^{\rm p} = \delta\langle r^2\rangle\,.
\end{equation}

The common problem of any many-body theory is how to calculate
the dynamical correlation energy $E_{\rm corr}$ or how to
take it into account in some effective way.
In nuclear physics a few self-consistent methods are very popular:

\begin{itemize}
\item[--] The HF method with effective density-dependent
interaction (e.g., with zero-range Skyrme~\cite{VAB}
or finite-range Gogny force~\cite{DeGo80}). In this method
the energy is calculated as a ground state expectation value
of an effective Hamiltonian $H_{\rm eff}$ taken over a Slater
determinant,
\begin{equation}
E[\hat \rho] =
\langle{\rm HF}\vert{\hat H}_{\rm eff}\vert{\rm HF}\rangle\,.
\end{equation}
\end{itemize}

\begin{itemize}
\item[--]
The relativistic mean field model~\cite{Wal,WalS} based on
a phenomenological relativistic Lagrangian which includes mesonic
and nucleonic degrees of freedom. In most applications to the ground
states of finite nuclei, the relativistic Hartree formalism with
static meson fields, and with no-sea approximation for nucleons, has
been used (see~\cite{Ring96} and references therein).
\end{itemize}

\begin{itemize}
\item[--]
The quasiparticle Lagrangian method (QLM) which is an
extension~\cite{KhSa82} of the Landau-Migdal quasiparticle concept.
In this method an effective local quasiparticle Lagrangian is
constructed taking into account the first-order energy-dependence of
the nucleon mass operator and the requirements imposed by
self-consistency relations~\cite{FK73}. The QLM can be reformulated
in terms of an effective Hamiltonian~\cite{KSZ87} in which the
(linear) energy dependence of the nucleon mass operator is
completely hidden so that this approach becomes equivalent to the EDF
method (see also the discussion of this point in~\cite{STF88}).
\end{itemize}

\begin{itemize}
\item[--] The method of the effective quasiparticle local energy
density functional (which is referred to as EDF here). The
possibility to use this method is based on the existence theorem of
Hohenberg and Kohn~\cite{HKohn}. With the Kohn-Sham quasiparticle
formalism~\cite{KSham}, this theorem allows one to write the nuclear
EDF in the form
\begin{equation}
E[\hat \rho] = {\rm Tr}(t\hat\rho) + E_{\rm int}[\rho]\,,
\end{equation}
where the first term is taken with the free kinetic energy operator
$t = p^2 /2m$ ($m$ is the bare nucleon mass). The ground state
energy is then determined by making this EDF stationary with
respect to infinitesimal variations of the single-particle wave
functions belonging to the class of the Slater determinants
from which the density matrix $\hat \rho$ (and the density $\rho$)
can be calculated self-consistently. This method is flexible in the
sense that the EDF does not (and should not) expose the same symmetry
properties for the effective force, which is the second variational
derivative of the EDF with respect to $\rho$, as the underlying
bare NN interaction or effective Skyrme-type forces. Particularly,
the force in the ph channel obtained from the EDF does not come out
to be antisymmetrized, and the effective force in the pp channel
needed for describing the nuclear pairing properties has to be
derived in a different way. As a result, the matrix elements of the
pp force, although they should be antisymmetrized, have no direct
connection, say by means of simple angular momentum recoupling like
the Pandya transformation, with those of the ph
force\footnote{This was explicitly demonstrated, already on the level
of the ``ladder'' approximation, in~\cite{KMS76}. We note the term
``force'', used to name the second variational derivative of the EDF
with respect to the density matrix, may be somewhat confusing in this
context. In fact, this derivative, for infinite systems, has strict
correspondence to the quasiparticle interaction amplitude on the
Fermi surface introduced in the Landau theory of Fermi
liquids~\cite{Land58}. The total quasiparticle scattering amplitude
is, of course, antisymmetric due to the Pauli principle resulting in
sum rules for the Landau parameters. These parameters should satisfy
the Pomeranchuk stability conditions~\cite{Chuk58} in order to
prevent the collapse of the system with respect to the low-energy ph
excitations and provide the ground state energy to be a minimum
rather than simply stationary. For finite systems, this means that
all the RPA solutions obtained with the self-consistent ph
interaction, taken as the second variational derivative of EDF at
the stationary point, should have real frequencies $\omega$ such that
$\omega^2>0$.}. This is, as discussed in the Introduction, in accord
with the philosophy of Migdal theory of finite Fermi
systems~\cite{Migdal}. (For an example of the phenomenological
nuclear EDF see~\cite{STF88,FTZ94}).
\end{itemize}

In most of the applications of the conventional functionals to finite 
nuclei the pairing correlations are introduced either at the BCS or
the HFB level (see, e.g., Ref.~\cite{Patyk99} and references therein). 
When the contact pairing interaction is used, the pairing part of
the functional is usually evaluated within a truncated space of the
quasiparticle levels imposing an energy cutoff which is often chosen 
with some freedom but not too far from the Fermi surface. A more
rigorous approach should be based on the general variational principle
for the cutoff functional which uses the same quasiparticle basis 
both in the ph and pp channel.    

We proceed as follows.
In nuclei with pairing correlations one works with approximate
state vectors which are not eigenstates of the particle number.
One assumes nonzero anomalous expectation values, ${\hat \nu}(1,2) =
\langle \Phi\vert\Psi(1)\Psi(2)\vert\Phi\rangle \neq 0$, and the
energy of a superfluid nucleus may be given by a functional of the
generalized density matrix $\widehat R$ containing both normal,
$\hat \rho$, and anomalous, $\hat \nu$, components (see Appendix A).
In analogy to eq.~(\ref{EHF}), the energy of a superfluid nucleus
can be written as
\begin{equation}
E[\widehat R] = \langle{\rm HFB}\vert{\widehat H}\vert{\rm HFB}
\rangle + E_{\rm corr}[\widehat R]\,  \label{EHFBC}
\end{equation}
with $\vert{\rm HFB}\rangle$ the HFB quasiparticle vacuum and
$E_{\rm corr}$ the dynamical correlation energy.

The generalized density matrix $\widehat R$, in turn, is
a functional of $\varrho^{\rm p}$ and therefore
\begin{equation}
E = E[\widehat R[\varrho^{\rm p}]] = E[\varrho^{\rm p}] \,.
\end{equation}

The method we follow here is based on an effective quasiparticle
energy functional of the generalized density matrix:
\begin{equation}
E[\widehat R] = E_{\rm kin}[{\hat\rho}] +
E_{\rm int}[{\hat\rho},{\hat\nu}]\,,
\label{EHFB}
\end{equation}
where
$ E_{\rm kin}[{\hat\rho}] = {\rm Tr}(t\hat \rho)\,, $ and
$ E_{\rm int}[{\hat\rho},{\hat\nu}]
= E_{\rm int(normal)}[{\hat\rho}] +
E_{\rm anomal}[{\hat\rho},{\hat\nu}]\,. $ The anomalous
energy $E_{\rm anomal}$ is chosen such that it vanishes in the
limit $\nu \to 0$.

It should be emphasized that the weak-pairing
approximation $\vert \bar \Delta \vert \ll \epsilon_{\rm F}$ is
assumed throughout this paper. That means we need to retain only
the first-order term $\sim \nu^2$ in the $\nu$-dependent, anomalous
part of the energy functional. One may then write
\begin{equation}
E_{\rm anomal}[{\hat \rho},{\hat \nu}]
= \quart \left({\hat \nu}^\dagger
{\hat{\mathcal F}}^{\rm pp}_{\rm a}[{\hat\rho}]{\hat\nu}\right)\,,
\label{Eanomal}
\end{equation}
where ${\hat{\mathcal F}}^{\rm pp}_{\rm a}$ is an antisymmetrized
effective interaction in the pp channel and the round brackets
(\dots) imply integration and summation over all variables.

To calculate the ground state properties, one can now use
the general variational principle with two constraints,
\begin{equation}
\langle{\rm HFB}\vert{\hat N(\mu)}\vert{\rm HFB}\rangle
\equiv N(\mu) = N\,,
\label{consN}
\end{equation}
\begin{equation}
{\widehat R}^2 = \widehat R\,,
\label{R2}
\end{equation}
leading to the variational functional of the form
\begin{equation}
I[\widehat R] = E[\widehat R] - \mu N(\mu)
- {\rm Tr}{\hat \Lambda}(\widehat R - {\widehat R}^2)\,,
\label{VFUNC}
\end{equation}
where $N$ is the particle number, $\mu$ the chemical potential,
and ${\hat \Lambda}$ the matrix of Lagrange parameters (see, e.g.
Ref.~\cite{RiSc80}, and Appendix A for more details).

The major difficulty in implementing this general variational
principle in practice is connected with the anomalous part of the
energy functional. The anomalous energy~(\ref{Eanomal})
can be calculated if one knows a solution of the gap
equation for the pairing field $\hat \Delta$ and the anomalous
density matrix $\hat \nu$. In general, the gap equation,
\begin{equation}
\hat \Delta = \half {\hat {\mathcal F}}^{\rm pp}_{\rm a}
{\hat \nu}\,,
\end{equation}
is nonlocal and its solution (starting, for example, from a realistic
bare NN interaction~\cite{CCS86,BCLL90,KKC96,EEHO96}), even for
uniform nuclear matter, poses serious problems. First attempts were
made very recently to construct an effective pairing interaction for
semi-infinite nuclear matter~\cite{BLSZ95} and for finite
nuclei~\cite{DBL95} with an approximate version of the Brueckner
approach. Those studies are in an initial stage, and so far they do
not give any guidance how to choose an effective pairing interaction,
in particular its density dependence, that could be used in nuclear
structure calculations. It is assumed that a simple universal
effective interaction in the pp-channel can be invented to correctly
describe the nuclear pairing properties. Our intention is to show
that, to a good approximation in the case of weak pairing
$\vert \bar \Delta \vert \ll \epsilon_{\rm F}$, the EDF method and
the general variational principle can be used with an effective
density-dependent contact pp-interaction.

We proceed with the formal development by using the Green's function
formalism as described in detail in Appendices A and B.
Here we give only a brief account of the main issues.

We introduce an arbitrary cutoff $\epsilon_{\rm c}$ in the energy
space, but such that $\epsilon_{\rm c} > \epsilon_{\rm F}$,
and split the generalized density matrix into two parts,
\begin{equation}
\widehat R = {\widehat R}_{\rm c} + \delta_{\rm c} {\widehat R}\,,
\end{equation}
where $\delta_{\rm c} {\widehat R}$ is related to the
integration over energies
$\vert \epsilon \vert > \epsilon_{\rm c}$.

The gap equation is renormalized to yield
\begin{equation}
\hat\Delta=\half\hat{\mathcal F}^\xi_{\rm a}{\hat\nu}_{\rm c}\,,
\end{equation}
where ${\hat\nu}_{\rm c}$ is the cutoff anomalous density matrix
and $\hat{\mathcal F}^\xi_{\rm a}$ is the effective
antisymmetrized pp-interaction in which the contribution coming
from the energy region $\vert \epsilon \vert > \epsilon_{\rm c}$
far from the Fermi surface is included by renormalization
(see Appendix A).

For homogeneous infinite matter with weak pairing
($\vert\bar\Delta\vert\ll\epsilon_{\rm F}$), it is shown that
the variational functional $E-\mu N$ does not change in
first order in $\vert\bar\Delta\vert^2/\epsilon_{\rm F}$
upon variation with respect
to $\delta_{\rm c} {\widehat R}$. The total energy of the system
and the chemical potential also remain the same in first order
in $\vert\bar\Delta\vert^2/\epsilon_{\rm F}$
if one imposes the particle number constraint~(\ref{consN}) for
the cutoff functional. To a good approximation, as discussed in
Appendix A, this should also be valid for finite (heavy) nuclei.

Such an outcome may be understood by noting that the major pairing
effects, if $\vert \bar \Delta \vert \ll \epsilon_{\rm F}$ which is
generally the case in nuclear matter and in finite nuclei, are
developed near the Fermi surface. In this connection we mention the
well-known fact that the pairing energy, in the BCS approximation,
is defined by a sum concentrated near the Fermi surface (see, for
example, Ref.~\cite{RiSc80}):
\begin{equation}
E_{\rm pair}\approx -\half {\bar n}{\bar\Delta}^2\,,  \label{EBCS}
\end{equation}
with $\bar n$ the average level density and $\bar \Delta$
the average energy gap in the vicinity of the Fermi surface.  In
infinite matter, this corresponds to the pairing energy per particle
(see Appendix B):
\begin{equation}
\frac{E_{\rm pair}}{N} = - \frac 3 8
\frac{\Delta^2(\vec p_{\rm F})}{\epsilon_{\rm F}}\,.
\end{equation}
It follows that, with the cutoff functional, this leading pairing
contribution to the energy of the system is exactly accounted for.

Thus we find that nuclear ground state properties can be described 
by applying the general variational principle to minimize the cutoff
functional which has exactly the same form as in eq.~(\ref{VFUNC}) 
with the constraint~(\ref{consN}) but with $\widehat R$ replaced by 
$\widehat R_{\rm c}$:
\begin{equation}
I_{\rm c}[\widehat R] = E_{\rm kin}[{\hat \rho}_{\rm c}] + 
E^{\rm c}_{\rm int}[\rho_{\rm c},\nu_{\rm c}]
- \mu_{\rm c} N_{\rm c}(\mu_{\rm c}) - {\hat \Lambda}(\widehat R 
- {\widehat R}^2)\,,  \label{FC}
\end{equation}
\begin{equation}
\langle{\rm HFB}\vert{\hat N_{\rm c}(\mu_{\rm c})}
\vert{\rm HFB}\rangle
\equiv N_{\rm c}(\mu_{\rm c}) = N\,.
\label{MuC}
\end{equation}
Here $E_{\rm kin}[{\hat \rho}_{\rm c}] = 
{\rm Tr}(t{{\hat\rho}_{\rm c}})\,$ and
$E^{\rm c}_{\rm int}[\rho_{\rm c},\nu_{\rm c}] = 
E^{\rm c}_{\rm int(normal)}[{\hat\rho}_{\rm c}]
+ E^{\rm c}_{\rm anomal}[{\hat\rho}_{\rm c},{\hat\nu}_{\rm c}]\,$ 
with
\begin{equation}
E^{\rm c}_{\rm anomal}[{\hat\rho_{\rm c}},{\hat\nu}_{\rm c}]
=\quart ({\hat \nu}^\dagger_{\rm c}
{\hat{\mathcal F}}^\xi_{\rm a}[{\hat\rho}_{\rm c}]
{\hat\nu}_{\rm c})\,.
\end{equation}

The above consideration has been carried out without any 
pre-assumptions concerning the density functional 
$E_{\rm int}[\widehat R]\approx 
E^{\rm c}_{\rm int}[\rho_{\rm c},\nu_{\rm c}]\,$. 
Now, recalling the Hohenberg-Kohn theorem~\cite{HKohn}, we specify 
that the density functional can be chosen to be of a local form, 
i.e. dependent on the normal and anomalous local real 
densities $\rho(\vec r, \tau)$ and $\nu(\vec r, \tau)$ defined in 
Appendix A by~(\ref{rhoc}) and~(\ref{nuc}), respectively. 
Then $E^{\rm c}_{\rm normal}$ is a functional of the normal 
densities (isoscalar, isovector, spin--orbital, etc.), and 
$E^{\rm c}_{\rm anomal}$ is a functional of the normal and 
anomalous densities, the latter acquiring the simple form
\begin{equation}
E^{\rm c}_{\rm anomal}[\rho_{\rm c},\nu_{\rm c}] = 
\sum_{\tau=\rm n,p}
\int\d {\vec r}\,\nu^*_{\rm c}(\vec r,\tau)
{\mathcal F}^\xi(\vec r,\tau;[\rho_{\rm c}]) \nu_{\rm c}
(\vec r,\tau)\,.  \label{Eano}
\end{equation}
The pairing potential is defined by
\begin{equation}
\Delta(\vec r,\tau) = \frac 12 
\frac {\delta E^{\rm c}_{\rm anomal}
[\rho_{\rm c},\nu_{\rm c}]}{\delta \nu_{\rm c}(\vec r,\tau)}\,.
\end{equation}
That corresponds to the gap equation which takes on now a very
simple multiplicative form:
\begin{equation}
\Delta(\vec r,\tau) = {\mathcal F}^\xi(\vec r,\tau;[\rho_{\rm c}])
\nu_{\rm c}(\vec r,\tau)\,.  \label{gapeq}
\end{equation}

The normal mean-field potential is given by
\begin{equation}
U(\vec r,\tau) = 
\frac{\delta E^{\rm c}_{\rm int}[\rho_{\rm c},\nu_{\rm c}]}
{\delta \rho_{\rm c}(\vec r,\tau)}\,.
\end{equation}
Due to the density dependence of ${\mathcal F}^\xi$ it includes 
also a contribution arising from the variation of the pairing 
interaction energy:
\begin{equation}
U_{\rm pair}(\vec r,\tau) = 
\frac{\delta E^{\rm c}_{\rm anomal}[\rho_{\rm c},\nu_{\rm c}]}
{\delta \rho_{\rm c}(\vec r,\tau)}\,.  \label{UPA}
\end{equation}

We emphasize that having found the gap function $\Delta(\vec r)$,
and the mean-field potential $U(\vec r)$, the Gor'kov 
equations~\cite{Gor} may be solved, for spherical nuclei, exactly 
by using the coordinate-space technique (see 
Refs.~\cite{BSTF87,STF88,FTTZ98} and Appendix C). The
generalized Green's function obtained this way can be integrated 
over energy up to $\epsilon_{\rm c}$ to yield both the normal and 
anomalous densities $\rho_{\rm c}$ and $\nu_ c$ which are used then 
to compute the energy of the system. This corresponds to a full HFB 
treatment of the nuclear ground state properties. It remains, of 
course, an art to find a ``good" local functional, in particular its
anomalous part.

\Section{The normal part of the functional}
\label{Enorm}
In this and the next section we describe in some detail the
energy-density functional which we use in the present calculations.
We shall omit in the following the cutoff index c for simplicity.
The functional is local in the sense that it contains only normal 
and anomalous densities, not the density matrix.
The interaction part of the energy density is
\begin{equation}
\varepsilon_{\rm{int}}=\varepsilon_{\rm{main}}+
\varepsilon_{\rm{Coul}} + \varepsilon_{sl} +
\varepsilon_{ss} + \varepsilon_{\rm anomal}\,.
\label{eint}
\end{equation}
The terms $\varepsilon_{\rm main}$ and $\varepsilon_{\rm Coul}$
have been described in Refs.~\cite{FTZ94,HSFT96}, where also the 
connection of the parameters involved with the characteristics of 
nuclear matter has been given. Two different versions of the 
spin--orbit part $\varepsilon_{sl}$ have been used: the one of 
Ref.~\cite{FTZ94}, and also, after~\cite{STF88}, the 
variant\footnote{As a rule, the convention $\hbar=1$ is used 
throughout the paper but $\hbar$ will appear in some expressions.
Hopefully this should not lead to any confusion.}
\begin{eqnarray}
\varepsilon_{sl}(\vec r)&=& {\frac 1 2} C_0 r_0^2\sum_{i,k={\rm p,n}}
\left\{
\kappa^{ik}[\nabla\rho^i \cdot \sum_{ss^\prime}
\langle\psi^{\dagger}(\vec r,s)[\vec p\times
       \vec\sigma]\psi(\vec r,s^\prime)\rangle^k \right.\nonumber \\
& + & \nabla\rho^k\cdot\sum_{ss^\prime}
\langle\psi^{\dagger}(\vec r,s)[\vec p\times
       \vec\sigma]\psi(\vec r,s^\prime)\rangle^i]
\nonumber\\
& + &\left. g_1^{ik}\sum_{\alpha,\beta}\sum_{ss^\prime}
\langle\psi^{\dagger}(\vec r,s)\sigma_\alpha
                  p_\beta\psi(\vec r,s^\prime)\rangle^i
\sum_{ss^\prime}\langle\psi^{\dagger}(\vec r,s)\sigma_\alpha
                  p_\beta\psi(\vec r,s^\prime)\rangle^k\right\}
\label{sodef}
\end{eqnarray}
where $\psi^\dagger$ and $\psi$ are
field operators, the brackets denote the ground state expectation
value, and a superscript $k$ indicates that a projection on 
particles of type $k$ is performed.

As given by~(\ref{sodef}), $\varepsilon_{sl}$
corresponds to the two-body spin--orbit interaction
\begin{equation}
{\mathcal F}^\omega_{sl} = 
C_0r_0^2(\kappa+\kappa^\prime\vec\tau_1\!\cdot\!\vec\tau_2)
[\nabla_1\delta(\vec r_1-\vec r_2)
\times(\vec p_1-\vec p_2)]\!\cdot\! (\vec\sigma_1+\vec\sigma_2)\,,
\label{soforce}
\end{equation}
and to the first--order velocity harmonic of the spin--dependent 
part of the Landau interaction amplitude
\begin{equation}
{\mathcal F}^\omega_1 = 
C_0r_0^2(g_1+g_1^\prime\vec\tau_1\!\cdot\!\vec\tau_2)
\delta(\vec r_1-\vec r_2)
(\vec\sigma_1\!\cdot\!\vec\sigma_2)(\vec p_1\!\cdot\!
\vec p_2)\,,  \label{g1force}
\end{equation}
which has to be symmetrized in such a way that the momentum 
operators ${\vec p}_{1,2}$ never act on the delta-function 
$\delta(\vec r_1-\vec r_2)$.

In these expressions, the multiplier $r_0^2$ has been introduced 
to make the parameters $\kappa$ and $g_1$ dimensionless.  It is 
given by $r^2_0=(3/8\pi\rho_0)^{2/3}$ where $\rho_0$ is the 
equilibrium density of one kind of particles in symmetric nuclear 
matter. The factor $C_0$ is the inverse density of states at the 
Fermi energy, $\epsilon_{0\rm F}=k_{0\rm F}^2/2m^*$, in saturated 
nuclear matter. It is given by $C_0=2\epsilon_{0\rm F}/3\rho_0=
\pi^2/k_{0\rm F}m^*$.

Defining spin--orbit densities by
\begin{eqnarray}
\rho^k_{sl}(\vec r)=\sum_{ss^\prime}
\langle\psi^{\dagger}(\vec r,s)({\vec \sigma}\!\cdot\!
             {\vec l})\psi(\vec r,s^\prime)\rangle^k\,, \nonumber
\end{eqnarray}
then in spherically symmetric nuclei, the spin--orbit energy 
density~(\ref{sodef}) can be simplified to~\cite{STF88}
\begin{equation}
\varepsilon_{sl}=C_0r_0^2\sum_{i,k={\rm n,p}}
\left ( {\frac 1 r} \rho_{sl}^i\kappa^{ik}{\frac{\partial\rho^k}
{\partial r}} +{\frac{1}{4r^2}} \rho_{sl}^i g_1^{ik}
\rho_{sl}^k\right )\,.
\label{sosph}
\end{equation}
In deformed nuclei, the expression for $\varepsilon_{sl}$ cannot be
written in such a simple form. In the axially symmetric case the
derivation can be performed in cylindrical coordinates; the resulting
formulae are given in~\cite{KTFZ95}.

We just mention some features of those parts of 
$\varepsilon_{\rm int}$ which are not given here in detail: The main 
contribution $\varepsilon_{\rm main}$ consists of a volume part and 
of a surface term both containing simple fractional-linear functions 
of the normal isoscalar density $\rho_+=\rho_{\rm n}+\rho_{\rm p}$. 
This leads to the effective density-dependent forces in the 
scalar-isoscalar and scalar-isovector channels (see 
e.g.~Refs.~\cite{STF88,FTZ94,HSFT96} for more details).
There is no momentum dependence in these parts, therefore we have 
the effective mass $m^*$ equal to the bare mass $m$. The surface 
term involves a finite, but small Yukawa range parameter 
$R=0.35$~fm; in the limit $R\rightarrow 0$ it could be written with 
the help of the Laplace operator acting on functions of the density.
It vanishes for constant density. The Coulomb part 
$\varepsilon_{\rm Coul}$ is taken in the usual form~\cite{VAB}
with an exchange contribution in the Slater approximation.

Note that in the more complicated parts of the force like those of
eqs.~(\ref{soforce}) and (\ref{g1force}) we did not introduce
any density dependence. In general, all of the constants $\kappa$,
$g_1$, and others to follow below could be density dependent as 
well. Until now there are no experimental data which would
make it necessary to introduce such a dependence for the laboratory
nuclei within or not too far from the stability valley. However,
going to the nucleon drip lines and beyond, a more complicated
density-dependent spin--orbit force might be of relevance.
This is indicated by the relativistic mean field models for very
neutron-rich nuclei~\cite{DHNS94} and by the exact Monte Carlo
methods for small pure neutron drops~\cite{PSCP96}. Both approaches
predict a much smaller (by a factor of 2--3) spin--orbit potential
than the usual Skyrme-type ansatz. 
The role played by the density dependence of the effective
spin--orbit interaction is left to be studied in future work.

The spin--spin term $\varepsilon_{ss}$ has hitherto been omitted
since its expectation value in spin-zero nuclei vanishes. We include
it here for completeness; it might become important for magnetic
excitations or polarization. A possible simple form of it is
\begin{equation}
\varepsilon_{ss}(\vec r)=\half C_0\sum_{i,k={\rm n,p}}g_0^{ik}
\langle\vec\sigma\rangle^i\!\cdot\!\langle\vec\sigma\rangle^k\,.
\label{ess} 
\end{equation}
Note that the strength $\propto g_0^{ik}$ can be considered as the
zeroth term in a decomposition of the spin--spin interaction
amplitude at the Fermi surface in Legendre polynomials of the
scattering angle. The next (first) term, proportional to $g_1^{ik}$,
has been included in the spin--orbit part of the functional; 
however, the normalization chosen in eq.~(\ref{sodef}) is different 
from that of Ref.~\cite{Migdal}.

As long as there is no static spin polarization 
($\langle\vec\sigma\rangle=0$) in the ground states 
of even nuclei and our interest is
only in electric and mass quantities we can continue to neglect
$\varepsilon_{ss}$. One delicate point should be addressed in this
connection. Usually, when deriving the Hartree-Fock equations, one
starts with a Hamiltonian containing a well behaved and properly
antisymmetrized interaction. In the case of contact forces,
antisymmetry prevents self-interaction of the particles, so there is
no need for precautions against that. Going over to an effective
force which, as in our case, is not antisymmetrized in the ph
channel, the self-interaction is not automatically excluded. Dealing
with a not-too-small number of particles, by fitting to the
experimental data the self-interaction is compensated (or accounted
for) in the average by the choice of the parameters.

This does not hold for the spin--spin part of the force. For an odd
nucleus, consisting of a core with $J=0$ plus one particle, there is
a nonvanishing spin density, and from eq.~(\ref{ess}) there is also
a contribution to the total energy which to lowest order is just the
self-interaction of the odd particle and should therefore be
excluded. On the other hand, considering the contribution of this
part of the functional to the single particle potential when
determining the wave function of the core, will give the polarization
of the core which is a physical quantity. --- Usually, on the level
of Hartree-Fock type calculations the spin--spin interaction is left
out; it may be considered in a second step by perturbation methods.

The basic set of functional parameters in the notations of
Refs.~\cite{STF88,FTZ94,BFKZ96} used in the present calculations is
the set DF3:
\begin{equation}
\left.  \begin{array}{ll}
a^{\rm v}_+ = -6.422,   &    a^{\rm v}_-=5.417,       \\
h^{\rm v}_{1+} = 0.163,\,\,h^{\rm v}_{2+} = 0.724,\qquad &
h^{\rm v}_{1-}=0,\,\,h^{\rm v}_{2-}=3.0, \\
a^{\rm s}_+=-11.1,\,\,h^{\rm s}_+ = 0.31, &
a^{\rm s}_-=-4.00,\,\,h^{\rm s}_- = 0,                \\
\kappa^{\rm pp} = 0.285,\,\,\kappa^{\rm pn} = 0.135,&
g_1^{\rm pp} = -g_1^{\rm pn} = -0.12,                 \\
r_0 = 1.147~{\rm fm}, & R = 0.35~{\rm fm}\,.            \\
\end{array}\right\}
\label{normset}
\end{equation}
This set set has been specifically deduced in~\cite{BFKZ96} to
reproduce not only the already known ground-state properties of magic
nuclei but also the very recent experimental data~\cite{MNPFS95} on
the single-particle energies near the ``magic cross'' at $^{132}$Sn.
The set~(\ref{normset}) corresponds to the following nuclear matter
characteristics at saturation: compression modulus $K_0 = 200$~MeV,
chemical potential (binding energy per nucleon) $\mu_0 = -16.05$~MeV,
asymmetry energy parameter $\beta_0 = 28.7$~MeV, saturation density
$2\rho_0 = 0.1582$~fm$^{-3}$ ($k_{0 \rm F} = 1.328$~fm$^{-1}$,
$\epsilon_{0\rm F} = 36.57$~MeV,
$C_0=\pi^2/k_{0\rm F}m=308.2$~MeV$\cdot$fm$^3$). As in
Ref.~\cite{STF88}, the functional contains zero-range isoscalar
spin--orbit interaction $\propto \kappa$ and velocity-dependent
spin-isospin interaction $\propto g_1^\prime$ (i.e. the first
Landau-Migdal harmonic in the $\sigma\tau$ channel). This is in
contrast to~\cite{FTZ94} where a finite range spin--orbit force had
been used.

With the above parameter set of the normal part of the density
functional the ground states of magic nuclei are described fairly
well. For example, the calculated rms charge radii
$\langle r^2_{\rm ch}\rangle^{1/2}$ of $^{40}$Ca, $^{48}$Ca and
$^{208}$Pb are 3.480~fm, 3.478~fm, and 5.500~fm, respectively, which
agree nicely with experimental values deduced very 
recently~\cite{Fri95} from the combined analysis of optical, muonic
and elastic electron scattering data:
$\langle r^2_{\rm ch}\rangle^{1/2} = 3.4767(8)$~fm  for $^{40}$Ca,
3.4736(8)~fm for $^{48}$Ca, and 5.5013(7)~fm for $^{208}$Pb.
The predictions for some other double magic nuclei are:
$\langle r^2_{\rm ch}\rangle^{1/2} = 3.721$~fm ($^{56}$Ni), 3.951~fm
($^{78}$Ni), 4.453~fm ($^{100}$Sn), and 4.705~fm ($^{132}$Sn).

\Section{The pairing part of the functional}
\label{spair}
The pairing energy density $\varepsilon_{\rm anomal}$ in
eq.~(\ref{eint}) is chosen in the form prescribed by~(\ref{Eano}):
\begin{equation}
\varepsilon_{\rm anomal}(\vec r) = \sum_{\tau=n,p}
{\mathcal F}^{\xi,\tau \tau}(\vec r;[\rho])
\vert \nu_\tau (\vec r)\vert^2\,.  \label{epair}
\end{equation}
As discussed in Refs.~\cite{ZRS87,ReZa88,ReZa89} and mentioned in the
introduction, effective three- or more-body forces are indispensable
if one wants to reproduce isotope shifts in charge radii which means
that $\varepsilon_{\rm anomal}$ must depend on the normal 
density~\cite{FTTZ94,FZ96}, and therefore we introduce
\begin{equation}
{\mathcal F}^{\xi,\,\rm nn} = {\mathcal F}^{\xi,\,\rm pp} 
= C_0f^{\xi}(x)\,,  \label{Fxi}
\end{equation}
The dimensionless strength $f^\xi$ is supposed to be, within the EDF
approach, a local functional of the isoscalar dimensionless density
$x = (\rho_{\rm n}+\rho_{\rm p})/2\rho_0$, with $\rho_{\rm n(p)}$
the neutron (proton) density; $C_0$ and $\rho_0$ are defined in the
previous section. We shall use the parametrization of $f^\xi$
suggested in Ref.~\cite{FZ96}:
\begin{equation}
f^\xi(x)=
f^\xi(x(\vec r))=f^\xi_{\rm ex} + h^\xi x^q(\vec r)
+f^\xi_\nabla r_0^2\left({\vec \nabla}x(\vec r)\right)^2\,.
\label{fxigrad}
\end{equation}
The parameter $f^\xi_{\rm ex}$ is negative and simulates an
attraction in the pp channel in the far nuclear exterior,
$h^\xi$ and $f^\xi_\nabla$ are taken to be positive~\cite{FZ96}.
The exponent $q$ in the second term is introduced to have a more
flexible parametrization. A repulsive short-range part of the
$G$-matrix effective interaction with a $x^{2/3}$ dependence has
been discussed, e.g., by Bethe many years ago~\cite{Bethe} while
developing a Thomas-Fermi theory for large finite nuclei. The choice
$q={\textstyle {\frac 2 3}}$ seems thus to be reasonable, and indeed
our calculations showed that some improvements in reproducing the
isotopic shifts may be achieved with this choice compared to the
linear case $q=1$. We shall use $q={\textstyle {\frac 2 3}}$ in the
present paper. The three-body force of Refs.~\cite{ReZa88,ReZa89}
corresponds to a linear dependence of ${\mathcal F}^\xi$ on $\rho$
($q=1$, $f^\xi_\nabla=0$). As shown in Ref.~\cite{FZ96}, the
self-consistent EDF (HF+BCS) calculations with the density-gradient
term $\propto f^\xi_\nabla$ in pairing force provide desirable size
of isotopic shifts and right order of odd-even staggering observed
in lead isotopes, the coupling of the proton mean field with neutron
pairing being the major effect in this case. The variation of the
anomalous energy~(\ref{epair}) incorporating the above force with
respect to normal densities gives the following contribution to the
central mean-field potential:
\begin{equation}
U^{{\rm n},{\rm p}}_{\rm pair} = \frac{C_0}{4 \rho_0}
\left[qh^\xi x^{q-1}-2r_0^2f^\xi_\nabla \left(\Delta x +
{\vec \nabla}x\cdot{\vec \nabla}\right)\right]
\left(\vert \nu_{\rm n}\vert^2 + \vert \nu_{\rm p}\vert^2\right)\,.
\label{Upairgrad}
\end{equation}
As will be shown in the next section, different choices of the 
parameters of
the particle--particle force~(\ref{fxigrad}) are possible to
reasonably describe the neutron separation energies and isotopic
shifts in charge radii. In particular, the following sets are
deduced for the lead isotopes:
\begin{equation}
\left. \begin{array}{lllr}
f^\xi_{\rm ex}=-0.56,\;\;&h^\xi=0,\;\;\; &f^\xi_\nabla=0  &\qquad 
(\rm a)\\
f^\xi_{\rm ex}=-1.20,\;\;&h^\xi=0.56,\;\;&f^\xi_\nabla=2.4&\qquad 
(\rm b)\\
f^\xi_{\rm ex}=-1.60,\;\;&h^\xi=1.10,\;\;&f^\xi_\nabla=2.0&\qquad 
(\rm c)\\
f^\xi_{\rm ex}=-1.79,\;\;&h^\xi=1.36,\;\;&f^\xi_\nabla=2.0&\qquad 
(\rm d)\\
f^\xi_{\rm ex}=-2.00,\;\;&h^\xi=1.62,\;\;&f^\xi_\nabla=2.0&\qquad 
(\rm e)\\
f^\xi_{\rm ex}=-2.40,\;\;&h^\xi=2.16,\;\;&f^\xi_\nabla=2.0&\qquad 
(\rm f)\\ \end{array}\,
\right\}\,.
\label{pppar}
\end{equation}
Let us, in the rest of this section, study the behavior of $\Delta$
as a function of density $\rho=2\rho_0x$ (or the Fermi momentum
$k_{\rm F}=(3\pi^2\rho/2)^{1/3} \equiv k_{0\rm F}x^{1/3}$) in
symmetric ($\rho_{\rm n}=\rho_{\rm p}$) uniform infinite nuclear
matter with pairing interaction of eqs.~(\ref{Fxi}), (\ref{fxigrad}),
and also discuss how the pairing affects the equation of state (EOS)
and the position of the equilibrium point. The density-gradient term
$\propto f^\xi_\nabla$ vanishes in this case, thus only the terms
with parameters $f^\xi_{\rm ex}$ and $h^\xi$ are left. For uniform
matter the gap equation~(\ref{gapeq}) reduces to
\begin{equation}
\Delta(x)=-{\mathcal F}^{\xi}(x)
\int\nolimits_{k\le k_{\rm c}}\,\frac{\d \vec k}{(2\pi)^3}
\frac{\Delta(x)}{2\sqrt{(\epsilon_k-\epsilon_{\rm F}(x))^2
+\Delta^2(x)}}\,,
\label{gapinf}
\end{equation}
where\footnote{Note that, in our approach, the upper limit in the
truncated momentum or energy space depends on density since 
$\epsilon_{\rm c}$ is chosen to be measured from the Fermi level
position determined by the chemical potential $\mu$ (see 
Appendix~C). At any given density, $\mu$ is introduced as a Lagrange
multiplier and therefore, in varying the cutoff functional, the 
phase space is kept fixed.} $k_{\rm c} = \sqrt{2m(\epsilon_{\rm F}
+\epsilon_{\rm c})}/\hbar$ and $\epsilon_k=\hbar^2k^2/2m$.
Canceling $\Delta(x)$ on both sides of this equation, to find a
nontrivial solution $\Delta \ne 0$, one gets 
\begin{equation} 
\frac 1 4 x^{1/3}f^{\xi}(x)\int_0^{t_{\rm c}(x)} \d t 
\frac{\sqrt t}{\sqrt{(t-1)^2+\delta^2(x)}}=-1\,,
\label{wpapp}
\end{equation}
where $\delta(x) = \Delta(x)/\epsilon_{\rm F}(x)$ and
$t_{\rm c}(x)=1+\epsilon_{\rm c}/\epsilon_{\rm F}(x)$
with $\epsilon_{\rm c}$ the energy cutoff measured from
the Fermi energy $\epsilon_{\rm F} = \epsilon_{0\rm F}x^{2/3}$.
The solution of the last equation in the weak pairing 
approximation, for $f^\xi(x)<0$, is given by (see, e.g., 
\cite{BSTF87,Khod97}):
\begin{equation}
\Delta(x)=8\epsilon_{0\rm F}\,x^{2/3}
\sqrt{\frac{s(x)-1}{s(x)+1}}\,\exp\left(s(x)-2+
\frac{2}{f^\xi(x)x^{1/3}}\right)\,,
\label{weak}
\end{equation}
with $s(x)=\sqrt{1+\epsilon_{\rm c}/\epsilon_{\rm F}(x)}\equiv
k_{\rm c}/k_{\rm F}$. In the far nuclear exterior when 
$x\rightarrow 0$, ${\mathcal F}^\xi$ should be determined by the 
free NN scattering. However, approaching the nuclear surface, 
finite range and nonlocal in-medium effects may already 
considerably modify the force even at low density, and therefore 
the best parametrization of an effective contact force at 
$x\ll 1$ might not necessarily agree with the vacuum value which 
could be extracted from the free NN interaction, as discussed by 
Migdal many years ago~\cite{Migdal}.

\begin{figure}[b]
\begin{center}
\leavevmode
\epsfxsize=22pc 
\epsfbox{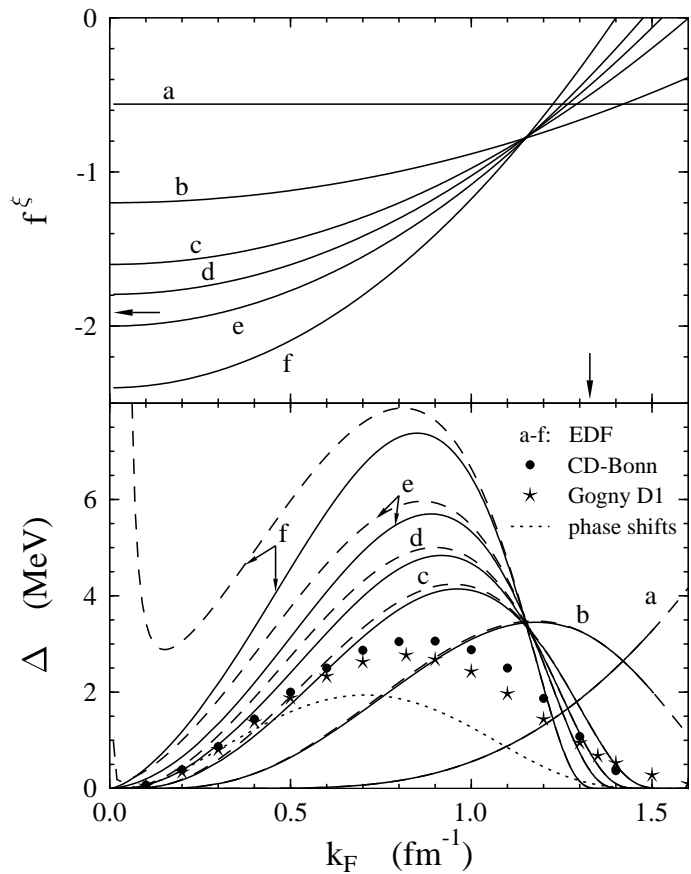} 
\end{center}
\caption{Dimensionless pairing strength (upper panel) and pairing 
gap (lower panel) in infinite nuclear matter as functions of the 
Fermi momentum. In the upper panel, the horizontal arrow shows 
the critical value of $f^\xi_{\rm ex}$ ($f^\xi_{\rm cr}=-1.912$) 
at which an $l=0$ spin singlet nucleon--nucleon bound state appears
at zero energy while the vertical one marks the Fermi momentum at 
saturation point for the functional DF3 without pairing, 
$k_{0\rm F}=1.328$~fm$^{-1}$. In the lower panel, full (dashed) 
curves (a)--(f) are obtained by solving eqs.~(\ref{gapinf}) 
and~(\ref{partnum}) (by using the weak pairing approximation, 
eq.~(\ref{weak})) with contact pairing force~(\ref{Fxi}), and an 
energy cutoff $\epsilon_{\rm c}=40$~MeV, and correspond, 
respectively, to the parameter sets (a)--(f) of eq.~(\ref{pppar}). 
The solid circles and stars are the solutions of the nonlocal gap 
equation with the CD-Bonn potential~\cite{EHJ98} and with the 
finite-range Gogny D1 force~\cite{Kuch89}, respectively. The 
dotted line shows the values of the pairing gap calculated using 
eq.~(\ref{delcot}) with the free NN scattering phase shifts 
(see text).} \label{f:f1} 
\end{figure}

Shown in Fig.~\ref{f:f1} are the values of dimensionless strength
$f^\xi$ (upper panel) and the results for $\Delta$ (lower panel) in
infinite matter with the parameter sets (a)--(f) of eq.~(\ref{pppar})
deduced for the lead chain. The calculations are performed with
cutoff $\epsilon_{\rm c}=40$~MeV and Fermi energy 
$\epsilon_{0\rm F}=36.57$~MeV of saturated nuclear matter from 
the parametrization~(\ref{normset}) of the normal part of the 
density functional.

One should notice that $\epsilon_{\rm F}$ entering the integrand of 
the gap equation~(\ref{gapinf}) can be expressed directly through 
density by $\epsilon_{\rm F}=\hbar^2k_{\rm F}^2/2m$ with $k_{\rm 
F}=(3\pi^2\rho/2)^{1/3}\equiv k_{0\rm F}x^{1/3}$ only if the pairing
is weak indeed so that the dependence of the Fermi energy (and the 
chemical potential $\mu$) on $\Delta$ can be disregarded. Otherwise 
one should introduce the particle number condition
\begin{equation}
x=\frac{2}{\rho_0}\int\nolimits_{k\le k_{\rm c}}\,
\frac{\d \vec k}{(2\pi)^3}n_k(x)\,, \label{partnum}
\end{equation}
where
\begin{equation}
n_k(x)=
\frac{1}{2}\left(1-\frac{\epsilon_k-\epsilon_{\rm F}(x)}
{\sqrt{(\epsilon_k-\epsilon_{\rm F}(x))^2+\Delta^2(x)}}\right)\,,
\label{nk}
\end{equation}
and solve the system of the two equations~(\ref{gapinf})
and~(\ref{partnum}) with respect to $\Delta$ and $\epsilon_{\rm F}$.
The results shown in Fig.~\ref{f:f1} by full lines correspond to 
such a direct solution while those shown by dashed lines to the 
weak pairing approximation, eq.~(\ref{weak}).

As seen in Fig.~\ref{f:f1}, the approximation~(\ref{weak}) works 
well in the entire range of $k_{\rm F}$ for the set (a), but for 
the other sets this is true only at $k_{\rm F}$ greater than 
$\approx$1.2~fm$^{-1}$ and also, for the sets (b), (c) and (d), at 
$k_{\rm F}$ less than 0.42, 0.14 and 0.042~fm$^{-1}$, respectively 
(in these regions the ratio $\Delta/\epsilon_{\rm F}$ does not 
exceed 0.1). For the sets (e) and (f), at lower densities, 
eq.~(\ref{weak}) can not be used any more to estimate the gap even 
by order of magnitude since, as seen from the behavior of the 
corresponding dashed lines, it becomes divergent. 

\begin{figure}[t]
\begin{center}
\leavevmode
\epsfxsize=22pc 
\epsfbox{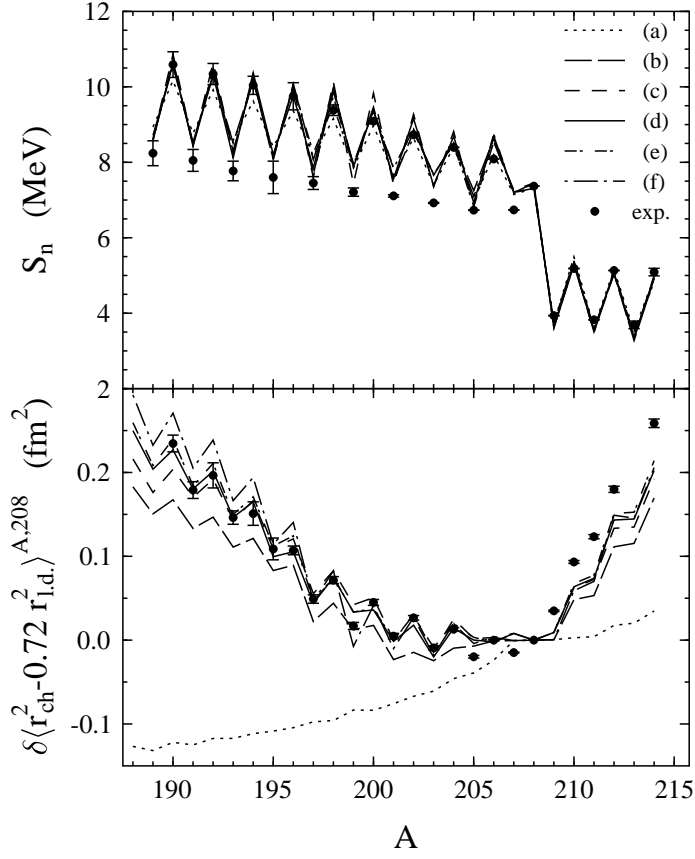} 
\end{center}
\caption{Upper panel: neutron separation energies $S_{\rm n}$ for
lead isotopes. Lower panel: differences of mean squared charge 
radii $\delta\langle r^2_{\rm ch}\rangle$ with respect to 
$^{208}$Pb as reference nucleus; 72\% of the corresponding liquid 
drop values (using $r_0=1.1$~fm) are subtracted to enhance the 
visibility of the small differences. The calculations are performed
with the functional DF3 and the coordinate-space technique. Curves 
(a)--(f) connect the points obtained with parameter sets (a)--(f) 
of the pairing force~(\ref{pppar}), respectively. Experimental data 
for $S_{\rm n}$, including those derived from systematic trends, are 
from~\cite{WA85,AW93}; data for $\delta\langle r^2_{\rm ch}\rangle$ 
are from~\cite{Ans86,Ott89,Dut91}.}  \label{f:f5a}
\end{figure}

All parameter sets~(\ref{pppar}) except (a) reproduce the neutron 
separation energies and the isotope shifts of charge radii of lead
isotopes reasonably well (see next section). For the sake of 
comparison also shown in Fig.~\ref{f:f1} are the values of the 
$^1S_0$ pairing gap in nuclear matter obtained for the CD-Bonn 
potential without medium effects (using free single-particle 
spectrum $\epsilon_k=k^2/2m$)~\cite{EHJ98} and for the Gogny D1 
force in the HFB framework~\cite{Kuch89}. The agreement between 
the two latter calculations is relatively good while both deviate 
noticeably from our predictions. The curve for contact 
density-independent pairing force, set (a), stands by itself with 
a positive derivative $\d \Delta(x)/\d x$ everywhere; no acceptable 
description of $\langle r^2_{\rm ch}\rangle$ could be obtained
in this case (see Fig.~\ref{f:f5a}). Interestingly, for the sets
(b)--(e) which reproduce satisfactorily both the neutron 
separation energies $S_{\rm n}$ and mean squared charge radii 
$\langle r^2_{\rm ch}\rangle$ there exists a pivoting point at 
$k_{\rm F}\approx 1.15$~fm$^{-1}$ (at $\approx$0.65 of the 
equilibrium density) with the same value of
$\Delta_{\rm piv}\approx 3.3$~MeV 
($f^\xi_{\rm piv}\approx -0.78$).

One can see in Fig.~\ref{f:f1} that for parameter sets with bigger 
absolute values of $f^\xi$ and $h^\xi$ the region where $\Delta(x)$ 
varies strongly moves towards lower densities, the slope becomes 
steeper and around $x\approx 1$ the pairing gap tends to be very 
small indeed. One may think that in finite nuclei the effect of the 
strong dependence of $\Delta$ on $x$, especially at small densities, 
would not influence the nuclear properties noticeably. But even if 
$\Delta$ would be concentrated only in the surface region and 
outside of the nucleus, the anomalous density, due to quantum
effects, would not vanish inside the nucleus where
$x\approx 1$. In this case the effect of density dependence should 
persist in the nuclear volume and should be more pronounced at 
larger values of $h^\xi$. This point was confirmed by our 
calculations (cf. Ref.~\cite{FZ96} and below).

Consider in more detail the behavior of $\Delta$ at very low 
densities when $k_{\rm F}\rightarrow 0$ assuming the weak pairing 
regime, i.e. $f^\xi_{\rm ex} > f^\xi_{\rm cr}$ where 
$f^\xi_{\rm cr}$ is a critical strength constant given by
\begin{equation}
f^\xi_{\rm cr}=-2\frac{k_{0\rm F}}{k_{0\rm c}}\,,
\label{fxicr}
\end{equation}
with $k_{0\rm c}=\sqrt{2m\epsilon_{\rm c}}/\hbar$. This constant 
is determined by the condition
\begin{equation}
1+\frac{f^\xi_{\rm cr}}{4\sqrt{\epsilon_{0\rm F}}}
\int_0^{\epsilon_{\rm c}}
\frac{\d \epsilon}{\sqrt{\epsilon}}=0\,, \label{gapx0}
\end{equation}
obtained from the gap equation~(\ref{gapinf}) at
$\Delta=\epsilon_{\rm F}=0$. For the chosen energy cutoff and
parametrization~(\ref{normset}) of our density functional we have
$f^\xi_{\rm cr}=-1.912$. To leading order, on the other hand, at
$k_{\rm F}\rightarrow 0$ from eq.~(\ref{weak}) we obtain
\begin{equation}
\Delta=c\epsilon_{\rm F}
\exp{\left(\frac{\pi}{2k_{\rm F}a}\right)}\,,\quad a<0\,,
\label{gapweak}
\end{equation}
where $c=8{\rm e}^{-2}\approx 1.083$ and where we have introduced 
the quantity 
\begin{equation}
a= \frac{\pi}{2k_{0\rm F}}
\left(\frac{\sqrt{2m\epsilon_{\rm c}}}{\hbar k_{0\rm F}}+
\frac{2}{f^\xi_{\rm ex}}\right)^{-1} \equiv
\frac{\pi}{4k_{0\rm F}}\left(\frac{1}{f^\xi_{\rm ex}}-
\frac{1}{f^\xi_{\rm cr}}\right)^{-1}\,. \label{scl}
\end{equation}
It can be shown that this quantity is nothing but the singlet 
nucleon--nucleon scattering length. To be more specific, consider 
first the two-neutron problem in the vicinity of the critical point, 
$f^\xi_{\rm ex} \approx f^\xi_{\rm cr}$, when the attraction is 
strong enough to produce a bound pair state (the scattering problem
at any $f^\xi$ will be considered below). The bound-state wave 
function for the relative motion of two neutrons in case of 
contact interaction $C_0f^\xi_{\rm ex}\delta(\vec r)$ is determined 
by (cf. Ref.~\cite{BE91}):
\begin{equation}
\psi(\vec r)=-C_0f^\xi_{\rm ex}G_0(\vec r, 
\epsilon_{\rm b})\psi(0)\,, \label{psinn}
\end {equation}
where $G_0$ is the free Green's function in the truncated space,
\begin{equation}
G_0(\vec r, \epsilon_{\rm b})=
\int\nolimits_{\epsilon_k^{\rm nn}\le\epsilon_{\rm c}^{\rm nn}}\,
\frac{\d \vec k}{(2\pi)^3}
\frac{{\mathrm e}^{{\mathrm i}\vec k \vec r}}{\epsilon_k^{\rm 
nn}-\epsilon_{\rm b}}\,, \label{G0}
\end{equation}
with $\epsilon_{\rm b}$ the binding energy and 
$\epsilon_k^{\rm nn}=\hbar^2k^2/m$ (the reduced mass is $m/2$).
In the rest system of a nucleus, the center-of-mass energy 
$\epsilon^{\rm nn}$ in the scattering problem would correspond
to the energy $\epsilon^{\rm nn}/2$ of each of the two nucleons
in the $s$-wave pairing problem. This implies that the cutoff
in the $k$-space in eq.~(\ref{G0}), 
$k_{\rm c}^{\rm nn}=\sqrt{m\epsilon_{\rm c}^{\rm nn}}/\hbar$, 
should be the same as in the gap equation. Thus, in the energy 
space, one has $\epsilon_{\rm c}^{\rm nn}=2\epsilon_{\rm c}$ and 
from~(\ref{psinn}) at $\vec r =0$ one obtains the equation to 
determine $\epsilon_{\rm b}$:
\begin{equation}
1=-\frac{1}{4\sqrt{\epsilon_{0\rm F}}}f^\xi_{\rm ex}
\int_0^{\epsilon_{\rm c}}\d \epsilon
\frac{\sqrt{\epsilon}}{\epsilon-\epsilon_{\rm b}/2}\,, \label{nneb}
\end{equation}
which reduces to
\begin{equation}
\frac{f^\xi_{\rm ex}}{f^{\xi {\rm nn}}_{\rm cr}}=
\left(1-\sqrt{\frac{-\epsilon_{\rm b}}{2\epsilon_{\rm c}}}
\arctan{\sqrt{\frac{2\epsilon_{\rm c}}{-\epsilon_{\rm b}}}}
\right)^{-1}\,, \label{atan}
\end{equation}
where $f^{\xi {\rm nn}}_{\rm cr}$ is a critical value of 
$f^\xi_{\rm ex}$ at which eq.~(\ref{psinn}) has a bound state 
solution $\epsilon_{\rm b}=0$. Comparing eq.~(\ref{nneb}) at 
$\epsilon_{\rm b}=0$ with eq.~(\ref{gapx0}) one finds that 
$f^{\xi {\rm nn}}_{\rm cr}$ coincides with the value defined
above by eq.~(\ref{fxicr}). Now, in the vicinity of 
$f^\xi_{\rm cr}$ we can set 
$\arctan{\sqrt{2\epsilon_{\rm c}/\vert\epsilon_{\rm b}\vert}}
\approx \pi/2$ and use $\epsilon_{\rm b}=-\hbar^2/ma^2_{\rm nn}$ 
in eq.~(\ref{atan}). Then it is easy to see that the solution for 
the scattering length $a_{\rm nn}$ is exactly the same as given 
by~(\ref{scl}). The expression~(\ref{gapweak}) for $\Delta$ agrees
with the results of Ref.~\cite{KKC96} based on a general analysis 
of the gap equation at low densities when $k_{\rm F}\vert a \vert 
\ll 1$. But we should stress that~(\ref{gapweak}) is valid only in 
the weak-coupling regime corresponding to negative $a$. In the 
opposite case the gap in the dilute limit has to be found in a 
different way.   

At $f^\xi_{\rm ex} > f^\xi_{\rm cr}$ the scattering length is 
negative, and from eq.~(\ref{gapweak}) it follows that at low
densities the pairing gap is exponentially small and eventually
$\Delta(k_{\rm F}\rightarrow 0)=0$. Such a weak pairing regime with
Cooper pairs forming in a spin singlet $l=0$ state exists up to the
critical point at which the attraction becomes strong enough to
change the sign of the scattering length. Then the strong pairing
regime sets in, eqs.~(\ref{weak}) and~(\ref{gapweak}) are not valid
any more, and $\Delta$ should be determined directly from the
combined solution of the gap equation~(\ref{gapinf}) and the particle
number condition~(\ref{partnum}). In the dilute systems,
$\epsilon_{\rm F}$ plays the role of the chemical potential $\mu$.
The latter is defined by 
$\mu=\epsilon_{\rm F}(k_{\rm F})+U(k_{\rm F})$ with $U(k_{\rm F})$ 
the HF mean field at the Fermi surface which is negligible for the 
fermion gas. At the critical point $\mu$ becomes negative 
and a bound state of a single pair of nucleons with the binding 
energy $\epsilon_{\rm b}=2\mu$ becomes possible~\cite{KK68,NSR85}. 
This can be easily seen from the gap equation~(\ref{gapinf})
written in the form
\begin{equation}
\left(\frac{k^2}{m}-2\mu\right)\phi_k=
-{\rm sgn}(\epsilon_k-\mu)\sqrt{1-\phi_k^2}
\int\nolimits_{k\,'\le k_{\rm c}}\,\frac{\d \vec k\,'}
{(2\pi)^3}{\mathcal F}^\xi\phi_{k\,'}\,, \label{phik}
\end{equation}
where we have introduced the functions
$\phi_k=\Delta/\sqrt{(\epsilon_k-\mu)^2+\Delta^2}$ and replaced
$\epsilon_{\rm F}$ by $\mu$. In the strong coupling regime, $\mu<0$,
and in the dilute limit, $\vert \phi_k\vert \ll 1$, this equation 
reduces to the Schr\"odinger equation for a single bound pair where 
$2\mu$ plays the role of the eigenvalue. It is equivalent to the 
coordinate-space equation~(\ref{psinn}). Thus 
$2\mu=\epsilon_{\rm b}$ where the binding energy $\epsilon_{\rm b}$ 
is determined from~(\ref{atan}). The pairing gap $\Delta$ at low 
densities in this regime can be found from the particle number
condition~(\ref{partnum}) which in the leading order now reads
\begin{equation}
x=\frac{3\Delta^2}{8\epsilon_{0\rm F}^{3/2}}
\int_0^{\epsilon_{\rm c}}\frac{\sqrt{\epsilon}\d \epsilon}
{(\epsilon+\vert \mu \vert)^2}\,. 
\end{equation}
In the vicinity of the critical point from this equation we find
\[\Delta^2=
\frac{16}{3\pi}x\epsilon_{0\rm F}^2\frac{1}{k_{0\rm F}a}\,,\]
which gives
\begin{equation}
\Delta=\frac{\hbar^2}{m}\left(\frac{2\pi\rho}{a}\right)^{1/2}\,,
\quad a>0\,.
\label{gapstrong}
\end{equation}

\begin{figure}[b]
\begin{center}
\leavevmode
\epsfxsize=22pc 
\epsfbox{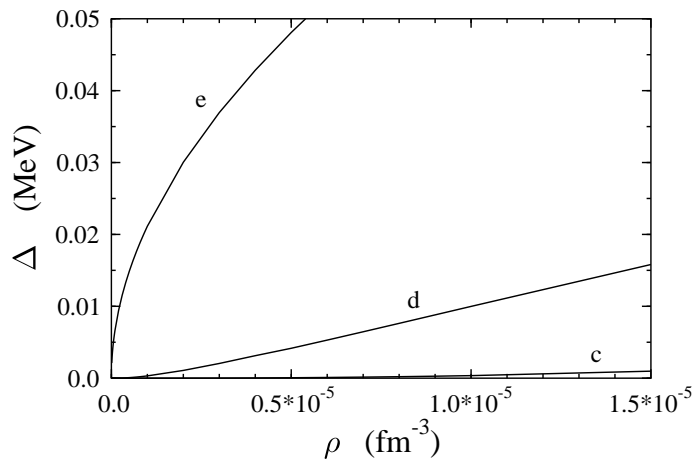} 
\end{center}
\caption{Pairing gap $\Delta$ as a function of $\rho$ at very low
densities. Curves (c)--(e) correspond to the parameter sets
(c)--(e) of eq.~(\ref{pppar}), respectively.} 
\label{f:f2}
\end{figure}

From this consideration it follows that, in the dilute case,
the energy needed to break a condensed pair goes smoothly from 
$2\Delta$ to $2\mu=\epsilon_{\rm b}$ as a function of the coupling
strength when the regime changes from weak to strong pairing. But 
as seen from~(\ref{gapweak}) and~(\ref{gapstrong}), the behavior of
$\Delta$ at low densities is such that the derivative 
$\d \Delta/\d \rho$ at $\rho\rightarrow 0$ as a function of 
$f^\xi_{\rm ex}$ exhibits a discontinuity from 0 to $\infty$. This 
is illustrated in Fig.~\ref{f:f2} where we have plotted the gap 
$\Delta(\rho)$ at very low densities for the sets (c)--(e) of 
eq.~(\ref{pppar}) embracing both regimes. We note also that the
analytical expressions~(\ref{gapweak}) and~(\ref{gapstrong}) give a
purely imaginary gap at the critical point when the scattering 
length changes sign 
($k_{\rm F}a\rightarrow \sqrt{2m\mu}/\hbar\rightarrow {\mathrm i}$
if $\mu$ becomes negative). In this connection it is instructive to 
write the weak coupling expression~(\ref{weak}) for $\Delta$ in the
following form~\cite{Fay99}:
\begin{equation}
\Delta(k_{\rm F})=c\epsilon_{\rm F}\exp
\left[-\frac{\pi}{2}\cot\delta(k_{\rm F})\right]\,, \label{delcot}
\end{equation}
where $c=8{\rm e}^{-2}$ and where we have introduced the Fermi 
level phase shift $\delta(k_{\rm F})$ defined by  
\begin{equation}
k_{\rm F}\cot\delta(k_{\rm F}) = 
-\frac{4k_{0\rm F}}{\pi}\left(\frac{1}{f^\xi(k_{\rm F})}+
\frac{k_{\rm c}(k_{\rm F})}{2k_{0{\rm F}}}\right)
-\frac{k_{\rm F}}{\pi}\ln\left(\frac{k_{\rm c}(k_{\rm F})-k_{\rm F}}
{k_{\rm c}(k_{\rm F})+k_{\rm F}}\right)\,,  \label{kcot}
\end{equation} 
with $k_{\rm c}(k_{\rm F})=\sqrt{k_{0\rm c}^2+k_{\rm F}^2}$. 
Eq.~(\ref{kcot}) corresponds to an exact solution of the nn 
scattering problem at the relative momentum $k=k_{\rm F}$ with the
states truncated by a momentum cutoff 
$k_{\rm c}=k_{\rm c}(k_{\rm F})$ for contact interaction 
$C_0f^\xi(k_{\rm F})\delta({\vec r})$ (see, e.g., 
Ref.~\cite{EBH97}). In the dilute limit, from eq.~(\ref{delcot}) 
one notices again that the pairing gap becomes pure imaginary at 
the critical point since $\cot\delta\rightarrow {\mathrm i}$ when 
$\vert a_{\rm nn}\vert \rightarrow \infty$. At very low densities, 
with the parametrization~(\ref{fxigrad}) of the pairing force and 
with the chosen density-dependent cutoff, eq.~(\ref{kcot}) reduces 
to
\begin{equation}
k_{\rm F}\cot\delta(k_{\rm F}) \approx -\frac{1}{a_{\rm nn}} + 
\frac{1}{2}r_{\rm nn}k_{\rm F}^2 - \frac{2k_{\rm F}}{\pi}
\left[\frac{k_{\rm F}}{2k_{0\rm c}}-\frac{2h^\xi}
{(f^\xi_{\rm ex})^2}\left(\frac{k_{\rm F}}
{k_{0\rm F}}\right)^{3q-1}\right]\,, \label{kcot0}
\end{equation}
where $a_{\rm nn}$ is the scattering length defined by
eq.~(\ref{scl}) and $r_{\rm nn}$ is the effective range,
$r_{\rm nn}=4/\pi k_{0\rm c}$. The first two terms in this equation
would describe low-energy behavior of the nn $s$-wave phase shift
through an expansion of $k\cot\delta$ in powers of the relative 
momentum $k=k_{\rm F}$ if the interaction were density-independent --
in our case, if the coupling strength and momentum cutoff were fixed
by $f^\xi=f^\xi_{\rm ex}$ and $k_{\rm c}=k_{0\rm c}$, respectively.
It follows that with a density-dependent effective force, such an
expansion contains additional terms which for the parametrization 
used here are of the same order as the effective range
term\footnote{Our effective pairing interaction with the choice
$q=1/3$ would lead in the dilute limit to the expression for $\Delta$
of the form of eq.~(\ref{gapweak}) but with a different prefactor $c$
depending on the value of $h^\xi$. If, furthermore, we define the
latter parameter by $h^\xi=(1+2\ln 2)(f^\xi_{\rm ex})^2/6$ we get in
the leading order $\Delta(k_{\rm F})=(2/{\rm e})^{7/3}
\epsilon_{\rm F}\exp(\pi/2k_{\rm F}a)$, i.e. the result obtained in
Ref.~\cite{GMB61} for a non-ideal Fermi gas by studying the
singularities of the interaction amplitude (vertex function $\Gamma$)
and taking into account the terms up to the second order in
$k_{\rm F}a$.}. This simply demonstrates that, for reproducing the
pairing gap, the effective interaction even at very low densities
need not necessarily coincide with the bare NN interaction.

When the Fermi momentum $k_{\rm F}$ approaches from below the upper
critical point at which the pairing gap closes,
$k_{\rm F}^{\rm cr}=k_{0{\rm F}}(-f^\xi_{\rm ex}/h^\xi)^{1/3q}$ 
defined by $f^\xi(k_{\rm F})=0$, we get from eq.~(\ref{delcot}) 
\begin{equation}
\Delta(k_{\rm F}) \approx c\epsilon_{\rm F}\exp
\left[-\frac{2k_{0{\rm F}}}
{3qf^\xi_{\rm ex}(k_{\rm F}-k_{\rm F}^{\rm cr})}\right]\,.
\label{delcr}
\end{equation}
Thus, at higher densities, the pairing gap becomes exponentially
small when $k_{\rm F}$ approaches $k_{\rm F}^{\rm cr}$, 
in agreement with
general analysis of the gap solutions~\cite{KKC96}. In weak coupling,
as we have already shown, $\Delta(k_{\rm F})$ is also exponentially
small at low densities, in the vicinity of $k_{\rm F}=0$. It is
noteworthy that near both critical points the pairing potential
$\Delta$ can be found from the behavior of the phase shift by using
eq.~(\ref{delcot}) with a smooth density-dependent prefactor
$c(k_{\rm F})$ (the details will be given elsewhere~\cite{FZ2000}).
In the present paper, as an illustration, we show in Fig.~\ref{f:f1}
by the dotted line the values of $\Delta(k_{\rm F})$ obtained from
eq.~(\ref{delcot}), with a prefactor $c=8{\rm e}^{-2}$, by using 
``experimental'' nn phase shifts, without electromagnetic
effects\footnote{We thank Rupert Machleidt for providing us with
these nn phase shifts.}. It is seen that $\Delta$ obtained this way 
at low densities closely follows the solution of the gap equation
with the CD-Bonn potential~\cite{EHJ98}. The nn phase shift passes
zero at the relative momentum $k\approx 1.71$~fm$^{-1}$, and the gap
should vanish at the corresponding Fermi momentum. Unfortunately,
the solutions for $\Delta$ are given in Ref.~\cite{EHJ98} only in
the region up to $k_{\rm F}=1.4$~fm$^{-1}$ which is rather far from
the upper critical point. 

For symmetric nuclear matter, with the functional DF3 used in the
present paper, the energy per particle is 
\begin{equation} 
\frac{E}{A}(x) = 
\frac{2}{\rho_0 x}\int\nolimits_{k\le k_{\rm c}} \frac{\d \vec 
k}{(2\pi)^3}\frac{\hbar^2k^2}{2m}n_k(x)+ \frac{1}{3}\epsilon_{0\rm 
F}a_+^{\rm v}f_+^{\rm v}(x)x+ \frac{3\Delta^2(x)}
{2f^\xi(x)x\epsilon_{0\rm F}}\,,   \label{EOS}
\end{equation} 
where $f_+^{\rm v}(x)=(1-h^{\rm v}_{1+}x)/(1+h^{\rm v}_{2+}x)$ with
the parameters given by~(\ref{normset}); here the ``particle--hole''
term $\propto f_+^{\rm v}$ vanishes in the dilute limit linearly in
density. The chemical potential is
\begin{equation}
\mu(x)=\epsilon_{\rm F}(x)+\frac{1}{3}\epsilon_{0\rm F}
a_+^{\rm v}[f^{{\rm v}\prime}_+(x)x^2+2f^{\rm v}_+(x)x]
+\frac{3f^{\xi\prime}(x)}{2f^{\xi 2}(x)}
\frac{\Delta^2(x)}{\epsilon_{0\rm F}}\,,
\label{chimpot}
\end{equation}
where the prime denotes the derivative with respect to the
dimensionless density $x$. The Fermi energy $\epsilon_{\rm F}(x)$ and
the pairing gap $\Delta(x)$ entering these equations are determined 
from~(\ref{gapinf}) and~(\ref{partnum}). The last two terms 
in~(\ref{chimpot}), even in strong pairing regime, vanish in the
dilute limit at least as $x^q$ if $0<q<1$ or linearly in $x$ if
$q\geq 1$ (in our case, the exponent in the density-dependent pairing
force is $q=\frac{2}{3}$). Thus, we see again that, in strong
coupling, in the leading order
$\mu=\epsilon_{\rm F}=\epsilon_{\rm b}/2 < 0$.  

\begin{figure}[t]
\begin{center}
\leavevmode
\epsfxsize=22pc 
\epsfbox{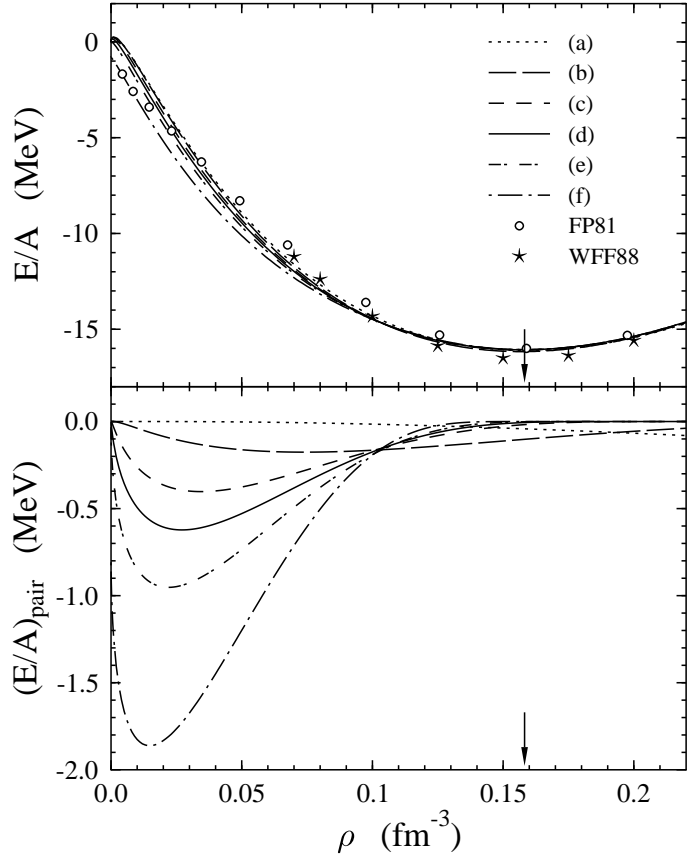} 
\end{center}
\caption{Energy per nucleon $E/A$ (top) and pairing contribution to
$E/A$ (bottom) in symmetric nuclear matter. Curves (a)--(e) are
calculated using eq.~(\ref{EOS}) and correspond to the strength
parameters (a)--(e) of eq.~(\ref{pppar}), respectively. Open circles
and stars are the calculations of Ref.~\cite{FP81} and
Ref.~\cite{WFF88}, respectively, for the UV14 plus TNI model. The
vertical arrows mark the saturation density for the functional DF3
without pairing, $2\rho_0=0.1582$~fm$^{-3}$.}
\label{f:f3} 
\end{figure}

\begin{figure}[t]
\begin{center}
\leavevmode
\epsfxsize=22pc 
\epsfbox{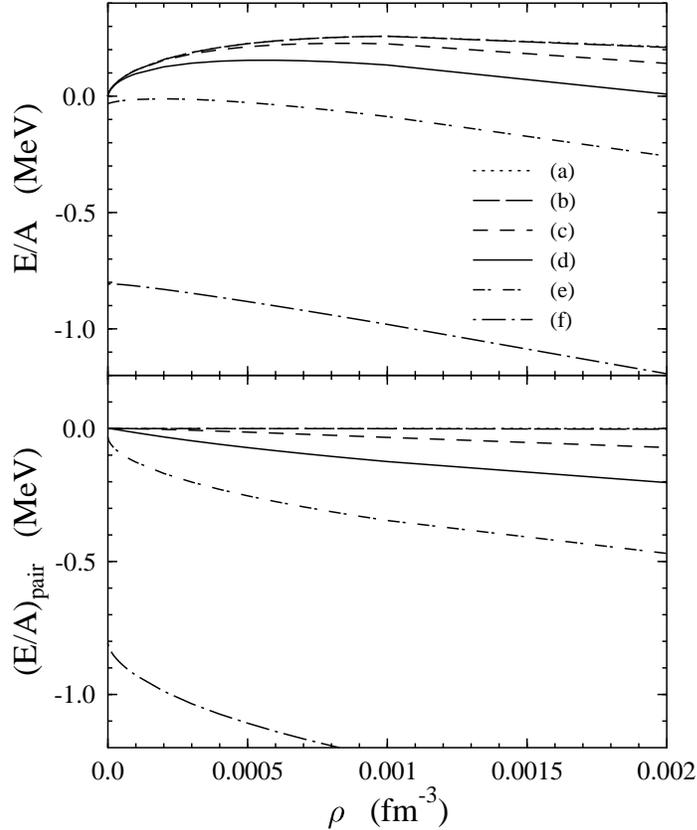} 
\end{center}
\caption{Energy per nucleon $E/A$ (top) and pairing contribution to
$E/A$ (bottom) for symmetric nuclear matter at low densities. The
notations are the same as in Fig.~\ref{f:f3}.}
\label{f:f4} 
\end{figure}

The calculated energy per nucleon as a function of the isoscalar
density $\rho$ is shown in the upper panel in Fig.~\ref{f:f3}
together with the results of the nuclear matter
calculations~\cite{FP81,WFF88} for the UV14 plus TNI model. It is
seen that DF3 gives qualitatively reasonable description of the
nuclear matter EOS and that pairing could contribute noticeably to
the binding energy especially at lower densities. In the lower
panel in Fig.~\ref{f:f3} we have plotted the pairing energy per
nucleon, $(E/A)_{\rm pair}$, obtained by subtracting from
eq.~(\ref{EOS}) the corresponding value of $E/A$ at $\Delta=0$.
This pairing energy is a sum of the positive contribution coming
from the kinetic energy, i.e. from the first term in~(\ref{EOS}),
and the negative anomalous energy -- the last term in~(\ref{EOS})
(an expression for $(E/A)_{\rm pair}$ in the case of weak pairing
is derived in Appendix~B). The pairing contribution is small for the
density-independent force with $f^\xi=0.56$, set (a) of
eq.~(\ref{pppar}), and increases, as expected, for the sets (b)--(f)
with a shift to lower densities as $f^\xi_{\rm ex}$ becomes
gradually more attractive. For the sets (e) and (f) the attraction
is strong, $f^\xi_{\rm ex} < f^\xi_{\rm cr}$. In these cases a
nonvanishing binding energy in the dilute limit is solely due to
Bose-Einstein condensation of the bound pairs, the spin-zero bosons,
when all the three quantities, $\mu$, $E/A$ and $(E/A)_{\rm pair}$, 
reach the same value $\epsilon_{\rm b}/2$ ($\epsilon_b = -0.0646$ and
$-1.616$~MeV for the set (e) and (f), respectively). This is
illustrated in Fig.~\ref{f:f4} where we have plotted $E/A$ and
$(E/A)_{\rm pair}$ as functions of $\rho$ at very low densities.
Analytically, for the kinetic energy term in~(\ref{EOS}) in the
leading order we find
\begin{equation}
\frac{E_{\rm kin}}{A} = \frac{3\Delta^2}{8x\epsilon_{0\rm F}^{3/2}}
\int_0^{\epsilon_{\rm c}}
\frac{\epsilon\sqrt{\epsilon}\d \epsilon}
{(\epsilon+\vert \mu \vert)^2}
\approx \frac{3\Delta^2}{2x\epsilon_{0\rm F}}
\left(-\frac{1}{f^\xi_{\rm cr}}-\frac{3\pi}{8}\frac{1}
{k_{0\rm F}a}\right)\,,  \label{eakin}
\end{equation}
where we have used the definition~(\ref{fxicr}) and the relation 
$\vert\mu\vert=\hbar^2/2ma^2$. Combining this with the last term 
in~(\ref{EOS}), one gets
\begin{equation}
\frac{E}{A}=-\frac{3\pi\Delta^2}{16x\epsilon_{0\rm F}}
\frac{1}{k_{0\rm F}a}\,.
\end{equation}
By using eq.~(\ref{gapstrong}), this expression reduces exactly to 
$-\hbar^2/2ma^2=\epsilon_{\rm b}/2$. It follows that the model just 
described is in fact parameter-free: in the strong coupling regime 
near the critical point, the ground state properties of nuclear 
matter in the dilute limit are completely determined by the 
scattering length.  

We have considered subsaturated nuclear matter with $^1S_0$ pairing
within our local EDF framework and demonstrated some results,
including extrapolation to the low density limit, with a few possible
parameter sets of the pairing force deduced from experimental data
for lead isotopes. At low densities, in the $T=0$ case of symmetric
$N=Z$ matter, the $^3S_1-^3D_1$ pairing correlations leading to a
Bose deuteron gas formation might be more important since the n--p
force is more attractive than that in the p--p or n--n pairing
channels (see Ref.~\cite{BLS95} and references therein). Thus, 
our approach, with $^1S_0$ pairing only, would be more relevant for
an asymmetric $N\neq Z$ case and for pure neutron systems. From this
point of view the best choice of the effective contact pairing force
for the EDF calculations seems to be the set (d) of eq.~(\ref{pppar})
since it gives the singlet scattering length
$a_{\rm nn}\approx -17.2$~fm which corresponds to a virtual state at
$\approx 140$~keV known experimentally\footnote{The contact
interaction in a truncated space of states, with two parameters 
$f^\xi_{\rm ex}$ and $k_{0\rm c}$, can be calibrated to produce not
only the empirical scattering length $a_{\rm nn}$ but also a realistic
effective range $r_{\rm nn}\approx 2.8$~fm which is directly related
to the momentum space cutoff by $r_{\rm nn}=4/\pi k_{0\rm c}$, see
Ref.~\cite{EBH97}. It would yield a rather low value for the energy
cutoff $\epsilon_{\rm c}\approx 9$~MeV. However, as we have already
mentioned, even at low densities the force can be considerably 
modified by nonlocality and polarization effects so that the energy 
dependence of the Fermi-level $s$-wave nn phase shift,
eq.~(\ref{kcot0}), at low relative momenta $k$ governed by the
effective range might be different from the free two-nucleon case.}.
As seen in Fig.~\ref{f:f1}, with this choice the behavior of $\Delta$
at low densities agrees well with the calculations based on realistic
NN forces. At higher densities, however, our predictions for $\Delta$
with the set (d) go much higher reaching a maximum of
$\approx 4.84$~MeV at $k_{\rm F}\approx 0.92$~fm$^{-1}$ while the 
calculations of Ref.~\cite{EHJ98} give a maximum of about 3~MeV at
$k_{\rm F}\approx 0.82$~fm$^{-1}$. With a bare NN interaction,
assuming charge independence and $m_{\rm n}=m_{\rm p}$ in the free
single-particle energies, the pairing gap would be, at given
$k_{\rm F}$, exactly the same both in symmetric nuclear matter and in
neutron matter. As shown in Refs.~\cite{WAP93,SCL96}, if one includes
medium effects in the effective pairing interaction, most importantly
the polarization RPA diagrams in the cross channel, the pairing gap
in neutron matter would be substantially reduced to values of the 
order of 1~MeV at the most. Whether such a mechanism works in the
same direction for symmetric nuclear matter is still an open
question. If the gap were smaller than that obtained with
the Gogny D1 force, which is also shown in Fig.~\ref{f:f1}, it would
be difficult to explain the observed nuclear pairing properties. The
effective contact density-dependent force~(\ref{fxigrad}) with our
preferable parameter set (d) of eq.~(\ref{pppar}) yields larger
pairing energy in nuclear matter than the Gogny force, but in finite
nuclei this is compensated by the repulsive gradient term. The
phenomenological pairing force used here contains dependence on the 
isoscalar density only since we have analyzed the existing data on
separation energies and charge radii for finite nuclei with a
relatively small asymmetry characterized by $(N-Z)/A\leq 0.25$. An
extrapolation to neutron matter with such a simple force would give
a larger pairing gap than for nuclear matter. This suggests that some
additional dependence on the isovector density
$\rho_{\rm n}-\rho_{\rm p}$ might be present in the effective pairing 
force. This possibility is planned to be tested in our future work,
with a more careful analysis of experimental data though the relevant
data base is not rich enough.          

Now consider how the density changes when the pairing gap appears in
nuclear matter. To leading order, one finds the following expression
for the energy per nucleon near the saturation point:
\begin{equation}
\frac E A = \frac{E_0}{A}+\frac{K_0}{18}\frac{(x-x_0)^2}{x_0^2}
+\beta(x)I^2-\frac 3 8 \frac{\Delta^2(x)}{\epsilon_{\rm F}(x)}\,
\label{Eaexp}
\end{equation}
where $K_0$ is the compression modulus at saturation density,
$\beta(x)I^2$ is the asymmetry energy, with
$I=(\rho_{\rm n}-\rho_{\rm p})/2\rho_0x \equiv (N-Z)/A$.
This expression is valid at $\vert x-x_0 \vert \ll 1,\,
\vert I \vert \ll 1$ and $\Delta \ll \epsilon_{\rm F}$.
Higher order effects connected with $I$-dependence of $K, \beta,
\Delta$ and $\epsilon_{\rm F}$ are neglected. The derivation of the
last term in~(\ref{Eaexp}) is discussed in detail in Appendix~B. Due
to this pairing term, the position of the equilibrium point may be
shifted to lower or higher densities depending on the behavior of
$\Delta(x)$ near $x=1$. If $\Delta$ does not depend on $x$ in the
vicinity of $x=1$ then the equilibrium density decreases due to
presence of $\epsilon_{\rm F}(x) \propto x^{2/3}$ in the
denominator of the pairing term (the system gains more binding
energy). This means an expansion of the system. The effect is
enhanced if $\Delta$ becomes larger during such an expansion, i.e.
when the derivative $\d \Delta(x)/\d x$ at $x=1$ is negative. This
point may be illustrated by the following simple consideration. At
equilibrium the pressure $P=0$ which means
\begin{equation}
\frac{\partial}{\partial x}(E/A) = 0\,.  \label{p}
\end{equation}
For saturated symmetric nuclear matter without pairing the
dimensionless density is, by definition, $x_0 = 1$. When $I \ne 0$
and $\Delta \ne 0$, from eq.~(\ref{Eaexp}) with condition~(\ref{p})
the new equilibrium density can be found (in units of $2\rho_0$),
which is the solution of the equation 
\begin{equation}
x-x_0 =
\frac{9x_0^2}{K_0}\left[\frac{\Delta(x)}{4\epsilon_{\rm F}(x)}
\left(3\frac{\d \Delta(x)}{\d x}-\frac{\Delta(x)}{x} \right)- 
\frac{\d \beta(x)}{\d x} I^2 \right]\,.  \label{infrho}
\end{equation}
From this equation one can see that the density should be sensitive
indeed to the derivatives of $\Delta$, and a negative slope in
$\Delta$ should cause a decrease of the density. The obtained
relations can be used to estimate the influence of pairing on the
charge radii for heavy nuclei as was done in Ref.~\cite{FTTZ94}. It
was shown that pairing interaction with strong $\rho$-dependence at
$x\approx 1$ might significantly change the equilibrium density. For
the parametrization used here the size of this effect is controlled
by the parameter $h^\xi$. The shift of the saturation point is
relatively small: $\vert\delta\rho\vert/2\rho_0 \leq 0.8$~\% for all 
the six sets of eq.~(\ref{pppar}) ($\delta\rho/2\rho_0 \approx 
+0.4$~\% and $\approx -0.4$~\% for the set (a) and (d),
respectively). In finite nuclei, the surface term $\propto
f^\xi_\nabla$ is equally important to produce a kink in the radius
evolution along isotope chain at magic neutron number and,
especially, to explain the observed odd-even staggering in 
$\langle r^2_{\rm ch}\rangle$~\cite{FZ96}. Although, as seen in 
Fig.~\ref{f:f1}, the calculations with bare NN interaction or with
Gogny force give a negative sign for $\d \Delta(x)/\d x$ near 
$x\approx 1$ ($k_{\rm F} \approx 1.33$~fm$^{-1}$), but the slope
might be not steep enough. This is the probable reason why the HFB
calculations with the Gogny force, which give a good description of
the global pairing properties of nuclei, could not reproduce the kink
in lead isotopes~\cite{Taj93}. The density dependence of the pairing
force leads to the direct coupling between the neutron anomalous 
density and the proton mean field as given by~(\ref{Upairgrad}). The
suppression of $\vert\nu_{\rm n}\vert^2$ in the odd neutron subsystem
because of the blocking effect influences the potential
$U^{\rm p}_{\rm pair}$ of the proton subsystem through the volume,
$\propto h^\xi$, and surface, $\propto f^\xi_\nabla$, couplings, and
this moves the behavior of the proton radii towards the desired
regime~\cite{FZ96}.

\Section{Numerical results for some isotope chains}
\label{disres}
The calculations for finite spherical nuclei are performed using the
coordinate-space technique which is described in detail in
Appendix~C. In the construction of the Gor'kov Green's functions, the
four linear-independent solutions $(u_i, v_i)\,,i=1$--4, were found
by the Numerov method with a radial step of 0.1~fm and physical
boundary conditions imposed for each $lj$ channel at the origin and
at $r=25$~fm. The contour in the complex energy plane with 
Im$\epsilon=\pm8$~MeV along the horizontal sections and with energy
cutoff $\epsilon_{\rm c}=40$~MeV measured from the Fermi level was
used (see Figs.~\ref{f:f24}, \ref{f:f25} in Appendix C). The
integration along the contour was performed by the Simpson method
with automatic step selection. The convergence of the iteration
procedure is controlled by a few criteria: for two successive steps
$i$ and $i+1$, the conditions $\vert\rho_{i+1}(r)-\rho_i(r)\vert < 
10^{-6}$~fm$^{-3}$ and $\vert\Delta_{i+1}(r)-\Delta_i(r)\vert <
50$~keV should be achieved for all $r$, the chemical potential $\mu$
should finally satisfy the condition $\vert N(\mu)-N\vert < 0.01$ and
the contribution $N_{lj}$ to the total particle number $N(\mu)$ from
the states with higher angular momenta $lj$ should not exceed 0.01.
Such criteria guarantee an accuracy not worse than 0.1\% for all the
calculated quantities of our interest. The mean square charge radii
were computed from ground state charge densities obtained by folding
the point nucleon distributions with nucleon charge form factors, 
including the relativistic electromagnetic spin--orbit correction, in
the same way  as in Ref.~\cite{Kim92}.

\begin{figure}[t]
\begin{center}
\leavevmode
\epsfxsize=22pc 
\epsfbox{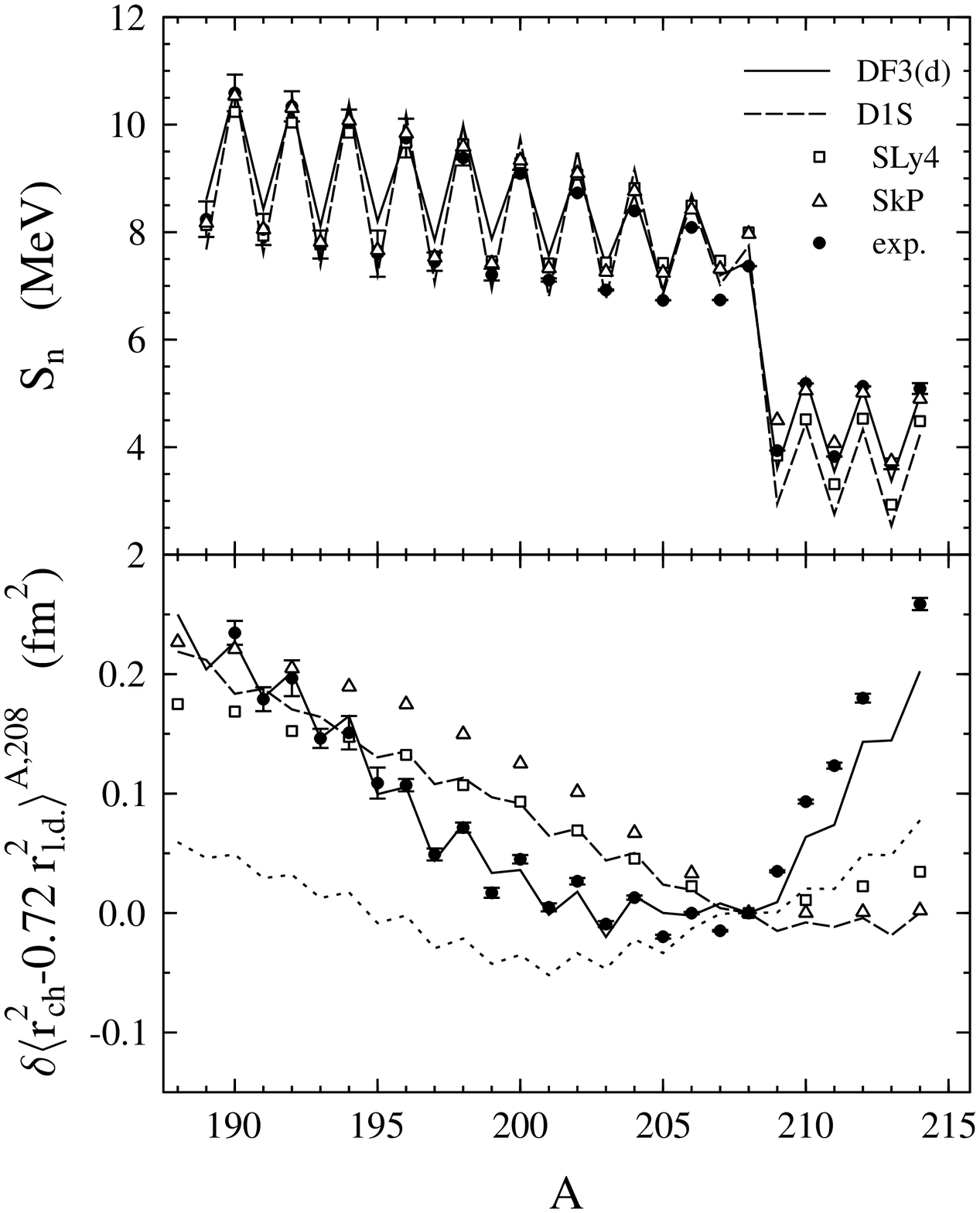} 
\end{center}
\caption{The same as in Fig.~\ref{f:f5a} but only for the set (d) of 
the pairing force in comparison with the HFB calculations with Skyrme 
forces SkP~\cite{DFT84} and SLy4~\cite{Chab97,SLy4}, and the Gogny 
D1S~\cite{BGG91} force. Also shown in the lower panel by the dotted 
line are the $\delta\langle r^2_{\rm ch}\rangle$ values obtained 
without gradient term in the pairing force 
($f^\xi_{\rm ex}=-1.79,\;h^\xi=1.66,\;f^\xi_\nabla=0$).} 
\label{f:f5b}
\end{figure}

Let us discuss first the results obtained for the lead chain. As
shown in the upper panel of Fig.~\ref{f:f5a}, the experimental 
neutron separation energies are reproduced equally well for all the 
chosen parameter sets of the pairing force. The curves (a)--(f) 
correspond to the sets marked with the same letters in 
eq.~(\ref{pppar}). The curve (a) is obtained for the ``constant'' 
pairing force, without density dependence. The other sets (b)--(f) 
differ from that by non-zero values of $h^\xi$ and $f^\xi_\nabla$ 
parameters. As already mentioned in Section~\ref{spair}, for each 
$f^\xi_{\rm ex}$ value it is possible to find such values of $h^\xi$ 
and $f^\xi_\nabla$ that the pairing energy for a given nucleus is 
practically the same. These parameters are kept fixed for all 
isotopes of the lead chain. 

In the lower panel of Fig.~\ref{f:f5a} the calculated isotope shifts
of mean squared charge radii, $\delta\langle r^2_{\rm ch}\rangle$,
for the lead chain with respect to $^{208}$Pb as a reference nucleus
are presented. These curves except (a) do not differ from each other
significantly and all of them reproduce qualitatively the kink and
the average size of odd-even staggering. The curve (a) corresponds
to a simple pairing delta-force; in this case the behavior of 
$\delta\langle r^2_{\rm ch}\rangle$ is rather smooth and neither kink 
nor staggering are reproduced. The variant (a) of eq.~(\ref{pppar}) 
was included just to give an example that without any density 
dependence in the contact pairing effective interaction it is also 
possible to describe the neutron separation energies but not the 
evolution of charge radii. With the parametrization of 
eq.~(\ref{fxigrad}) used here for the pairing force, the size of 
the staggering and the kink in charge radii of Pb isotopes can be 
reproduced only if the ``gradient'' parameter $f^\xi_\nabla$ is 
$\approx 2$. Without gradient term the results for radii would be 
in between the curves (a) and (b) shown in the lower panel of 
Fig.~\ref{f:f5a} (an example is given in the lower panel of 
Fig.~\ref{f:f5b}). On the whole, as seen in Fig.~\ref{f:f5a}, the 
set (d) yields a somewhat better description than the other sets, 
particularly for the lighter Pb isotopes. For this reason, and due 
to the fact that the set (d) corresponds to the correct value of 
the singlet scattering length in the dilute limit (see previous 
Section), this set will be taken as our preferable choice, and the 
corresponding functional will be referred to 
as DF3(d) in the following.

In Fig.~\ref{f:f5b} we compare the DF3(d) predictions for Pb 
isotopes with the HFB results obtained for the Gogny D1S 
force~\cite{BGG91} and for two state-of-the-art models based on 
Skyrme density-dependent forces SkP~\cite{DFT84}and SLy4~\cite{SLy4}.
The calculations with Gogny force for the odd isotopes are 
done in the blocking approximation, and thus we can show the results 
both for even and odd Pb isotopes\footnote{We thank Jean-Fran\c cois 
Berger, Jacques Decharg\'e and Sophie Peru for providing us with 
these results.}. The Skyrme-HFB calculations\footnote{We thank Jacek 
Dobaczewski for providing us with a numerical file of these 
Skyrme--HFB calculations.} were done, unfortunately, without 
blocking, and therefore we show the radii only for even isotopes;
the $S_n$ values for odd isotopes were obtained by correcting the 
no-blocking HFB energy by adding an average pairing gap, this 
was considered to be a very good approximation. The details of the 
HFB+SkP and HFB+SLy4 calculations can be found in~\cite{Miz99} (see 
also the references therein), and here we only briefly discuss how 
the pairing correlations have been implemented in these models. The 
HFB+SkP calculations have been done with the same SkP force in the ph 
and pp channels, and therefore in these calculations the pairing 
force has density dependence $x^{1/6}$. The energy cutoff has been 
chosen by including the quasi-particle states up to the energy equal 
to the depth of the effective single-particle potential~\cite{DFT84}. 
This implies that the cutoff depends not only on the particle numbers 
$N$ and $Z$ but also on the quantum numbers $j$ and $l$; for $l$=0 it 
varies in the range of about 40--50\,MeV. The HFB+SLy4 calculations 
have been done with the density-dependent zero-range pairing force in 
the pp channel, and the SLy4 force in the ph channel. The form of 
this pairing force is identical to the velocity-independent piece of 
the Skyrme interaction (see Ref.~\cite{DNW95}) with parameters
$V_0 = t_0 = -2488.913$~MeV\,fm$^3$ and 
$V_3 = 19990$~MeV\,fm$^{3+1/6}$, and it also has the $x^{1/6}$ 
density dependence. The prescription for the energy cutoff has been 
identical as in the HFB+SkP calculations. We remark that such a
prescription is much different from that which we use here; with 
such a recipe it is not easy to construct the corresponding
effective contact pairing force with a fixed phase-space cutoff, 
hence no simple extrapolation to the pairing properties of uniform 
nuclear matter. 

One can see in the upper panel of Fig.~\ref{f:f5b} that the HFB+D1S, 
HFB+SkP, SkLy4 and our DF3(d) calculations describe the neutron 
separation energies with more or less the same quality, some
deviations are observed in the region above $^{207}$Pb. For the
lighter Pb isotopes, the Gogny force slightly overestimates the 
odd-even effect in $S_n$ while with DF3(d) it is slightly 
underestimated.      

In the lower panel of Fig.~\ref{f:f5b} it is seen that the 
Skyrme-HFB calculations with SkP and SLy4 forces give too small 
kink in radii. Calculations with the Gogny D1S force for 
$\delta\langle r^2_{\rm ch}\rangle$ could not reproduce the 
kink either. The latter model yields sizeable staggering in radii,
but the effect is too small, at least by a factor of two smaller in
amplitude than experimentally observed; moreover, its sign in 
isotopes below $A$=194 is reversed with respect to the observations.
As for the Skyrme functionals, one suspects that they could not give 
the correct size of staggering as well. Such a conclusion may be
supported by our DF3 calculations with parameters 
$f^\xi_{\rm ex}=-1.79,\;h^\xi=1.66,\;f^\xi_\nabla=0$, i.e. with the
pairing force of~(\ref{pppar}) which contains only a $x^{2/3}$ 
dependence, without gradient of density. The external strength 
constant $f^\xi_{\rm ex}=-1.79$ is taken from our preferable set 
(d), and the parameter $h^\xi=1.66$ is found to get practically the 
same $S_n$ values as with the variant (d) itself. The results for the 
radii are shown in the lower panel of Fig.~\ref{f:f5b} by the dotted 
line. It is seen that the staggering appears but it is too small 
indeed. With a $x^{1/6}$ dependence, which is used with the SkP and 
SLy4 parametrization, the effect would be even smaller. Thus, to 
reproduce the scale of the odd-even effects in radii (and the scale 
of the kinks) within the local energy-density functional approach, 
the blocking mechanism should be enhanced by some peculiar density 
dependence of the effective pairing force, the linear dependence on 
$\rho^\sigma$ with $\sigma$=1, 1/3, 1/6... could not give the 
desirable size.        

\begin{figure}[b]
\begin{center}
\leavevmode
\epsfxsize=22pc 
\epsfbox{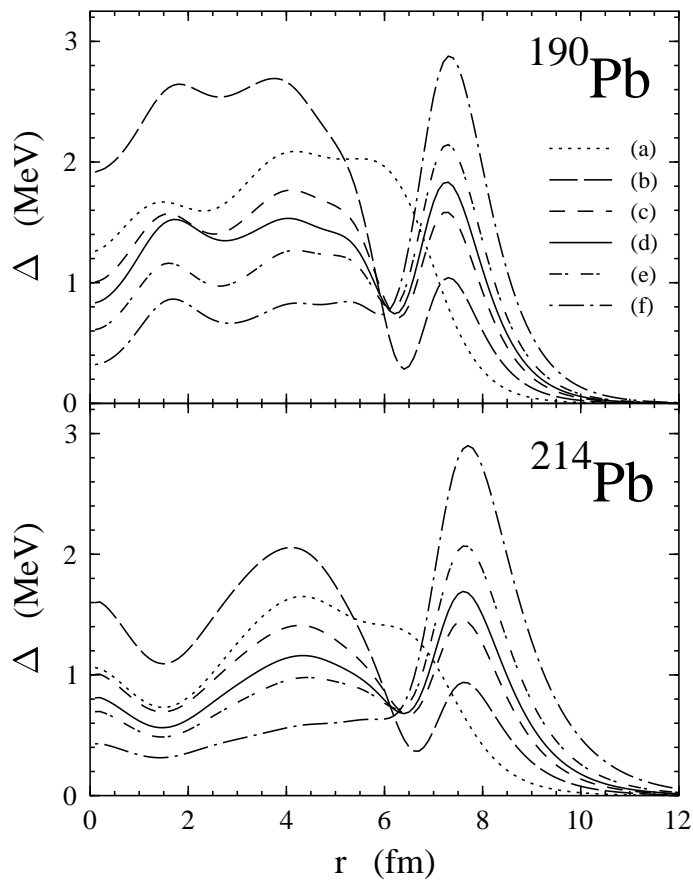} 
\end{center}
\caption{Pairing gap $\Delta$ for $^{190}$Pb (top) and $^{214}$Pb 
(bottom) as a function of the radial coordinate $r$. The curves are
marked in the same way as in Fig.~\ref{f:f5a}.} 
\label{f:f6}
\end{figure}

\begin{figure}[t]
\begin{center}
\leavevmode
\epsfxsize=22pc 
\epsfbox{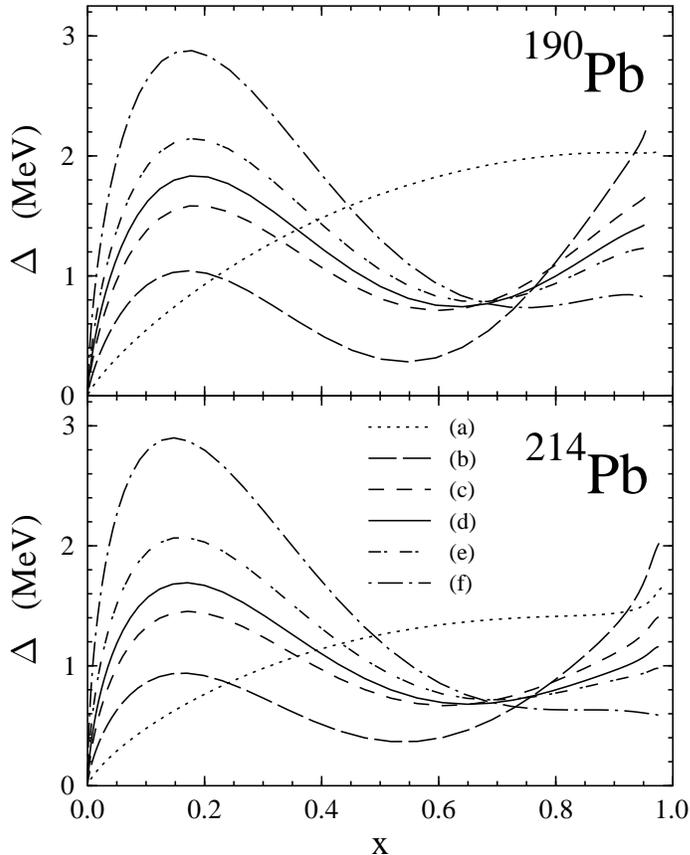} 
\end{center}
\caption{Pairing gap $\Delta$ for $^{190}$Pb (top) and $^{214}$Pb 
(bottom) as a function of the dimensionless isoscalar density $x$. 
The curves are marked in the same way as in Fig.~\ref{f:f5a}.} 
\label{f:f7}
\end{figure}

This fact clearly demonstrates the importance of including odd-mass
nuclei in the fitting procedure of force constants. Now, in general,
a complete description of the ground state of an odd nucleus is quite
difficult which reflects itself e.g. in the difficulty to reproduce
experimental magnetic moments. These depend sensitively on the
time-reversal-odd components of the polarization induced by the odd
particle. However, the operator $r^2$ is time-reversal-even, and the
small, not-well-known, odd components of the polarization enter the
expectation value only quadratically. If, with given interaction,
$\langle r^2\rangle$ is not well described by a HFB-type ground
state, there is no hope to reproduce it with a more sophisticated
state vector. As can be seen from Figs.~\ref{f:f5a} and~\ref{f:f5b}, 
it is possible to reproduce the order of magnitude of the staggering 
of $\langle r^2\rangle$ with density-dependent pairing 
force~(\ref{fxigrad}) containing a gradient term.

\begin{figure}[t]
\begin{center}
\leavevmode
\epsfxsize=22pc 
\epsfbox{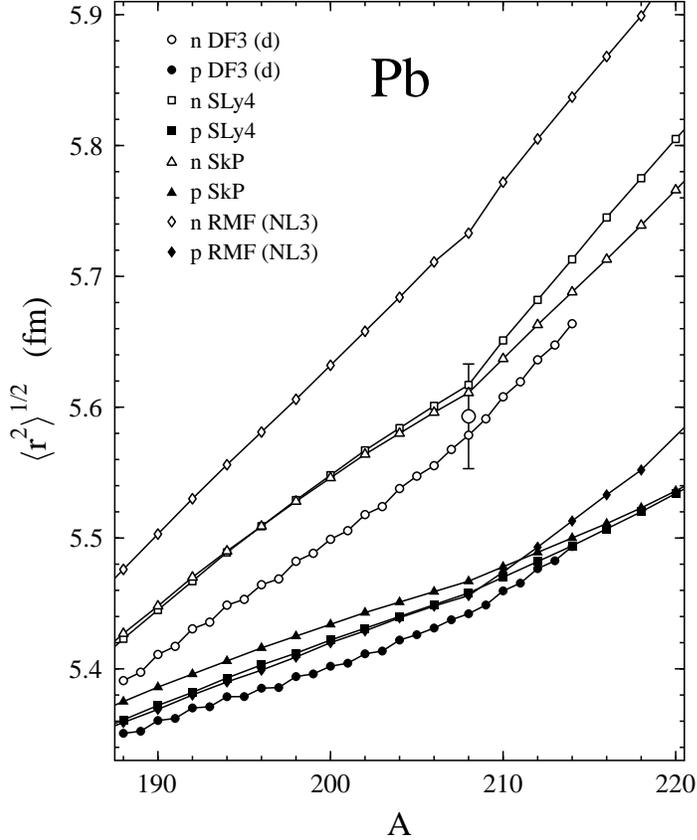} 
\end{center}
\caption{Root-mean-squared neutron (n, open symbols) and proton 
(p, solid symbols) radii for lead isotopes calculated with
functional DF3 and parameter set (d) of the pairing 
force~(\ref{pppar}) (circles), in comparison with the Skyrme--HFB 
calculations for the forces SLy4~\cite{SLy4} (squares) and 
SkP~\cite{DFT84} (triangles), and with the RMF--HBCS 
calculations~\cite{LRR99} (diamonds). The big open circle with
error bars shows the ``experimental'' rms neutron radius in 
$^{208}$Pb (see text).}  \label{f:f8}
\end{figure}

\begin{figure}[b]
\begin{center}
\leavevmode
\epsfxsize=22pc 
\epsfbox{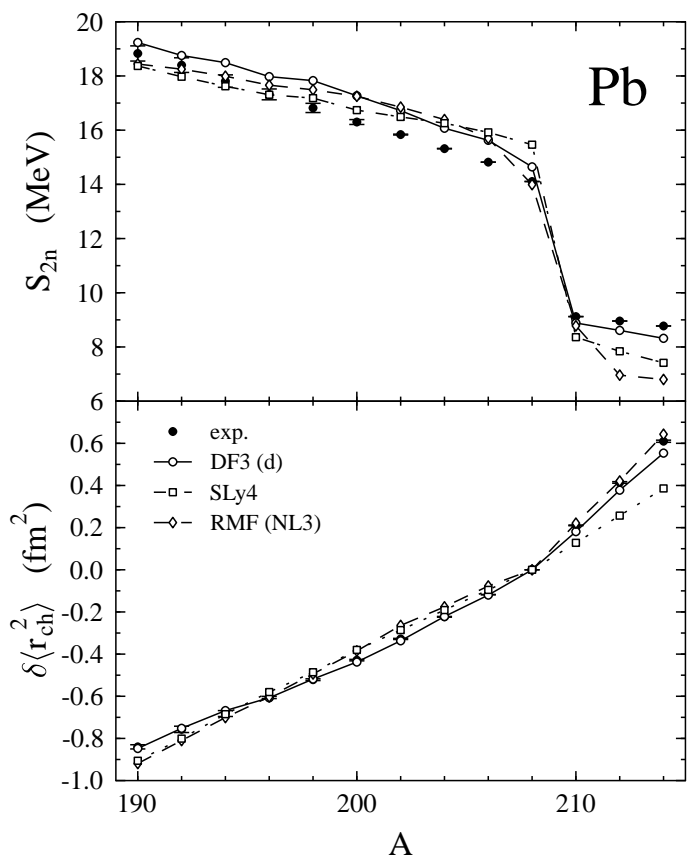} 
\end{center}
\caption{Two-neutron separation energies (upper panel) and
differences of mean squared charge radii with respect to $^{208}$Pb
(lower panel) for even lead isotopes. The open circles connected by 
the solid lines are obtained with functional DF3 and parameter set 
(d) of the pairing force~(\ref{pppar}). The open squares (diamonds) 
connected by the dotted (dashed) lines are from the Skyrme-SLy4 
(RMF--HBCS) calculations. Solid circles: experimental data for 
$S_{2\rm n}$, including those derived from systematic trends, 
from~\cite{AW93}, and data for $\delta\langle r^2_{\rm ch}\rangle$ 
from~\cite{Ans86,Ott89,Dut91}.}  \label{f:f9}
\end{figure}

Similar arguments apply to the nuclei off magic numbers. Looking only
at even neutron isotopes, the slope of $\langle r^2\rangle$ as a
function of $N$ changes at the magic numbers, producing a ``kink''.
With simple interactions in the pp-channel, this is not reproduced in
HFB-type calculations\footnote{A kink has, however, been obtained in
relativistic mean-field calculations, with simple pairing in the Pb
chain~\cite{Sha93}, at the expense of significantly too weak binding
of the neutron single particle states above the Fermi level of
$^{208}$Pb which increases the polarization effects through
neutron--proton interaction (see Ref.~\cite{RF95} for a detailed
discussion of this point, and also our commentary to Figs.~\ref{f:f8}
and~\ref{f:f9} of the present paper).}. This relative increase of
$\langle r^2 \rangle$ off magic numbers is thought to be connected
with ``dynamical deformation'' due to ground state fluctuations of
the surface. These ground state fluctuations of the surface, 
being time-independent, actually can not be separated from a static 
diffuseness of the surface, and are present in an independent
particle model too. In Section~\ref{rpacorr} below we will give some 
estimates of the contribution of RPA-type ground state correlations 
to this effect. It emerges that the main part of the effect 
should---and can---be obtained on the HFB level.

In Fig.~\ref{f:f6} the $r$-dependence of $\Delta$ for two isotopes
$^{190}$Pb and $^{214}$Pb (the ends of the measured chain) is shown.
One can see that with increasing external attraction $\propto
f^\xi_{\rm ex}$, $\Delta$ becomes smaller in the volume and gradually
concentrates in the outer part of the nuclear surface. This is
demonstrated more clearly in Fig.~\ref{f:f7} were $\Delta$ is plotted
as a function of dimensionless isoscalar density $x$. The curves
(c)--(f) are crossing very nearly at the same point which has
approximately the same coordinates in both cases: $x_{\rm piv}\approx
0.7$ and $\Delta_{\rm piv}\approx 0.7$--$0.8$~MeV. This 
``pivoting'' point corresponds to the one seen in Fig.~\ref{f:f1} for
$\Delta$ in nuclear matter but in finite nuclei the value of 
$\Delta_{\rm piv}$ turns out to be lower by a factor of 4 due to the 
repulsive gradient term $\propto f^\xi_\nabla$.

It is of interest to compare the predictions obtained with different 
mean-field approaches not only for the proton radii but also for the
neutron ones. An example is given in Fig.~\ref{f:f8} where the
results for the lead isotopes obtained with the functional DF3(d)
are shown in comparison with the HFB+SkP and HFB+SLy4 calculations, 
and with recently published relativistic mean-field Hartree
calculations~\cite{LRR99} based on the effective force NL3 and the 
constant pairing gap prescription of Ref.~\cite{MNix92} within the 
Bardeen-Cooper-Schriffer formalism (RMF--HBCS). It is seen that the
EDF model with gradient pairing yields staggering in the evolution
of the proton and neutron radii, and produces a kink in both of them
at A=208. The Skyrme-type functionals produce very small kinks in 
proton radii. The calculations with the RMF--HBCS model are given in
Ref.~\cite{LRR99} for even-even nuclei only, but one would guess that
this model is not able to produce noticeable odd-even staggering too.
At the same time, as seen in Fig.~\ref{f:f8}, both Skyrme--HFB 
and the RMF--HBCS functionals predict bigger neutron radii than our
EDF calculations, and all of them yield a distinctive kink in the
evolution of $\langle r^2\rangle_{\rm n}$ at $A=208$. One observes
rather close agreement between the proton radii obtained with all
considered models but a wide spread in the neutron radii. The 
RMF--HBCS neutron radii are particularly large. We remark 
at this point that the ``experimental'' rms neutron radius in 
$^{208}$Pb of 5.593~fm deduced from the analysis of the high-energy 
polarized proton scattering in Ref.~\cite{BFG89} is significantly 
lower than the prediction of the RMF(NL3) model; this fact has been
already mentioned in Refs.~\cite{Patyk99,DNW96x}. But one should 
bear in mind that the result of this analysis is model-dependent, 
and the errors in the extracted neutron radii could be quite large. 
The difference between neutron and proton rms radii for $^{208}$Pb,
+0.14$\pm$0.04~fm, has been also deduced in Ref.~\cite{BFG89},  
and we just put the corresponding error bars for the  
$\langle r^2 \rangle_n^{1/2}$ value in this nucleus as shown in 
Fig.~\ref{f:f8}. 

Considering general trends in the behavior of a given sequence of
nuclei, we notice that, because of the strong pn attraction, the
evolution of the proton radius should be driven by that of the
neutron one through self-consistency, and vice versa. Moreover, in
neutron-rich nuclei, an anticorrelation should exists between neutron
radii and neutron separation energies: the larger
$\langle r^2\rangle_{\rm n}$, the smaller $S_{2\rm n}$. One may
suspect that too big neutron radii with a kink at $N=126$, which is
correlated with a kink in proton radii, would mean too small
$S_{2\rm n}$ values beyond the shell closure. This is indeed the case 
as clearly demonstrated in Fig.~\ref{f:f9}. This figure displays 
two-neutron separation energies $S_{2\rm n}$ (upper panel) and
differences of mean square charge radii
$\delta\langle r^2_{\rm ch}\rangle$ with respect to $^{208}$Pb (lower
panel) for the even lead isotopes. The results obtained with the
RMF(NL3), Skyrme-SLy4 and EDF-DF3(d) models are compared there with 
experiment. It is seen that all these models reproduce $S_{2\rm n}$
values for lighter Pb isotopes below $A=208$ fairly well, with more
or less the same quality, the deviation from experiment being within
1~MeV or so (one should bear in mind that $S_{2\rm n}$ for the
$A=190$--$198$ lead isotopes are derived in Ref.~\cite{AW93} from
systematic trends, and the models discussed here describe these
trends reasonable well). But for the heavier lead isotopes the
predictions are different. Our EDF calculations with gradient pairing
agree with experimental $S_{2\rm n}$ values within 0.5 MeV while
Skyrme-SLy4 and RMF(NL3) models give deviations up to 1.5 and
2.0~MeV, respectively. It is also seen in the lower panel of
Fig.~\ref{f:f9} that the isotope shifts in charge radii for lighter
Pb isotopes are nicely reproduced with our EDF approach, the
$\delta\langle r^2_{\rm ch}\rangle$ values for the $A>208$ nuclei
and the kink being slightly underestimated. The Skyrme-SLy4 and
RMF(NL3) models work worse and yield results of equal quality for the
lighter isotopes, but produce much different kinks and 
$\delta\langle r^2_{\rm ch}\rangle$ values beyond $A=208$. The kink
obtained with the RMF model incorporating simple BCS pairing, with
constant pairing gap, is impressive indeed, and good description of
this anomaly in the isotope shifts of Pb nuclei has been reputed as
a considerable success of the relativistic mean-field approach. The
above consideration, which has much in common with that of
Ref.~\cite{RF95}, points out, however, the importance of
simultaneous description of both physical quantities, i.e. the 
energetic ($S_{2\rm n}$) and geometrical ( $\delta\langle
r^2_{\rm ch}\rangle$) differential observables, in order to 
get a deeper insight into the nature of this anomaly in the
isotope shifts.

\begin{figure}[t]
\begin{center}
\leavevmode
\epsfxsize=22pc 
\epsfbox{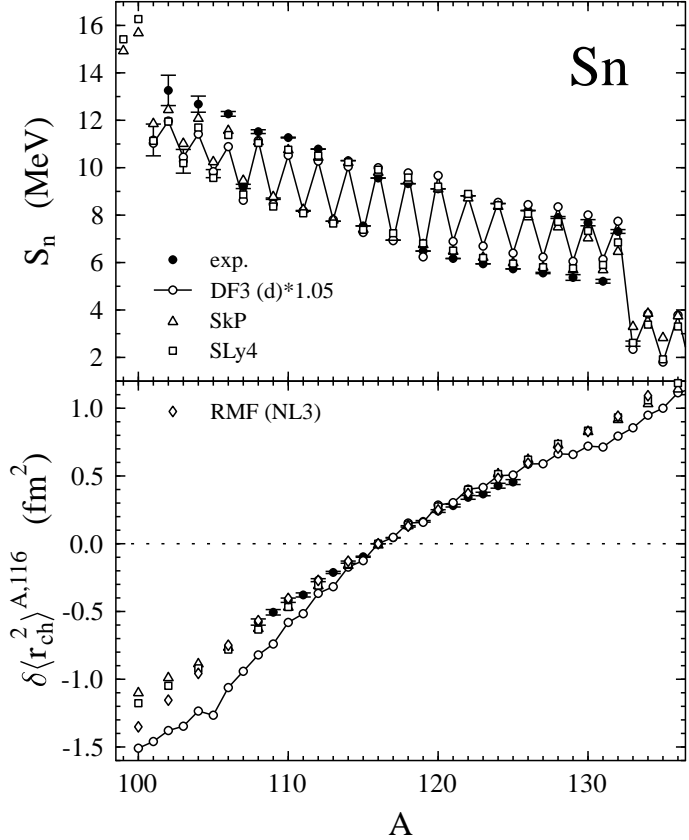} 
\end{center}
\caption{Neutron separation energies (top) and differences of mean
square charge radii with respect to $^{116}$Sn (bottom) for tin 
isotopes. Open circles: EDF calculation with the set (d) of the 
pairing force, eq.~(\ref{pppar}). Open triangles (squares): 
Skyrme--HFB calculation with the force SkP (SLy4). In the lower 
panel, the open diamonds correspond to the RMF--HBCS $\delta\langle
r^2_{\rm ch}\rangle$) values~\cite{LRR99}. Solid circles: 
experimental data from~\cite{AW93} (for $S_{\rm n}$) and~\cite{Ebe87} 
(for $\delta\langle r^2_{\rm ch}\rangle$).} \label{f:f10} 
\end{figure}

\begin{figure}[t]
\begin{center}
\leavevmode
\epsfxsize=22pc 
\epsfbox{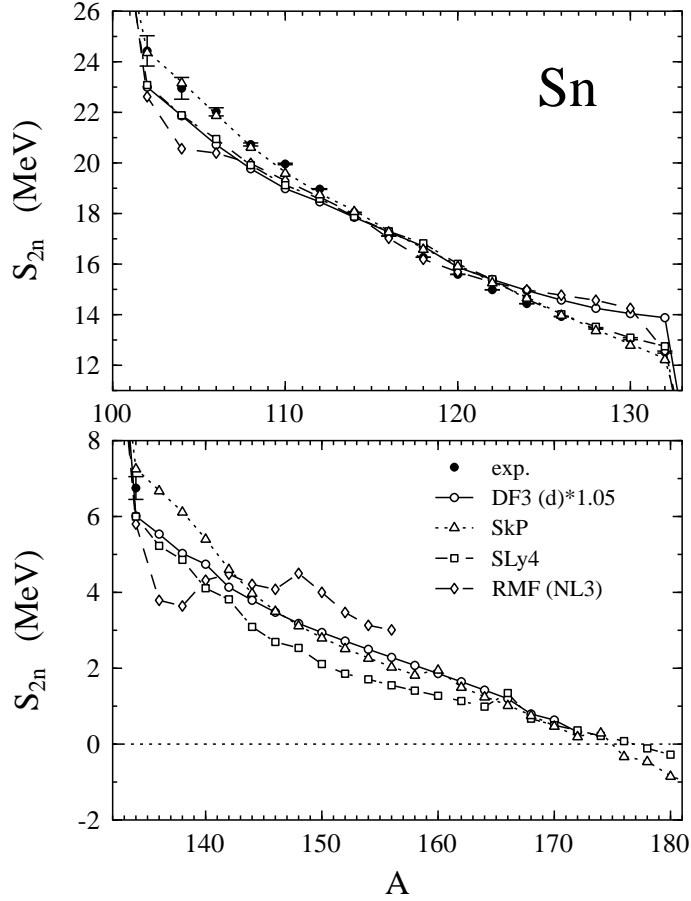} 
\end{center}
\caption{Two--neutron separation energies for even tin isotopes.
Open circles: EDF--DF3 calculation with the set (d) of the pairing
force, eq.~(\ref{pppar}), with a scaling factor of 1.05. Open 
triangles (squares): Skyrme--HFB calculation with the force SkP 
(SLy4). Open diamonds: RMF--HBCS calculation~\cite{LRR99}. Solid
circles: experimental data from~\cite{AW93}.}  \label{f:f11} 
\end{figure}

The calculated neutron separation energies and differences of mean 
squared charge radii with respect to $^{116}$Sn for tin isotopes 
with mass number from A=99 to A=136 are shown in Fig.~\ref{f:f10} 
in comparison with experimental data and with the Skyrme--HFB
calculations based on the SkP and SLy4 forces. In the lower panel of
this figure, we also show the RMF--HBCS $\delta\langle r^2_{\rm ch}
\rangle$ values~\cite{LRR99}. The results for radii from these three
models are available for even isotopes only. Of the six parameter 
sets~(\ref{pppar}) of the pairing force, only the results for our 
preferable set (d), with a scaling factor of 1.05, are shown. 
Similar to the lead chain, the $S_{\rm n}$ values for tin isotopes 
are reproduced reasonably well. However, on both ends of the 
considered chain, near the magic nuclei $^{100}$Sn and 
$^{132}$Sn, the common trend for all mean-field calculations 
presented in the upper panel of Fig.~\ref{f:f10} is that the size 
of the odd-even effect in neutron separation energies becomes 
noticeably smaller than the experimental one. As seen in the lower 
panel in Fig.~\ref{f:f10}, the evolution of the charge radii in the
Skyrme-type and RMF--HBCS calculations is rather smooth, the SkP,
SLy4 and RMF functionals slightly overestimate
$\delta\langle r^2_{\rm ch}\rangle$ in the heavier tin isotopes 
above $A=120$. The desirable size of staggering in the behavior of 
$\delta\langle r^2_{\rm ch}\rangle$ for tin isotopes could be
obtained with gradient pairing as shown by open circles in this 
figure. At the same time, the EDF calculations with functional DF3
significantly underestimate $\delta\langle r^2_{\rm ch}\rangle$ in
lighter isotopes below A=114. This fact seems to be correlated with
the weakening of the pairing when approaching A=100 and points out
the need of both more careful adjustment of the normal isovector 
part of the energy-density functional and, probably, invoking the
dependence of the pairing force on the isovector density 
$\rho_-=\rho_{\rm n}-\rho_{\rm p}$.

\begin{figure}[t]
\begin{center}
\leavevmode
\epsfxsize=22pc 
\epsfbox{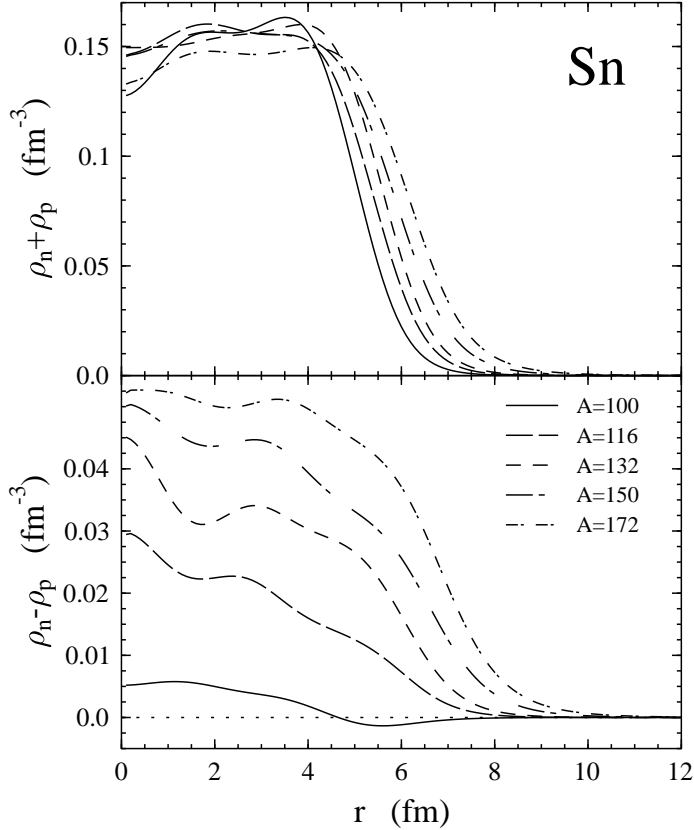} 
\end{center}
\caption{Isoscalar (top) and isovector (bottom) densities for some 
tin isotopes.}  \label{f:f12} 
\end{figure}

\begin{figure}[t]
\begin{center}
\leavevmode
\epsfxsize=22pc 
\epsfbox{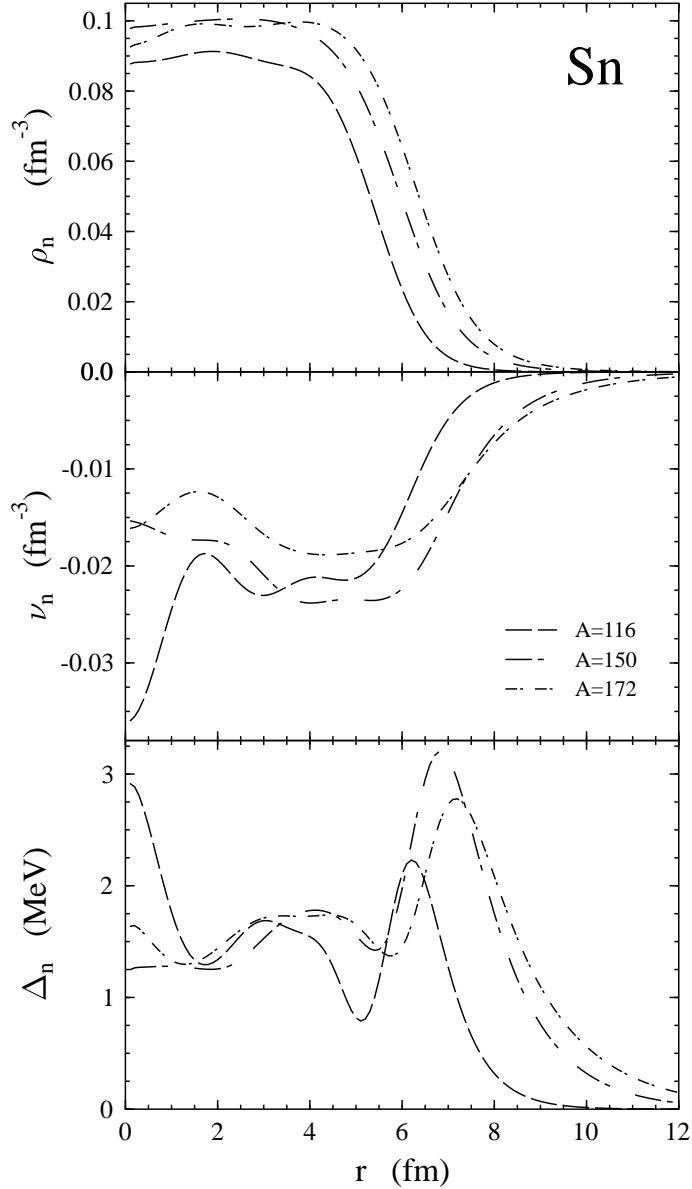} 
\end{center}
\caption{Neutron normal density (top), anomalous density (middle)
and pairing gap (bottom) in three selected tin isotopes.}
\label{f:f13} 
\end{figure}

\begin{figure}[t]
\begin{center}
\leavevmode
\epsfxsize=22pc 
\epsfbox{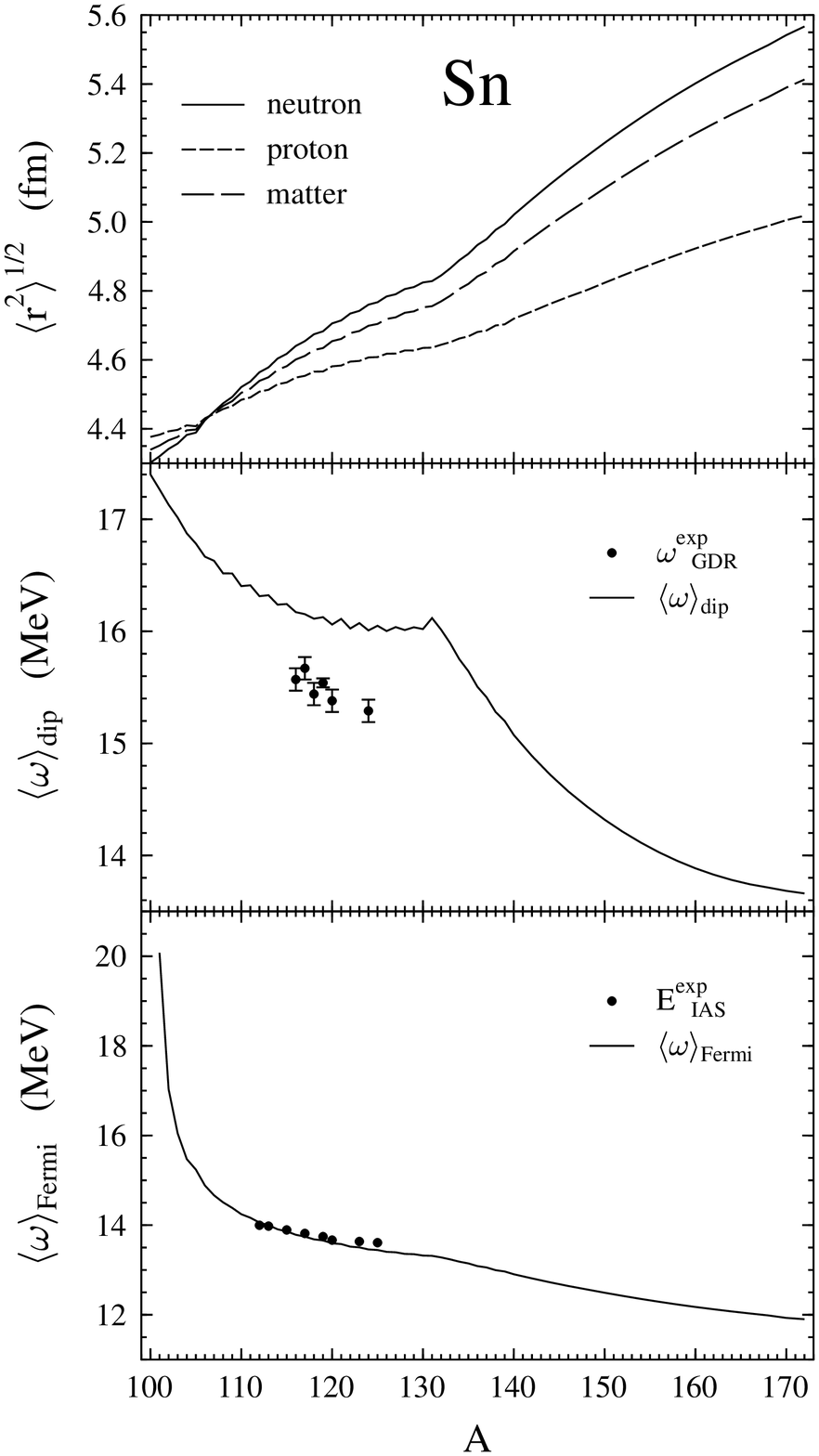} 
\end{center}
\caption{Neutron, proton and matter rms radii (top), mean energy of
isovector dipole (middle) and Fermi charge-exchange (bottom) 
transitions for tin isotopes. The solid dots for 
$\langle\omega\rangle_{\rm dip}$ correspond to the experimental 
data for the maximum of the dipole photoabsorption Lorentz curves, 
Refs.~\cite{Lep74,Ful69}, while those for
$\langle\omega\rangle_{\rm Fermi}$ 
to the experimental positions of the isobaric analog states with 
respect to the daughter nuclei, Ref.~\cite{ENSDF}.}  \label{f:f14} 
\end{figure}

The calculated two--neutron separation energies for even tin isotopes
up to the neutron drip line are shown in Fig.~\ref{f:f11}. Again, the
set (d) of the pairing force with a scaling factor of 1.05 was used
in the EDF calculations with the parametrization DF3. One can observe
noticeable differences between the Skyrme--HFB calculations with the
SkP and SLy4 forces, the RMF--HBCS predictions, and the EDF results.
The RMF-HBCS calculations~\cite{LRR99} are performed with deformed
code, and available for the tin isotopes only up to $A=156$; these
calculations produce some irregularities in the $S_{2\rm n}$ values
near $A=100$ and $A=132$ (the nuclei beyond $^{140}$Sn are predicted
to be well deformed). At the same time, the three other spherical
microscopic calculations give surprisingly similar predictions when
approaching the neutron drip line, the heaviest bound tin nucleus
being $^{172}$Sn, $^{174}$Sn and $^{176}$Sn for the EDF, SkP and
SLy4, respectively. One should notice that the Skyrme--HFB
calculations were performed with a discretized continuum in a
spherical box of finite size~\cite{DFT84} which allows to
artificially obtain a stable solution for the nuclear ground state
even for positive value of the chemical potential. Such a solution
is not possible with the coordinate-space technique used in the
present paper because of the physical boundary conditions imposed
for the scattering states. 

The evolution of the isoscalar and isovector densities along the tin
isotope chain is illustrated in Fig.~\ref{f:f12}. It is seen that
isovector density $\rho_-=\rho_{\rm n}-\rho_{\rm p}$ increases both
in the volume and at the surface as the neutron number $N$ becomes
larger. The isoscalar (matter) density tends to be slightly decreased
with $N$ in the volume and became more diffused at the surface. The
matter distributions in the two magic nuclei, $^{100}$Sn and
$^{132}$Sn, have a steeper slope at the surface compared to the
non-magic isotopes. The radial dependence of the neutron pairing
field $\Delta(r)$, of the anomalous density $\nu(r)$ and of the 
normal neutron density $\rho_{\rm n}(r)$ for the stable nucleus
$^{116}$Sn, for the unstable isotope $^{150}$Sn and for the drip-line
nuclide $^{172}$Sn are shown in~Fig.~\ref{f:f13}. One can see that in
the heavier isotopes, the anomalous density becomes more diffuse at
the surface with a longer tail compared to the normal densities, in
agreement with the discussion given in Appendix~C. It is also seen 
that the pairing field $\Delta$ in the outer part of the nuclear 
surface has a prominent maximum which moves outwards and gets a 
larger tail near the drip line. The concentration of $\Delta$ 
outside of the half density point is related to the specific 
cancellation between the external attraction and the repulsive 
gradient term in the effective pairing force~\cite{FZ96}. 

The staggering phenomenon observed in the behavior of charge radii
plotted as a function of neutron number is the one of the prominent
odd-even effects which has been systematically measured and widely
discussed. Similar effects are expected to exist for other quantities
such as neutron and matter radii, centroid energies of multipole
excitations (position of the giant resonances), etc.

Shown in the upper panel of Fig.~\ref{f:f14} are the
root-mean-square proton, neutron and matter radii for the tin
isotope chain calculated with the functional DF3 and with the set 
(d) of the pairing force. In the middle panel of this figure we show
the mean energies $\langle\omega\rangle_{\rm dip}$ of dipole
isovector excitations. These are calculated as the square root of 
the ratio $m_3/m_1$ with $m_1$ and $m_3$ the linear and cubic 
energy-weighted sum rule, i.e. the first and the third moment of the
corresponding RPA strength distribution, respectively~\cite{TF82}:
\begin{equation}
\langle\omega\rangle_{\rm dip}=\left[-{{\hbar^2}\over{3m}}
{A\over{NZ}} \int \d \vec r \d \vec r\,'\,
{{\partial\rho_{\rm n}}\over{\partial\vec r}}
{\mathcal F}^{\rm np}(\vec r,\vec r\,')
{{\partial\rho_{\rm p}}\over{\partial\vec r\,'}}
\right]^{1/2}\,.
\label{odip}
\end{equation}
Here  ${\mathcal F}^{\rm np}$ is the effective neutron--proton
interaction obtained from the energy-density functional as the
second variational derivative 
$\delta^2 E_{\rm int}/\delta\rho_{\rm n}\delta\rho_{\rm p}$. 
In the lower panel of Fig.~\ref{f:f14} we show the mean energies
$\langle\omega\rangle_{\rm Fermi}$ of the charge-exchange $0^+$
excitations, i.e. the Fermi transitions, in tin nuclei with $N>Z$.
These energies are calculated within the sum rule approach as the
ratio $(m_1^++m_1^-)/(m_0^+-m_0^-)$ with $m_1^+$ and $m_1^-$ the
first moment of the strength distribution of the Fermi transitions
in the $\beta^+$ and $\beta^-$ channel, respectively
(energy-weighted sum rules) and with $m_0^+$ and $m_0^-$ the
corresponding non-energy weighted sum rules. The resulting
expression for $\langle\omega\rangle_{\rm Fermi}$ is given by 
(see, for example, \cite{PF83}): 
\begin{equation}
\langle\omega\rangle_{\rm Fermi}={1\over{N-Z}}\int \d \vec r\,
U_{\rm Coul}(\vec r)\left(\rho_{\rm n}(\vec r)-
\rho_{\rm p}(\vec r)\right)\,,  \label{suru}
\end{equation}
where $U_{\rm Coul}$ is the Coulomb mean field potential.

It is seen in Fig.~\ref{f:f14} that the rms neutron, proton and 
matter radii as functions of the mass number $A$ reveal a kink 
at magic $^{132}$Sn, and at larger $A$ the difference between rms 
neutron and proton radii starts to increase more rapidly. The 
staggering in radii is present mostly in the region between the two 
magic nuclei, from $^{100}$Sn to $^{132}$Sn, and this effect is 
practically washed out beyond $A=140$. The mean energy of dipole 
transitions occurs to be in anticorrelation with such a behavior in 
radii, and this seem to be in qualitative agreement with 
experimental data (one should mention that eq.~(\ref{odip}) 
overestimates the position of the giant dipole resonance by 
$\approx 0.5$~MeV because of the cubic energy-weighted sum rule 
$m_3$ used in the derivation of $\langle\omega\rangle_{\rm dip}$). 
One also observes a distinct kink in the behavior of 
$\langle\omega\rangle_{\rm dip}$ at $A=132$. Beyond this magic 
number the mean dipole energy decreases rather fast and then nearly 
saturates when approaching $A=172$. This might be connected with an 
enhancement of the low-energy dipole transitions~\cite{Ham96} and 
also with possible appearance of the so-called soft dipole 
mode~\cite{Fay91} in nuclei near the neutron drip line. The 
anticorrelations in the staggering behavior of 
$\langle\omega\rangle_{\rm dip}$ and the rms radii
$\langle r^2\rangle^{1/2}_{\rm n}$,
$\langle r^2\rangle^{1/2}_{\rm p}$ can easily be understood by
considering the influence of pairing on the gradients of neutron and
proton densities entering eq.~(\ref{odip}). The odd-even effect in
$\langle\omega\rangle_{\rm dip}$ is constructive and more pronounced
in the $A\leq 132$ region where both gradients strongly overlap.
Beyond $^{132}$Sn, the larger differences between neutron and proton
rms radii imply lower overlap between neutron and proton density
gradients at the nuclear surface, hence a smaller mean dipole energy.
The situation with mean energy of the Fermi charge-exchange
transitions is different. From eq.~(\ref{suru}) one expects that the 
correlation between $\langle\omega\rangle_{\rm Fermi}$ and the
neutron and proton rms radii should be destructive. As seen in the
lower panel in Fig.~\ref{f:f14}, the staggering in the evolution of
$\langle\omega\rangle_{\rm Fermi}$ with $A$ is very weak indeed and
almost invisible. The kink at the magic mass number $A=132$ is also
much less pronounced compared to the dipole case. Remarkable enough, 
the theoretical self-consistent sum-rule predictions for
$\langle\omega\rangle_{\rm Fermi}$ in tin isotopes are in excellent
agreement with the available experimental data on the position of
the isobaric analog states.  

\begin{figure}[t]
\begin{center}
\leavevmode
\epsfxsize=22pc 
\epsfbox{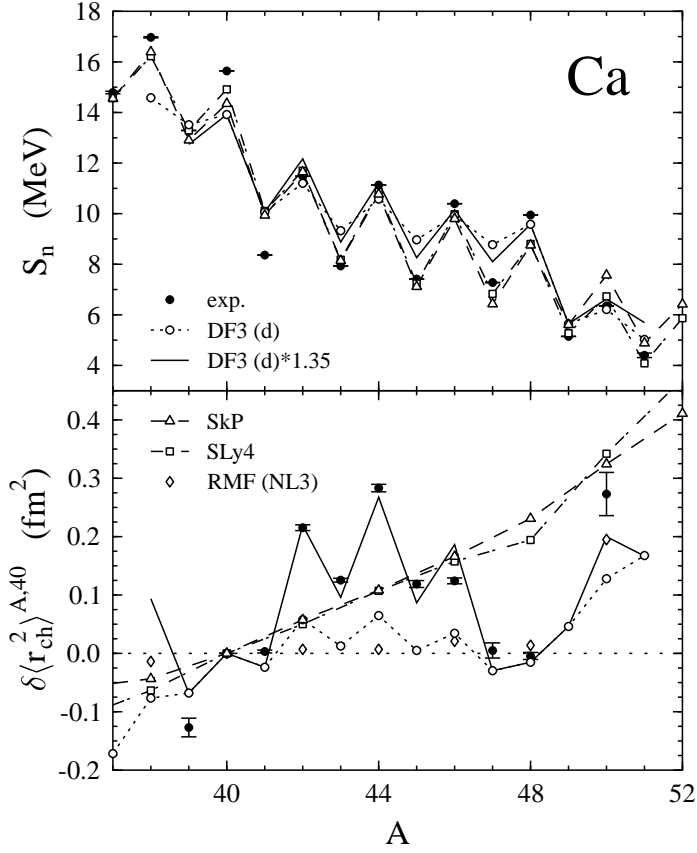} 
\end{center}
\caption{Neutron separation energies (top) and differences of mean
squared charge radii with respect to $^{40}$Ca (bottom) for calcium
isotopes. Open circles: the EDF calculation with the set (d) of the
pairing force, eq.~(\ref{pppar}). Solid lines connect the EDF 
results also obtained for the set (d) but with a scaling factor of 
1.35. Open triangles (squares): Skyrme--HFB calculation with the 
force SkP (SLy4). In the lower panel, the open diamonds correspond 
to the RMF--HBCS $\delta\langle r^2_{\rm ch}\rangle$ 
values~\cite{LRR99} (for even isotopes only). Solid circles: 
experimental data from~\cite{AW93} (for $S_{\rm n}$) 
and~\cite{Palm84,Verm92,Verm96} (for 
$\delta\langle r^2_{\rm ch}\rangle$).} \label{f:f15}
\end{figure}

\begin{figure}[t]
\begin{center}
\leavevmode
\epsfxsize=22pc 
\epsfbox{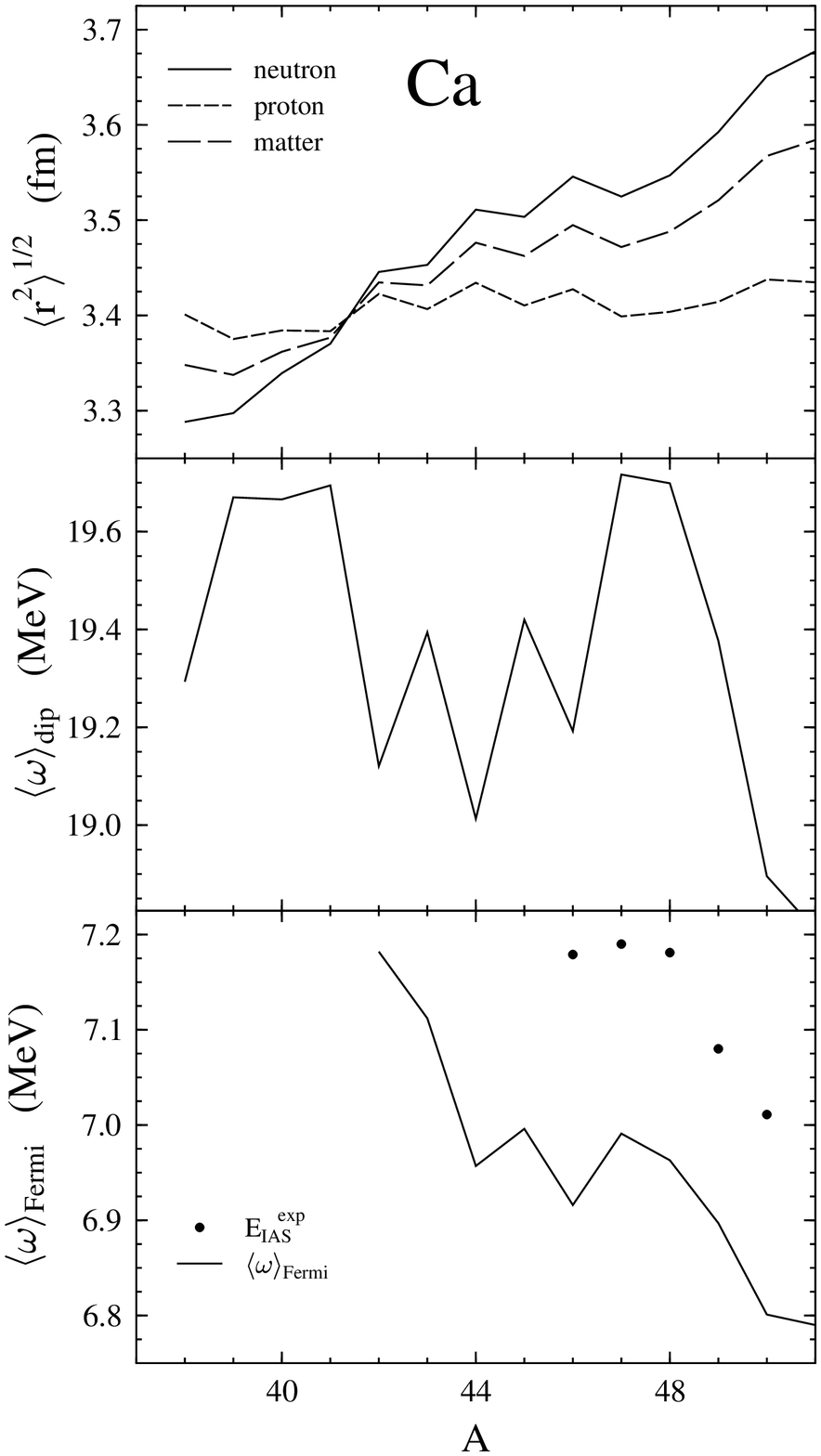} 
\end{center}
\caption{Neutron, proton and matter radii (top), mean energy of
isovector dipole (middle) and Fermi charge-exchange (bottom)
transitions for calcium isotopes. The solid dots for 
$\langle\omega\rangle_{\rm Fermi}$ correspond to the experimental
positions of the isobaric analog states with respect to the daughter
nuclei, Ref.~\cite{ENSDF}.}  \label{f:f16} 
\end{figure}

Fig.~\ref{f:f15} displays the neutron separation energies and isotope
shifts in mean square charge radii for the calcium isotope chain. In 
this case, the EDF calculations with the preferable set (d),
eq.~(\ref{pppar}), deduced from the Pb isotopes yield noticeable but 
rather small staggering both in $S_{\rm n}$ and 
$\delta\langle r^2_{\rm ch}\rangle$. A possible way to get the 
desirable size of odd-even effects is to enhance the pairing 
correlations by introducing a scaling factor 1.35 for the pairing 
force (all parameters being scaled in the same way). It is seen 
then in Fig.~\ref{f:f15} that the neutron separation energies are 
reproduced fairly well, with more or less the same quality as with 
the Skyrme SkP and SLy4 functionals, though the effect of the 
scaling is rather mild. The effect on the behavior of charge radii
is more drastic: the 
$\delta\langle r^2_{\rm ch}\rangle$ values are increased by a factor 
of 3 to 4 yielding a very good agreement with experiment. This
example demonstrates that the anomalous $A$-dependence of charge 
radii in Ca isotopes could be correctly described within the EDF
approach with renormalized pairing force. However, the possibility 
to determine a single parametrization of the effective interaction 
in the pp channel, which would be valid both for heavy and light 
nuclei, remains an open problem. In any case, we got an indication
that the staggering in $\delta\langle r^2_{\rm ch}\rangle$ of Ca
isotopes, which is a challenge for nuclear models, could be
explained with a peculiar density dependence of the pairing force.  
One can see in Fig.~\ref{f:f15} that the Skyrme SkP and SLy4 
functionals produce a nearly smooth increase of radii with $A$, with
weak irregularities below and above $A=48$, but still with rather
sizeable kinks at double magic $^{48}$Ca. The RMF--HBCS predictions
for $\delta\langle r^2_{\rm ch}\rangle$, Ref.~\cite{LRR99}, 
are much different: the charge radius is nearly a constant up to 
$^{48}$Ca, with a sudden increase for $^{50}$Ca. Interestingly, the 
functional DF3 yields, in agreement with experiment, a small 
negative isotope shift in charge radii between $^{40}$Ca and 
$^{48}$Ca, which does not depend on pairing. 

The EDF calculations, as demonstrated in Fig.~\ref{f:f16}, show 
that the trends and the staggering effects in the mean energy of 
isovector dipole excitations and charge-exchange Fermi transitions 
for calcium isotopes are in strong anticorrelation with evolution 
of the ground state mean squared radii. Similar to the case of the 
tin isotopes discussed above, the anticorrelation in 
$\langle\omega\rangle_{\rm Fermi}$ is weaker than in 
$\langle\omega\rangle_{\rm dip}$. Unfortunately, there are no 
systematic experimental data on dipole excitations in Ca isotopes 
at our disposal, from which the $\langle\omega\rangle_{\rm dip}$ 
values could be deduced. As for the mean energies of the Fermi 
transitions, which are calculated with the self-consistent EDF sum 
rule approach and plotted as a function of $A$ in the lower panel 
in Fig.~\ref{f:f16}, they may be compared with the experimental 
positions of the isobaric analog states. It is seen that the sum 
rule expression, eq.~(\ref{suru}), underestimates the energy of the 
analog states in calcium isotopes by some 200~keV but reproduces 
qualitatively the observed trend.   
 
As, in the past, RPA ground state correlations have been invoked to
explain details of the isotopic dependence of nuclear charge radii,
we estimate the influence of RPA ground state correlations.

\Section{Ground state fluctuations}
\label{rpacorr}
An important point is the issue of RPA-type ground state 
correlations. In several papers \cite{Ott89,EsB83,BaB85} the 
influence of the average mean square deformations on the mean 
square charge radii has been discussed in a phenomenological way 
which is based on the following sum rule:  Take the identity for 
the multipole operator $Q_{\lambda,\mu}=\int \d {\vec r}
r^\lambda Y_{\lambda,\mu}(\Omega)\psi^\dagger({\vec r})
\psi({\vec r})$
\begin{equation}
\sum_\mu\langle 0\vert\;\vert Q_{\lambda,\mu}\vert^2 \vert 0
\rangle=\sum_{\mu,s}\vert\langle s\vert Q_{\lambda,\mu}\vert 0
\rangle\vert^2\;.  \label{sum0}
\end{equation}
The right hand side is equal to the summed transition strength
$\sum_s \!B({\rm E}\lambda;\!{0\!\!\to \!\! s})$,
the left hand side is the expectation value of the squared multipole 
operator $Q_\lambda$ in the ground state. In semiclassical nuclear 
models, this expectation value is proportional to the square of the 
deformation of multipolarity $\lambda$. Thus, even for spherical 
nuclei the rms deformation does not vanish. In the liquid drop model 
this is interpreted as due to zero point surface vibrations, but it 
is to be emphasized that there is no observable time dependence, and 
no breaking of the spherical symmetry due to this ``dynamical'' 
deformation.  Experimentally, the effect can not be distinguished 
from diffuseness of the surface arising from any other mechanism.

Our method (as well as any HF or HFB calculation) incorporates only 
the dynamical deformation effects corresponding to a gas of 
noninteracting particles (magic nuclei) or of a gas with only 
pairing (nonmagic nuclei). To include the RPA ground state 
correlations, the calculated charge radii should be increased by
\begin{equation}
\delta \langle r_{\rm ch}^2\rangle =
\bar\delta \langle r_{\rm ch}^2\rangle_{\rm RPA}
-\bar\delta \langle r_{\rm ch}^2\rangle_{\rm HFB}.  \label{fcorr}
\end{equation}
However, the parameters have been adjusted to reproduce the 
experimental density distribution and therefore the average RPA 
correlations are included due to the choice of parameters. In a high 
precision fit, this should be corrected for, by subtracting the 
value given by (\ref{fcorr}) for the reference nucleus.

A microscopic method to calculate the corrections to
$\langle r^2_{\rm ch}\rangle$ due to low-energy surface vibrations 
has been proposed in~\cite{ZPS89}. We use the liquid drop model and 
eq.~(\ref{sum0}) to estimate the influence of isoscalar collective 
excitations on the charge radii, as proposed by Esbensen and 
Bertsch~\cite{EsB83}.

As already noted in~\cite{EsB83}, the inclusion of noncollective 
states in eq.~(\ref{sum0}), which can not be considered to be 
surface oscillations will not do much harm, because the contribution 
of these states will be essentially the same to
$\bar\delta\langle r^2_{\rm ch}\rangle_{\rm RPA}$ and
$\bar\delta\langle r^2_{\rm ch}\rangle_{\rm HFB}$ and will drop out.
According to~\cite{Ott89,Boh52,ReSo71}, for multipolarity 
$\lambda >0$
\begin{equation}
\bar\delta\langle r^2_{\rm ch}\rangle \equiv \langle r^2
\rangle_{\rm d}-\langle r^2\rangle_{\rm s}=
{5\over{4\pi}}\langle r^2\rangle_{\rm s}\sum_\lambda \langle
\beta_\lambda^2\rangle\,,  \label{radef}
\end{equation}
where the suffixes  d and s mean (dynamically) deformed and
spherical, respectively.

Within the same model
\begin{equation}
S^0({\rm E}\lambda)\equiv\sum_s B(E\lambda;0\!\rightarrow\! s) =
\sum_\mu\langle 0\vert\;\vert Q_{\lambda,\mu}\vert^2 \vert 0\rangle
=\left({{3ZeR_0^\lambda}\over{4\pi}}\right)^2\langle\beta_\lambda^2
\rangle\,.  \label{quamo}
\end{equation}

With the help of~(\ref{quamo}) and~(\ref{sum0}) the sum over 
$\beta_\lambda^2$ in eq.~(\ref{radef}) can be replaced by the sum 
over the total $B({\rm E}\lambda)$-values, and we reach the 
following expression:
\begin{equation}
\delta\langle r_{\rm ch}^2\rangle={{4\pi}\over{3Z^2e^2}}
\sum_{\lambda} {1\over{R_0^{\lambda-2}}}
[S_{\rm RPA}^0({\rm E}\lambda) - S_{\rm HFB}^0({\rm E}\lambda)]\,.
\label{findel}
\end{equation}
Of course, for $\lambda =1$ one has to exclude the spurious 
translation. For $\lambda = 0$ we have to use a compressible drop 
model, and obtain similarly
\begin{equation}
\delta\langle r_{\rm ch}^2\rangle={{140\pi}\over
{108Z^2e^2R_0^2}}
[S_{\rm RPA}^0({\rm E}0) - S_{\rm HFB}^0({\rm E}0)]\,.
\label{findel0}
\end{equation}
where $S({\rm E}0)$ is calculated according to eq.~(\ref{quamo})
with $Q_{00}=r^2Y_{00}$.

\begin{table}[t]
\begin{description}
\item[Table 1.] Excitation energies (MeV) and transition 
probabilities $(e^2~\cdot $fm$^{2L})$ for low-lying collective 
states. Experimental data are taken from Refs.~\cite{TabCa1,TabCa2} 
(for Ca), \cite{TabSn} (for Sn) and~\cite{TabPb} (for Pb).
\end{description}
$$ 
\begin{tabular}{|c|c|c|c|c|c|c|c|c|}
     \hline
 Isotope &
 \multicolumn{2}{c|}{exp for $2^+_1$} &
 \multicolumn{2}{c|}{theory for $2^+_1$} &
 \multicolumn{2}{c|}{exp for $3^-_1$} &
 \multicolumn{2}{c|}{theory for $3^-_1$} \\ \cline{2-9}
 & & & & & & & & \\ [-4mm]
 & $E_x$ & {\em B(E2)} & $E_x$ & {\em B(E2)} &
   $E_x$ & {\em B(E3)} & $E_x$ & {\em B(E3)} \\
 \hline\hline
 $^{40}$Ca &  -   &     -              &  -   &    -
           & 3.74 & $1.24 \cdot 10^4$ & 3.79 & $1.06 \cdot 10^4$ \\
 $^{42}$Ca & 1.53 &                    & 3.15 & $0.53 \cdot 10^2$
           & 3.45 &                    & 4.47 & $1.05 \cdot 10^4$ \\
 $^{44}$Ca & 1.16 & $ \!\! 4.75\pm 2.65 \cdot 10^2 \!\! $ & 3.00 & $0.71
 \cdot 10^2$
           & 3.31 &                    & 4.70 & $0.93 \cdot 10^4$ \\
 $^{46}$Ca & 1.35 & $1.78 \cdot 10^2$ & 2.22 & $0.48 \cdot 10^2$
           & 3.61 &                    & 4.73 & $0.66 \cdot 10^4$ \\
 $^{48}$Ca & 3.83 & $0.86 \cdot 10^2$ & 3.18 & $0.45 \cdot 10^2$
           & 4.51 & $0.67 \cdot 10^4$ & 4.46 & $0.38 \cdot 10^4$ \\ \hline

 $^{100}$Sn &  -   &     -              & 4.34 & $0.11 \cdot 10^4$
            &  -   &     -              & 6.00 & $1.00 \cdot 10^5$ \\
 $^{108}$Sn & 1.21 &                    & 1.76 & $0.17 \cdot 10^4$
            &      &                    & 3.70 & $0.74 \cdot 10^5$ \\
 $^{112}$Sn & 1.26 & $0.24 \cdot 10^4$ & 1.73 & $0.20 \cdot 10^4$
            & 2.35 & $0.87 \cdot 10^5$ & 3.36 & $0.80 \cdot 10^5$ \\
 $^{116}$Sn & 1.29 & $0.195 \cdot 10^4$ & 1.71 & $0.17 \cdot 10^4$
            & 2.27 & $0.60 \cdot 10^5$ & 3.17 & $0.83 \cdot 10^5$ \\
 $^{120}$Sn & 1.17 & $0.29 \cdot 10^4$ & 1.67 & $0.14 \cdot 10^4$
            & 2.40 & $0.90 \cdot 10^5$ & 3.24 & $0.84 \cdot 10^5$ \\
 $^{124}$Sn & 1.13 & $0.17 \cdot 10^4$ & 1.74 & $0.11 \cdot 10^4$
            & 2.62 & $0.60 \cdot 10^5$ & 3.49 & $0.79 \cdot 10^5$ \\
 $^{132}$Sn &  -   &     -       & 4.32 & $0.07 \cdot 10^4$
            & 4.04 &     -       & 4.94 & $1.06 \cdot 10^5$ \\ \hline

 $^{190}$Pb &  -   &     -              & 1.31 & $5.40 \cdot 10^3$
            &  -   &     -              & 2.43 & $0.42 \cdot 10^6$ \\
 $^{200}$Pb & 1.03 &                    & 1.23 & $3.40 \cdot 10^3$
            &      &                    & 3.01 & $0.46 \cdot 10^6$ \\
 $^{204}$Pb & 0.90 & $1.66 \cdot 10^3$ & 1.24 & $1.70 \cdot 10^3$
            & 2.61 &                    & 3.14 & $0.52 \cdot 10^6$ \\
 $^{206}$Pb & 0.80 & $1.15 \cdot 10^3$ & 1.20 & $0.92 \cdot 10^3$
            & 2.65 & $0.64 \cdot 10^6$ & 3.14 & $0.54 \cdot 10^6$ \\
 $^{208}$Pb & 4.08 & $3.18 \cdot 10^3$ & 4.87 & $2.42 \cdot 10^3$
            & 2.61 & $0.61 \cdot 10^6$ & 3.00 & $0.58 \cdot 10^6$ \\
 $^{210}$Pb & 0.80 & $0.51 \cdot 10^3$ & 1.29 & $0.40 \cdot 10^3$
            & 1.87 & $0.47 \cdot 10^6$ & 2.42 & $0.41 \cdot 10^6$ \\
 $^{212}$Pb & 0.81 &                    & 1.32 & $0.89 \cdot 10^3$
            & 1.82 &                    & 2.17 & $0.49 \cdot 10^6$ \\ \hline
\end{tabular} 
$$
\end{table}

\begin{table}[t]
\begin{description}
\item[Table 2.] $\delta\langle r^2_{\rm ch}\rangle$ (in fm$^2$) for different
multipolarities for $^{40}$Ca, $^{44}$Ca, and $^{48}$Ca isotopes calculated
within the framework of self consistent QRPA.
\end{description}
$$\mbox{%
\begin{tabular}{|c|c|c|c|}
     \hline
 $J^\pi$ &
 \multicolumn{3}{c|}{$\delta\langle r^2_{\rm ch}\rangle$} \\ \cline{2-4}
 & & & \\ [-4mm]
 & $^{40}$Ca & $^{44}$Ca & $^{48}$Ca \\
 \hline\hline
 $0^+$           & -0.0062 & -0.0046 & -0.0043 \\
 $2^+$           &  0.0322 &  0.0597 &  0.0278 \\
 $3^-$           &  0.2604 &  0.1463 &  0.0432 \\
 $4^+$           &  0.0111 &  0.0149 & -0.0293 \\
 $\sum_{2^+,3^-}$ &  0.2926 &  0.2060 &  0.0710 \\
 $\sum_{\rm tot}$     &  0.2975 &  0.2163 &  0.0374 \\
\hline
\end{tabular}}$$
\end{table}

\begin{figure}[b]
\begin{center}
\leavevmode
\epsfxsize=22pc 
\epsfbox{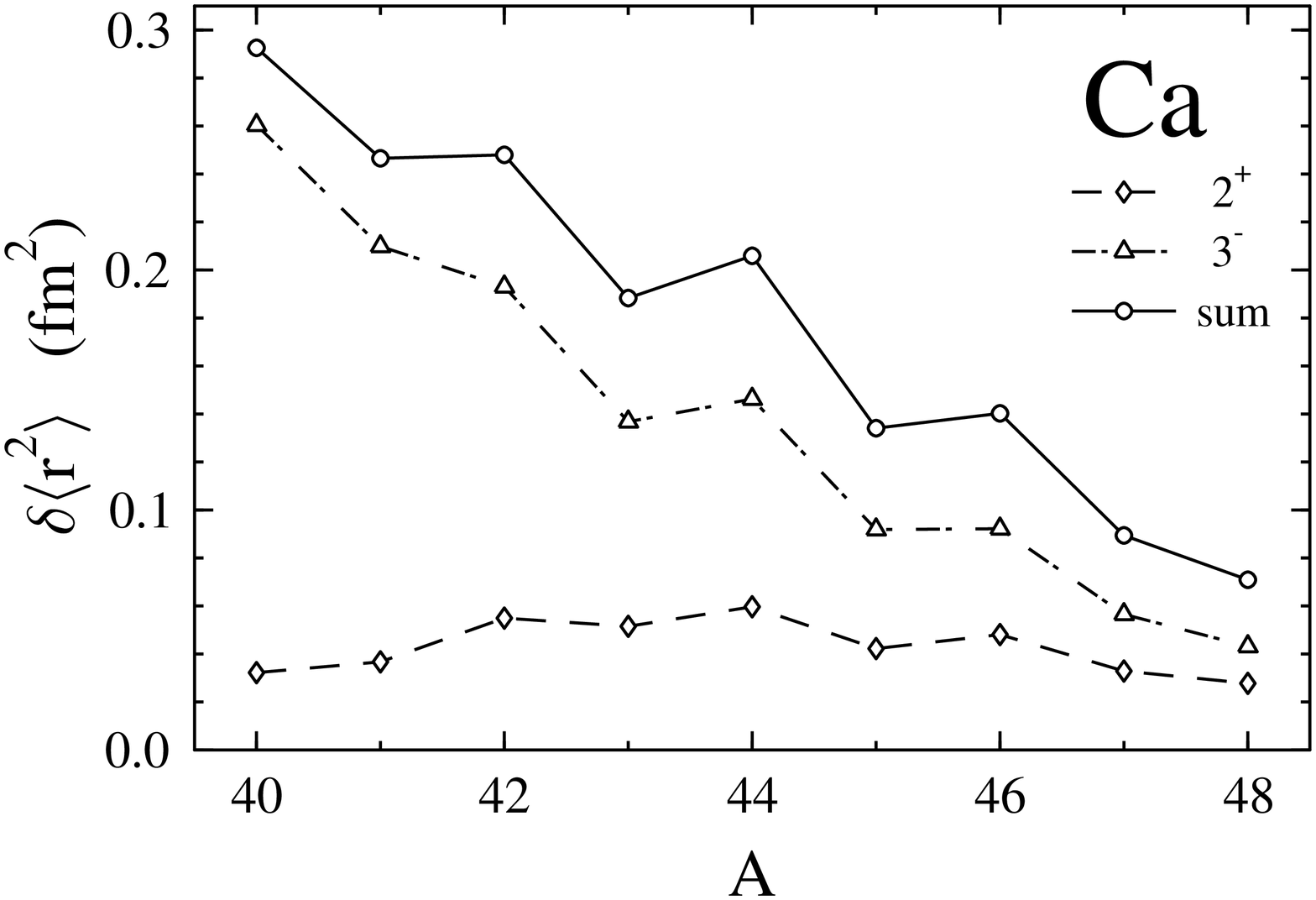} 
\end{center}
\caption{Contribution of RPA ground state correlations to the
squared charge radii in calcium isotopes. Shown are the $2^+$
(diamonds) and the $3^-$ (triangles) contributions and the sum of
both (circles). It is seen that $^{40}$Ca is quite soft for 3$^-$
deformation and gets a big contribution from the corresponding 
ground state fluctuations.}  \label{f:f17}
\end{figure}

The excited states have been calculated in quasiparticle-RPA (QRPA)
which has been used in the form (we adopt the same symbolic 
notations as in Ref.~\cite{Migdal}):
\begin{equation}
\hat{V}= e_q \hat{V}_0 + \hat{F}\hat{A}\hat{V}\,,\quad
S_{\rm RPA}^0 =-{1\over\pi}{\rm Im}(e_q\hat{V}_0\hat{A}\hat{V})\,,
\label{qrpa1}
\end{equation}
where $e_q$ is the quasiparticle local charge with respect to the 
external field $V_0$. For a long-range electric field, $V_0\propto 
Q_{\lambda,\mu}$, from gauge invariance one gets 
$e_q=1$~\cite{Migdal}. The $\hat{A}$ matrix is written in detail 
in~\cite{BTF95}.

As in our formalism the effective interaction amplitudes $\hat{F}$ 
are obtained as the second functional derivatives of $E_{\rm int}$ 
with respect to the corresponding densities, one can see here that 
due to $\rho$-dependence of $\varepsilon_{\rm anomal}$, 
eq.~(\ref{epair}), there are mixed terms for the effective 
interaction, $F^{\omega \xi}$ and $F^{\xi \omega}$, which do not 
vanish, and which couple the ph with the pp or hh channel 
($\tau$-indices are neglected for simplicity):
\begin{equation}
{\mathcal F}^{\omega \xi}={{\textstyle \delta^2 E_{\rm int}
[\rho, \nu]}\over{\textstyle \delta \rho \delta \nu}} =
{{\delta^2}\over{\delta \rho \delta \nu}}
\int \varepsilon_{\rm anomal}([\rho, \nu];\vec r)\d \vec r\,,
\label{qrpa4}
\end{equation}

A comparison of QRPA results for the first collective $2^+$ and 
$3^-$ states with experiment is presented in Table~1 for selected 
nuclei from the Ca, Sn and Pb isotope chains. The influence of the 
spin--orbit part in the ph interaction has been investigated 
in~\cite{FTZ94}; the results given here have been obtained without 
this spin--orbit contribution. Comparison with~\cite{FTZ94} shows 
that here the changes in the functional did not improve the 
agreement with experimental data.

\begin{figure}[t]
\begin{center}
\leavevmode
\epsfxsize=22pc 
\epsfbox{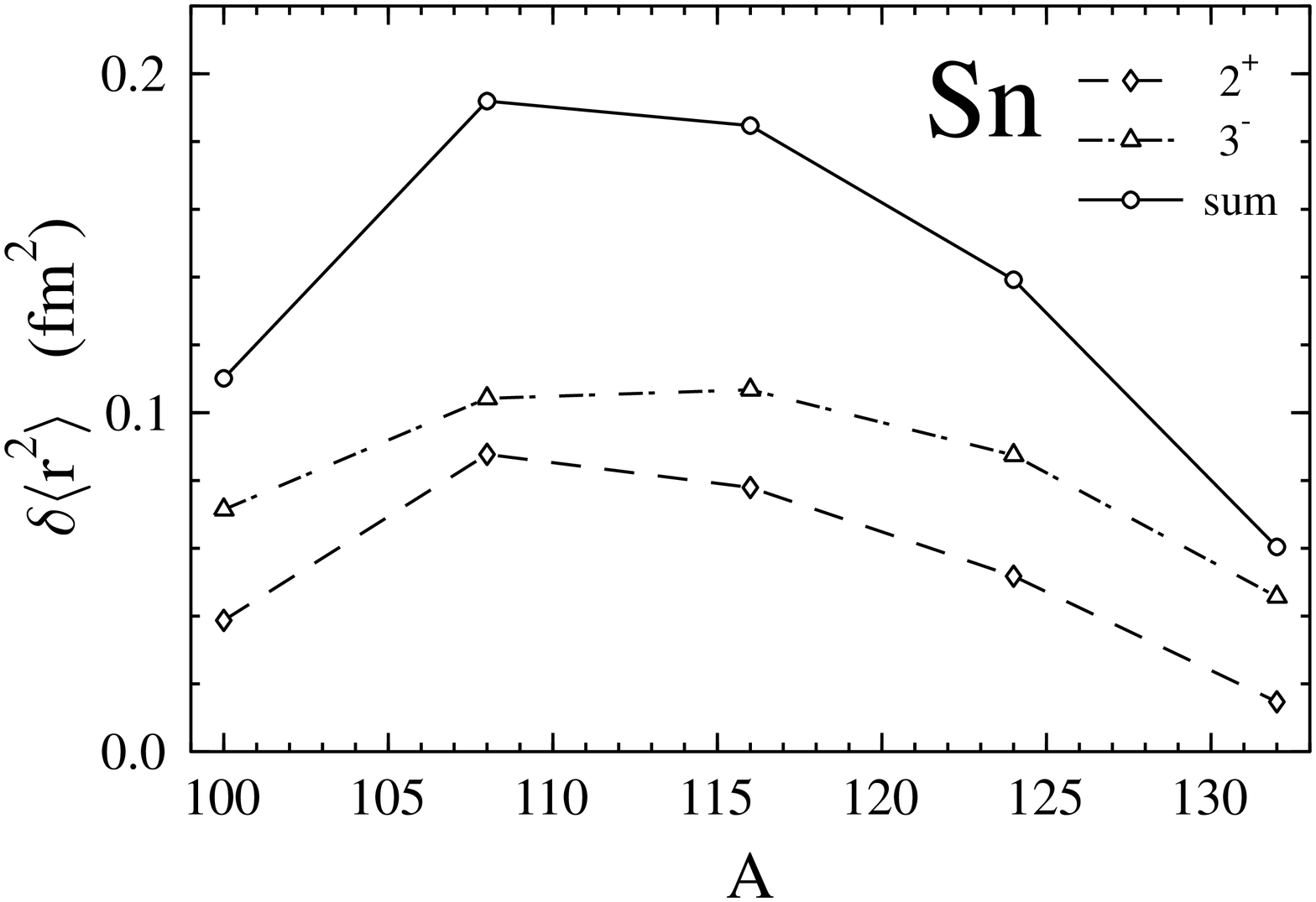} 
\end{center}
\caption{The same as Fig.~\ref{f:f17}, for selected isotopes from
the tin chain.}
\label{f:f18}
\end{figure}

\begin{figure}[t]
\begin{center}
\leavevmode
\epsfxsize=22pc 
\epsfbox{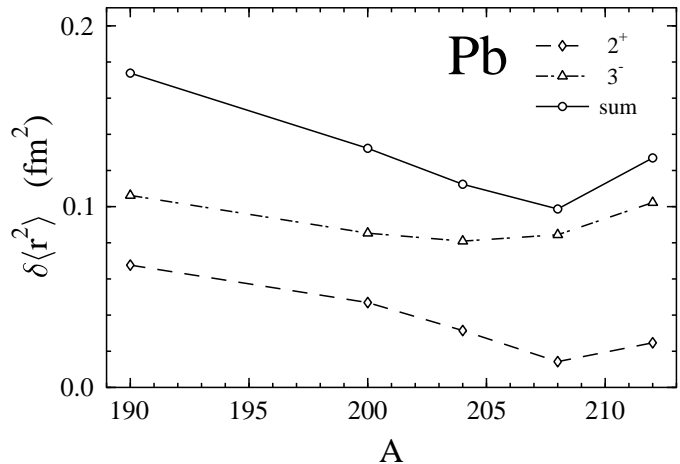} 
\end{center}
\caption{The same as Fig.~\ref{f:f18} for lead.}
\label{f:f19}
\end{figure}

We restricted our consideration to $2^+$ and $3^-$ states for all 
isotope chains because the other multipolarities contribute much 
less to the mean square charge radii~\cite{ZPS89}. This has been 
checked by computing also
$\delta\langle r^2_{\rm ch}\rangle$ from $0^+$ and $4^+$ states
for three calcium isotopes. In all these calculations the functional 
parameter set of Ref.~\cite{FTTZ94} have been used. These results 
are presented in Table~2 in comparison with the calculations for 
$2^+$ and $3^-$ multipolarities. One can see a here significant 
difference between $2^+$ and $3^-$ states on one hand and $0^+$ and 
$4^+$ states on the other hand. Therefore it is sufficient to 
include only strong phonons which contribute significantly to the 
mean square radii. This result is very similar to that 
of Barranco and Broglia~\cite{BaB85} which they obtained with the 
approach of Esbensen and Bertsch~\cite{EsB83} for the calculation of
$\delta\langle r^2_{\rm ch}\rangle$ values. However, it should be 
mentioned that they used a non self consistent model with more 
freedom in the choice of parameters.

Let us now look at the results of our calculation for the calcium, 
tin and lead isotope chains which are presented in Figs.~\ref{f:f17},
\ref{f:f18}, and~\ref{f:f19}, respectively. One can see here that 
for calcium isotopes $3^-$ states give the main contribution to 
$\delta\langle r^2_{\rm ch}\rangle_{\rm tot}$, especially for 
$^{40}$Ca (Fig.~\ref{f:f17}). The $2^+$ contribution is not very 
big, in contrast to the result of \cite{BaB85}. This is connected 
with the fact that we could not obtain the experimental values 
for excitation energies and especially for $B({\rm E}2)$ values in 
calcium isotopes in the middle of the chain. Presumably for good 
reproduction of the experimental values one has to go beyond the 
QRPA in a model which allows to take into account more complex 
configurations than 1p1h. Nevertheless, it is shown that there is 
considerable influence of ground state correlations on the calcium 
mean square charge radii, at least in second order in the phonon 
amplitudes. Combining the EDF results shown in Fig.~\ref{f:f17} with
those in Fig.~\ref{f:f15} one sees that the staggering in charge 
radii of calcium isotopes could be enhanced by the calculated ground
state correlation corrections; however, this would produce too big 
negative isotope shift between $^{40}$Ca and $^{48}$Ca thus 
deteriorating the good agreement with experiment seen in 
Fig.~\ref{f:f15} alone.     

In Figs.~\ref{f:f18} and~\ref{f:f19} the ground state correlation
contribution to $\delta\langle r^2_{\rm ch}\rangle$ is shown for a 
few isotopes of the tin and lead chains. These nuclei are more stiff,
and QRPA works much better (see Table~1). The results show that here
the ground state correlation contribution is only a small fraction 
of the value obtained without correlations. Moreover, the general 
trends look roughly like the trends of the curves for 
$\delta\langle r^2_{\rm ch}\rangle$ from Figs.~\ref{f:f5a}, 
\ref{f:f5b} and~\ref{f:f10}, which therefore are enhanced but not 
qualitatively changed by the correlations.

\Section{Summary and conclusions}
\label{sumconc}
In this work the energy density functional method for superfluid 
nuclei has been described in detail. The generalized variational 
principle has been formulated for the local functional with a 
contact effective force ${\mathcal F}^\xi$ in the 
particle--particle channel and cutoff energy
$\epsilon_{\rm c} \ge \epsilon_{\rm F}$. For spherical finite 
systems, the coordinate-space technique, involving the integration
in the complex energy plane of the Green's functions obtained by 
solving the Gor'kov equations exactly, with physical boundary 
conditions both for bound and scattering quasiparticle states has 
been used. The technique has been extended to odd nuclei by using 
a uniform filling approximation.       

Within the framework of this EDF approach, the combined analysis of
differential observables including odd-even mass differences and 
odd-even effects in charge radii was performed for a few isotopic 
chains. The staggering and kinks in 
$\delta\langle r^2_{\rm ch}\rangle$ were shown to be very sensitive
to the density dependence of the effective pairing force. The 
observed isotopic shifts were successfully described with the force
containing a density-gradient pairing term. With such a
parametrization, as the external attraction $\propto f^\xi_{\rm ex}$ 
increases, the pairing field $\Delta(r)$ becomes smaller in the 
volume and concentrates in the outer part of the nuclear surface. 

A few possible parameter sets extracted from the lead chain were used
to study the ground state properties of uniform nuclear matter with 
$s$-wave pairing, in particular the behavior of the energy gap at the
Fermi level as a function of density. The deduced sets, which give a
satisfactory description of both the neutron separation energies 
$S_{\rm n}$ and isotope shifts $\delta\langle r^2_{\rm ch}\rangle$ 
in charge radii, yielded about the same value of 
$\Delta \approx 3.3$~MeV at $k_{\rm F}=1.15$~fm$^{-1}$ (at
$\approx 0.65$ of the equilibrium density). It was found that some 
sets in the dilute limit correspond to the strong coupling regime. 
The situation when the regime changes from weak to strong pairing 
was considered in detail, and, in strong coupling, the properties 
of dilute matter were shown to be completely determined in leading 
order by the singlet scattering length $a_{\rm nn}$. In weak 
coupling, near gap closure, the pairing gap $\Delta_{\rm F}$ is 
shown to be directly expressed through the Fermi level phase shifts.
The density-dependent cutoff contact pairing interaction 
corresponding to the realistic value of the singlet NN scattering 
length was found to be a preferable choice between the deduced 
parameter sets to be used in the EDF calculations.        

The isotopic chain of tin isotopes from the proton to the neutron 
drip line was calculated, and the evolution of the normal and 
anomalous (pairing) density distributions with $A$ was analyzed. 
The parametrization DF3 of the normal part of the functional and 
the density-dependent pairing interaction provided fairly good 
description of odd-even effects in $S_{\rm n}$ and 
$\delta\langle r^2_{\rm ch}\rangle$; however, in the region of 
lighter tin isotopes the slope in 
$\delta\langle r^2_{\rm ch}\rangle$ plotted as a function of the 
mass number $A$ come out steeper than in experiment. This might 
indicate that dependence on isovector density 
$\rho_-=\rho_{\rm n}-\rho_{\rm p}$ should be incorporated in the 
effective pairing interaction. Anomalous behavior of charge radii 
in Ca isotopes was reproduced in excellent agreement with experiment
within the EDF approach by scaling the pairing effective interaction
which was extracted from the lead chain by a factor of 1.35.
   
It was shown with the self-consistent sum rule approach that the 
density-dependent pairing induces sizeable staggering and kinks in 
the evolution of the mean energies of multipole excitations along 
isotopic chains. These effects were found to be in anticorrelation 
with the behavior of the ground state radii.

Using quasiparticle-RPA multipole strength distributions obtained 
with the self-consistent interaction, the contribution of phonons 
(ground state correlations) to the differences of charge radii was 
calculated by the method proposed by Esbensen and 
Bertsch~\cite{EsB83}. It was shown that these phonon corrections, 
at least for tin and lead isotopes, are quite small in comparison 
with the values provided by the main HFB contribution, and do not 
lead to qualitatively new effects.

The results obtained in the present paper were compared with the 
conventional Skyrme--HFB, Gogny--HFB and relativistic mean field 
Hartree-BCS calculations. On the whole, a better description of the 
differential observables was achieved with the EDF method, the most
important effect which was successfully reproduced is the staggering
in charge radii. Up to now, this effect was practically missed in all 
other mean-field models. The famous kink in the isotope shifts of Pb 
isotopes has been also well described by our EDF calculations, with 
density-dependent gradient pairing, and the crucial point is that 
the neutron separation energies are reproduced as well. These results
support the conclusion drawn by Reinhard and Flocard~\cite{RF95} 
that this is apparently not the case in the RMF model.
These authors showed that the description of the kink in Pb chain 
is not proprietary to the relativistic model, it can be also produced 
with the Skyrme functionals by generalizing the spin-orbit term. 
However, even with their ``best'' set SkI4, Reinhard and Flocard have
found a large gap in the single-particle neutron spectrum similar to 
that of the RMF model. Our conclusion is that the physics behind the 
anomaly in the Pb radii is not solely due to the spin-orbit (which, 
moreover, has nothing to do with the staggering!), and that the 
simultaneous reproduction of both the energetic and geometrical 
observables is important to settle the problem.
   
Unfortunately, a universal parametrization of the pairing force is
still lacking, and for various isotope chains, especially for lighter
nuclei, the parameters of ${\mathcal F}^\xi$ had to be chosen
somewhat different. It seems to be necessary to use a more refined
parametrization, e.g. with additional dependence of
${\mathcal F}^\xi$ on $\rho_-$. Also, more attention should be payed
to the normal part of the density functional in order to choose more
carefully its parameters and improve the description of the bulk
properties of nuclei such as their masses and absolute values of 
radii~\cite{Fayans98,FZmbx}. These questions will be investigated
in future work.

\ack{We thank J.-F.~Berger, J.~Dobaczewski, V.V.~Khodel and 
E.~Saperstein for useful discussions. Partial support of this work by 
the Deutsche Forschungsgemeinschaft and by the Russian Foundation for 
Basic Research (project 98-02-16979) is gratefully acknowledged.}

\newpage
\setcounter{equation}{0}
\renewcommand{\theequation}{A.\arabic{equation}}
\setcounter{section}{0}
\section*{Appendix A. General variational principle and
ground state energy in local approximation}

In this appendix we show that, in the case of weak pairing 
$\vert\Delta\vert \ll \epsilon_{\rm F}$, the general variational
principle for a local density functional with an energy cutoff
$\epsilon_{\rm c} > \epsilon_{\rm F}$ can be used to describe the
ground state properties of superfluid systems. This allows to
perform a full HFB-like calculation with a renormalized gap equation
and an effective pairing $\delta$-force.

We start with the usual definitions and equations which follow from
the general variational principle. The total energy $E$ of a
superfluid system, as given by eqs.~(\ref{EHFBC})--(\ref{EHFB}), 
is a functional of the generalized one-body density
matrix $\widehat R$. This matrix can be expressed in terms of the
ground state expectation values of the pair products of the 
Landau-Migdal quasiparticle creation and annihilation operators 
$\psi^\dagger,\; \psi$:

\begin{eqnarray}
{\widehat R}(1,2) & = &\left(\matrix{
\langle \psi^\dagger(2)\psi(1)\rangle
&    \langle \psi(2)\psi(1)\rangle          \cr
\langle \psi^\dagger(2)\psi^\dagger(1)\rangle
&    \langle \psi(2)\psi^\dagger(1)\rangle  \cr
}\right)\nonumber\\ 
& = & \left(\matrix{
\rho(1,2)     &  \nu(1,2) \cr 
- \nu^* (1,2) &  \delta(1,2) - \rho^*(1,2)\cr
}\right)\,. 
\label{GDM}
\end{eqnarray}
Here the coordinate space representation is used. The numbers in 
brackets stand for the set of all relevant arguments and indices,
e.g. $(1)=(\{\vec r,s_z,\tau_3\}_1)$. In the following,
$\int \d 1$ means integration over the continuous and summation
over the discrete variables of this set.

The Bogolyubov quasiparticle creation and annihilation operators 
$\beta^\dagger_\alpha$, $\beta_\alpha$ are defined by the  
transformation
\begin{equation}
\left(\begin{array}{c} \beta_\alpha
\\ \beta^\dagger_\alpha \end{array}\right)
= \int \d 1 \left(\begin{array}{cc} U^*_\alpha(1) & V^*_\alpha(1)\\
V_\alpha(1)  &  U_\alpha(1) \end{array}\right)
\left(\begin{array}{c} \psi(1)\\\psi^\dagger(1)\end{array}\right)\,.
\end{equation}
This can be inverted to express the Landau-Migdal quasiparticle
operators in terms of those of Bogolyubov 
\begin{equation}
\left(\begin{array}{c} \psi(1) \\ \psi^\dagger(1)\end{array}\right)
= \sum_\alpha \left(\begin{array}{cc} U_\alpha(1)
&  V^*_\alpha(1) \\
V_\alpha(1)  &  U^*_\alpha(1) \end{array}\right)
\left(\begin{array}{c} \beta_\alpha
\\ \beta^\dagger_\alpha \end{array}\right)\,.
\label{psibet}
\end{equation}

The generalized Bogolyubov quasiparticle density matrix 
$\widehat Q$ is given by
\begin{equation}
{\widehat Q}_{\alpha \alpha^\prime} = \left(\matrix{
\langle \beta^\dagger_{\alpha^\prime} \beta_\alpha\rangle
& \langle \beta_{\alpha^\prime} \beta_\alpha\rangle           \cr
\langle \beta^\dagger_{\alpha^\prime} \beta^\dagger_\alpha\rangle
& \langle \beta_{\alpha^\prime} \beta^\dagger_\alpha\rangle   \cr
}\right)\;.
\label{QMAT}
\end{equation}
By demanding now the ground sate $\vert \Phi_0 \rangle$ to be the
Bogolyubov quasiparticle vacuum, 
$\beta_\alpha \vert \Phi_0 \rangle =0$,
the matrix $\widehat Q$ becomes diagonal in the
$\alpha$--representation:
\begin{equation}
\label{GQD}
{\widehat Q}_{\alpha \alpha^\prime}
 = \delta_{\alpha\alpha^\prime}Q\,, \quad {\rm with} \quad
Q = \left(\begin{array}{cc}
0 & 0 \\ 0 & 1 \end{array}\right)\,,
\end{equation}
and for (\ref{GDM}) one obtains a ``supermatrix'' formula:
\begin{equation}
{\widehat R}(1,2)  = \sum_{\alpha\alpha^\prime}
{\mathcal W}_\alpha(1)
{\widehat Q}_{\alpha \alpha^\prime} 
{\mathcal W}^\dagger_{\alpha^\prime} (2)
= \sum_\alpha {\mathcal W}_\alpha(1) Q 
{\mathcal W}^\dagger_\alpha (2)\,,  \label{GR}
\end{equation}
with the matrix ${\mathcal W}$ of the Bogolyubov transformation
(\ref{psibet}):
\begin{equation}
{\mathcal W}_\alpha(1) = \left(\begin{array}{cc}
U_\alpha(1) &  V^*_\alpha(1) \\
V_\alpha(1)  &  U^*_\alpha(1)\end{array}\right)\,.
\end{equation}
For the generalized density matrix we thus find
\begin{equation}
{\widehat R}(1,2) = \sum_\alpha\left(\matrix{
V_\alpha^*(1)V_\alpha(2) & V_\alpha^*(1)U_\alpha(2) \cr
U_\alpha^*(1)V_\alpha(2) & U_\alpha^*(1)U_\alpha(2) \cr}
\right)\,.
\label{RGEN}
\end{equation}
The matrix ${\mathcal W}$ obeys the generalized closure
and orthogonality relations:
\begin{equation}
\sum_\alpha {\mathcal W}_\alpha(1) {\mathcal W}^\dagger_\alpha (2)
= \left(\begin{array}{cc}
1 & 0 \\ 0 & 1 \end{array}\right)\delta(1,2)
\equiv{\hat I}\delta(1,2) \,,
\label{COMP}
\end{equation}
\begin{equation}
\int \d 1 {\mathcal W}^\dagger_\alpha(1) 
{\mathcal W}_{\alpha^\prime}(1)
= {\hat I} \delta_{\alpha\alpha\prime} \,.  \label{ORT}
\end{equation}
The generalized density matrix $\widehat R$ is Hermitian and
from eqs.~(\ref{GQD}--\ref{ORT}) follows its important
property~(\ref{R2})
which, in the coordinate space representation, reads
\begin{equation}
\int \d 3{\widehat R}(1,3){\widehat R}(3,2) = {\widehat R}(1,2)\,.
\label{RR}
\end{equation}
This property reflects the fact that the 
ground state is a HFB quasiparticle vacuum.

Minimizing the energy and imposing the average particle number 
conservation (\ref{consN}) and the relation (\ref{RR}) 
as constraints, we arrive at the variational principle 
$\delta I[\widehat R] = 0\,$ with the variational functional
of eq.~(\ref{VFUNC}) which we rewrite here as
\begin{equation}
I[\widehat R] = E[\widehat R] - \mu {\rm Tr}{\hat \rho} -
{\rm Tr}({\hat\Lambda}({\widehat R} - {\widehat R}^2))\,.
\label{II}
\end{equation}
The trace of a supermatrix $\hat A$, in the coordinate space 
representation, is defined by
\begin{equation}
{\rm Tr} \hat A
= {\rm Tr}\left(\begin{array}{cc}
{\hat A}^{11} & {\hat A}^{12} \\
{\hat A}^{21} & {\hat A}^{22}
\end{array}\right) = \int \d 1 \sum_i A^{ii}(1,1)\,.
\end{equation}
The total energy of the system, eqs.~(\ref{EHFBC})--(\ref{EHFB}),
is the functional 
\begin{eqnarray}
E[\widehat R] & = & {\rm Tr}(t{\hat\rho}) + E_{\rm int}[\widehat R]
= {\rm Tr}(t{\hat\rho}) + E_{\rm int(normal)}[\hat\rho]
+ E_{\rm anomal}[\widehat R]\nonumber\\ 
& = &  E_{\rm normal}[\hat\rho] + E_{\rm anomal}[\widehat R]\,,
\label{TENE}
\end{eqnarray}
where $t$ is the free kinetic energy operator and 
$E_{\rm int}[\widehat R]$ is the interaction energy containing 
both normal and anomalous terms. Because of condition (\ref{ORT}), 
the matrix $\Lambda_{\alpha \alpha^\prime}$, in the Bogolyubov 
quasiparticle space, may be taken to be diagonal:
\begin{equation}
\Lambda_{\alpha \alpha^\prime}
=\left(\begin{array}{cc}
E_\alpha & 0 \\ 0 & E_\alpha \end{array}\right)
\delta_{\alpha \alpha^\prime}
\equiv {\hat I} E_\alpha \delta_{\alpha \alpha^\prime}\,,
 \end{equation}
where $E_\alpha$ is the set of Lagrange multipliers.
In coordinate space representation one gets
\begin{equation}
\hat\Lambda (1,2)
=  \sum_\alpha E_\alpha{\mathcal W}_\alpha (1)
{\mathcal W}^\dagger_\alpha (2)\,.
\end{equation}
The variational equation reads
\begin{equation}
\delta\left(E[\rho,\nu] - \mu {\rm Tr}\rho - 
{\rm Tr}({\hat\Lambda}({\widehat R} - {\widehat R}^2))\right) = 0\,.
\label{VEQ}
\end{equation}
For the variation of the first two terms in this equation we have
\begin{equation}
\delta(E[\rho,\nu] - \mu {\rm Tr}\rho)
= {\rm Tr}(\hat{\mathcal H} \delta{\widehat R})
= \int \d 1 \d 2 \hat{\mathcal H}(1,2) \delta{\widehat R}(2,1)\,,
\label{TrHR}
\end{equation}
where the effective quasiparticle Hamiltonian 
$\hat {\mathcal H}$ is given by
\begin{equation}
\hat{\mathcal H}(1,2) = \left(\matrix {
h(1,2) - \mu\delta(1,2)   &  \Delta(1,2) \cr
-\Delta^*(1,2)  & \mu\delta(1,2) - h^*(1,2)  \cr}
\right)\,,
\label{EQPH}
\end{equation}
with
\begin{equation}
h(1,2) = \frac {\delta E[\rho,\nu]}{\delta\rho(2,1)}\,, \quad
\label{hdef}
\Delta(1,2) = \frac {\delta E[\rho,\nu]}{\delta\nu^\dagger(2,1)}
= \frac {\delta E[\rho,\nu]}{\delta\nu^*(1,2)} \,.
\end{equation}
The variation of the generalized density matrix,
\begin{equation}
\delta{\widehat R}(1,2)  = \left(\begin{array}{cc}
\delta\rho(1,2) & \delta\nu(1,2) \\ 
- \delta\nu^* (1,2) & -\delta \rho^*(1,2)\end{array}\right)\,, 
\end{equation}
may be written in the form
\begin{equation}
\delta{\widehat R}(1,2) = \sum_{\alpha\alpha^\prime}
{\mathcal W}_\alpha(1)
\delta{\widehat Q}_{\alpha \alpha^\prime}
{\mathcal W}^\dagger_{\alpha^\prime}(2)\,,
\end{equation}
where $\delta{\widehat Q}_{\alpha \alpha^\prime}$ is the variation
of the Bogolyubov quasiparticle density matrix~(\ref{QMAT}).
The choice of the matrix elements
$\delta{\widehat Q}_{\alpha \alpha^\prime}$
as independent variables leads to the HFB equations in a simple way. 
The variation of~(\ref{RR}) gives
\begin{equation}
\delta(\widehat R (1,2) - {\widehat R}^2(1,2)) =
\sum_{\alpha\alpha^\prime}
{\mathcal W}_\alpha(1)
\left(\begin{array}{cc}
1 & 0 \\ 0 & -1 \end{array}\right)
\delta{\widehat Q}_{\alpha \alpha^\prime}
{\mathcal W}^\dagger_{\alpha^\prime}(2)\,.
\end{equation}
Eq.(\ref{TrHR}) may thus be written as
\begin{eqnarray}
{\rm Tr}(\hat{\mathcal H} \delta{\widehat R})
 & = & \sum_{\alpha\alpha^\prime}{\rm Tr}
{\mathcal W}^\dagger_{\alpha^\prime}\hat{\mathcal H}
{\mathcal W}_\alpha
\delta{\widehat Q}_{\alpha \alpha^\prime} \nonumber\\
 & = & \sum_{\alpha\alpha^\prime}\int \d 1\d 2 \sum_{abcd}
{{\mathcal W}^\dagger}^{ab}_{\alpha^\prime}(1)
{\hat{\mathcal H}}^{bc}(1,2){\mathcal W}^{cd}_\alpha(2) 
\delta{\widehat Q}^{da}_{\alpha \alpha^\prime}\,,  \label{v1}
\end{eqnarray}
and the variation of the last term in~(\ref{VEQ})
may be expressed as 
\begin{equation}
\delta{\rm Tr}{\hat\Lambda}({\widehat R} - {\widehat R}^2))
= \sum_\alpha {\rm Tr}
\left(\begin{array}{cc}
E_\alpha &  0 \\  0  &  -E_\alpha \end{array}\right)
\delta{\widehat Q}_{\alpha \alpha}\,. 
\label{v2}
\end{equation}
Thus, the variation equation (\ref{VEQ}) becomes
\begin{equation}
\sum_{\alpha\alpha^\prime}{\rm Tr}\left[
{\mathcal W}^\dagger_{\alpha^\prime}\hat{\mathcal H}
{\mathcal W}_\alpha
- \delta_{\alpha^\prime \alpha}\left(\begin{array}{cc}
E_\alpha &  0 \\  0  &  -E_\alpha \end{array}\right)\right]
\delta{\widehat Q}_{\alpha \alpha^\prime} = 0\,.
\end{equation}
Since the elements $\delta{\widehat Q}_{\alpha \alpha^\prime}$
are independent, we obtain the matrix equation
\begin{equation}
\hat{\mathcal H}{\mathcal W}_\alpha
= {\mathcal W}_\alpha\left(\begin{array}{cc}
E_\alpha &  0 \\  0  &  -E_\alpha \end{array}\right)\,,
\end{equation}
which is equivalent to the HFB equations
\begin{equation}
\int \d 2 \,\hat{\mathcal H}(1,2)
\left(\begin{array}{c} U_\alpha(2)  \\
V_\alpha(2) \end{array}\right)
=E_\alpha
\left(\begin{array}{c}
U_\alpha(1)  \\V_\alpha(1)  \end{array}\right)\,,
\label{SEQ}
\end{equation}
\begin{equation}
\int \d 2 \,\hat{\mathcal H}(1,2)
\left(\begin{array}{c} V^*_\alpha(2) \\
U^*_\alpha(2)\end{array}\right)
= - E_\alpha \left(\begin{array}{c}
V^*_\alpha(1) \\
U^*_\alpha(1)\end{array}\right)\,.
\label{SEQ2}
\end{equation}

The matrix of second variational derivatives of the energy 
functional is the supermatrix of the effective quasiparticle 
interaction amplitudes $\hat {\mathcal F}$. In the pp--channel 
one finds 
\begin{equation}
{\mathcal F}^{\rm pp}(1,2;3,4) = \frac {\delta \Delta(1,2)}
{\delta \nu(3,4)}\,.
\end{equation}
We assume that ${\mathcal F}^{\rm pp}$ is a functional of the
normal quasiparticle density matrix $\hat\rho$ and that the term 
$E_{\rm anomal}$ in (\ref{TENE}), which depends explicitly on 
the anomalous density matrix, may be written as
\begin{equation}
E_{\rm anomal}[\widehat R] = \quart
\int \d 1\!\cdot\!\cdot\!\cdot\!\d 4 
\nu^\dagger(2,1){\mathcal F}^{\rm pp}_{\rm a}(1,2;3,4;[\hat 
\rho])\nu(3,4)\,, \label{EPP}
\end{equation}
with the antisymmetrized pp--interaction 
\begin{equation}
{\mathcal F}^{\rm pp}_{\rm a}(1,2;3,4) = 
{\mathcal F}^{\rm pp}(1,2;3,4) - {\mathcal F}^{\rm pp}(1,2;4,3)\,.
\end{equation}

The gap equation reads
\begin{equation}
\Delta(1,2) = \half\int \d 3\d 4{\mathcal F}^{\rm pp}_{\rm a}
(1,2;3,4;[\hat \rho]) \nu(3,4)\,.   \label{Gapeq}
\end{equation}
The expression for the anomalous energy~(\ref{EPP}) may thus be
written in the form
\begin{equation}
E_{\rm anomal}=\half\int\d 1\d 2 \nu^\dagger(2,1)\Delta(1,2)\,.
\label{EA2}
\end{equation}

To proceed further and perform the renormalization procedure it is 
convenient to use the Green's function formalism (see Appendix B).
From (\ref{RGEN}) and eqs.~(\ref{SG})--(\ref{SFP}) it follows that
the generalized density matrix $\widehat R$ may be represented 
by a contour integral of the generalized Green's function 
$\widehat G$ in the complex energy plane: 
\begin{equation}
{\widehat R}(1,2)=\int\nolimits_C\frac{\d \epsilon}{2\pi\mathrm{i}}
{\widehat G}(1,2;\epsilon)\,.
\end{equation}
The integral is performed along a contour $C$ which goes from 
$-\infty$ to 0 below the real $\epsilon$-axis (Im$\,\epsilon = -0$), 
crosses this axis at $\epsilon=0$ and goes back to $-\infty$ above
it (${\rm Im}\,\epsilon = +0$), with the energy variable $\epsilon$
measured from the chemical potential $\mu$. The contour is shown 
schematically in Fig.~\ref{f:f24}, see Appendix C.

The normal and anomalous density matrices are then given by
\begin{equation}
\rho(1,2)=\int\nolimits_C\frac{\d \epsilon}{2\pi\mathrm{i}}
G_s(1,2;\epsilon) = \sum_{(E_{\alpha}>0)} 
V^*_{\alpha}(1)V_{\alpha}(2)\,,
\end{equation}
\begin{equation}
\nu(1,2)=\int\nolimits_C\frac{\d \epsilon}{2\pi\mathrm{i}}
F(1,2;\epsilon) = \sum_{(E_{\alpha}>0)} 
V^*_{\alpha}(1) U_{\alpha}(2)\,.  \label{nuF}
\end{equation}
The summation (or integration) in these expressions 
extends, in principle, to infinity, 
the convergence being dependent on the properties of the effective 
interaction. It is well known that in the case of a local force the 
anomalous density matrix $\nu(1,2)$ diverges at 
$\vec r_1 \to \vec r_2$. In the gap equation 
(\ref{Gapeq}) with $\hat \nu$ expressed by (\ref{nuF}), 
the integration over the region far from the Fermi surface
can be avoided by using the standard renormalization procedure 
with an arbitrary energy cutoff \cite{Migdal}.
But to calculate the energy of the system, in particular 
its (negative) anomalous part (\ref{EA2}), one needs to know 
the ``untruncated" density matrices. On the other hand, as it is
also well known, the total pairing energy $E_{\rm pair}$ contains 
a (positive) contribution from the kinetic energy term so that the
sum of the two contributions converges more rapidly and 
$E_{\rm pair}$ gets a finite value even in the case of a local
pairing field (at least in infinite matter, see Appendix B). 
We shall show that the energy density of the uniform
infinite system may be calculated exactly in the leading order in 
$\Delta^2(p_{\rm F})/\epsilon_{\rm F}$ by using a renormalized gap 
equation with an arbitrary cutoff $\epsilon_{\rm c}$ but such that 
$\epsilon_{\rm c} > \epsilon_{\rm F}$, and by applying the
variational principle $\delta I[\widehat R] = 0\,$ to the functional
$I[\widehat R]$ given by~(\ref{II}) but considered now as a
functional of the corresponding cutoff generalized density matrix
${\widehat R}_{\rm c}$. 

We introduce an energy cutoff $\epsilon_{\rm c} > \epsilon_{\rm F}$.
The contour $C$ is split into two parts
$\int_C = \int_{C_{\rm c}} + \int_{C_B}\,,$
where $C_{\rm c}$ lies below and above the real 
$\epsilon$-axis for $ -\epsilon_{\rm c} < {\rm Re}\,\epsilon < 0$ 
and the contour $C_B$ lies below and above it for
$ {\rm Re}\,\epsilon < -\epsilon_{\rm c} $\,. Then we have
\begin{equation}
\nu(1,2) = \nu_{\rm c}(1,2) + \delta_{\rm c} \nu(1,2)\,,
\end{equation}
where
\begin{equation}
\nu_{\rm c}(1,2) = \sum_{0< E_\alpha < \epsilon_{\rm c}}
V^*_\alpha(1)U_\alpha(2)=\int\nolimits_{C_{\rm c}}\frac{\d \epsilon}
{2\pi\mathrm{i}}F(1,2; \epsilon)\,,
\end{equation}
and
\begin{equation}
\delta_{\rm c}\nu(1,2) = \sum_{E_\alpha > \epsilon_{\rm c}}
V^*_\alpha(1) U_\alpha(2)=\int\nolimits_{C_B}\frac {\d \epsilon}
{2\pi\mathrm{i}}F(1,2;\epsilon)\,.
\end{equation}
The renormalized gap equation may be written in the form
\begin{equation}
\Delta(1,2) = \half\int\d 3\d 4{\mathcal F}^\xi_a(1,2;3,4)
\nu_{\rm c}(3,4)\,.
\end{equation} 
where the effective scattering amplitude ${\mathcal F}^\xi$ is to
be found from the equation 
\begin{eqnarray}
{\mathcal F}^\xi(1,2;3,4)
& = & {\mathcal F}^{\rm pp}(1,2;3,4)\nonumber\\ 
& + & \int \d 5\!\cdot\!\cdot\!\cdot\!\d 8
{\mathcal F}^{\rm pp}(1,2;5,6)B(5,6;7,8){\mathcal F}^\xi(7,8;3,4)\,,
\end{eqnarray}
with
\begin{equation}
B(1,2;3,4) = \int\nolimits_{C_B}\frac{\d \epsilon}{2\pi\mathrm{i}}
G_0(1,3;\epsilon){\bar G}_s(4,2;\epsilon)\,.
\end{equation}
Thus the anomalous energy (\ref{EA2}) may be represented by a sum 
of two terms:
\begin{eqnarray}
E_{\rm anomal}
& = & E^{\rm c}_{\rm anomal} +\delta_{\rm c} 
E_{\rm anomal}\nonumber\\
& \equiv & \frac 1 2 \int \d 1\d 2 
\int\nolimits_{C_{\rm c}}\frac{\d \epsilon}{2\pi\mathrm{i}}
F^\dagger(2,1;\epsilon)\Delta(1,2)\nonumber\\
& + & \frac 1 2 \int \d 1\d 2 
\int\nolimits_{C_B}\frac {\d \epsilon}{2\pi\mathrm{i}}
F^\dagger(2,1;\epsilon)\Delta(1,2)\,.  \label{dano}
\end{eqnarray}
The normal density matrix is written in a similar way as a sum
\begin{equation}
\rho(1,2) = \rho_{\rm c}(1,2) + \delta_{\rm c} \rho(1,2)\,,
\end{equation}
where
\begin{equation}
\rho_{\rm c}(1,2) = \sum_{0< E_\alpha < \epsilon_{\rm c}}
V^*_\alpha(1) V_\alpha(2)
= \int\nolimits_{C_{\rm c}}\frac{\d \epsilon}{2\pi\mathrm{i}}
G_s(1,2; \epsilon)\,,  \label{RHOC}
\end{equation}
and
\begin{equation}
\delta_{\rm c}\rho(1,2)  =  \sum_{E_\alpha > \epsilon_{\rm c}}
V^*_\alpha(1) V_\alpha(2) 
=\int\nolimits_{C_B}\frac{\d \epsilon}{2\pi\mathrm{i}}
G_s(1,2;\epsilon)\,.
\end{equation}
Using the definition of the single particle Hamiltonian $h$,
eq.~(\ref{hdef}), the amount of the total energy related to 
$\delta_{\rm c}\rho$ in first order may be found by varying 
the normal part of the energy functional with respect to the 
normal density:
\begin{equation}
\delta_{\rm c}E_{\rm normal}=\int \d 1\d 2 h(1,2)
\delta_{\rm c}\rho(2,1)\,.  \label{dno}
\end{equation}
This includes also, if ${\mathcal F}^{\rm pp}$ depends on 
$\rho$, a contribution from the variation of $E_{\rm anomal}$.

With $\delta_{\rm c} E_{\rm anomal}$ from (\ref{dano}) and      
$\delta_{\rm c}E_{\rm normal}$ defined by (\ref{dno}) we obtain
the total change of the variational functional related to the 
cutoff:
\begin{equation}
\delta_{\rm c}(E-\mu N)=\int\nolimits_{C_B}\frac{\d \epsilon}
{2\pi\mathrm{i}}
(hG_s + \frac 1 2 F^\dagger\Delta - \mu G_s)\,.
\end{equation}
This can be written in another form using the relation 
\begin{equation}
\Delta F^\dagger = (\epsilon - h + \mu)G_s -1 
\end{equation}
which follows from the Gor'kov equations~(\ref{FP}):  
\begin{eqnarray}
\delta_{\rm c}\left(E - \mu N(\mu)\right)
& = & \frac 12{\rm Tr}\int\nolimits_{C_B}\frac{\d \epsilon}
{2\pi\mathrm{i}}(\epsilon + h -\mu)G_s\nonumber\\ 
& = & \frac 12{\rm Tr}\int\nolimits_{C_B}\frac{\d \epsilon}
{2\pi \mathrm{i}} 
(\epsilon + h -\mu)G_0\Delta F^\dagger\,.
\end{eqnarray}
To get the second equality in this equation we have used 
the relation\\ 
${\rm Tr}\int\nolimits_{C_B} \d \epsilon (\epsilon + h -\mu)G_0 = 0\,$  
which holds because the Green's function $G_0$ is diagonal 
in the basis of eigenfunctions of $h$ (see eq.~(\ref{SG0})) 
and has no singularities embraced by the contour $C_B$.

In lowest order in $\Delta$, from (\ref{FP}) one has 
$F^\dagger={\bar G}_0\Delta^* G_0$ and we get 
\begin{eqnarray}
\delta_{\rm c}\left(E -\mu N(\mu)\right)
& = & \frac 1 2 {\rm Tr}\int\nolimits_{C_B} \frac {\d \epsilon}
{2\pi\mathrm{i}}
(\epsilon + h -\mu)G_0\Delta{\bar G}_0\Delta^* G_0\nonumber\\
& = & \sum_{\epsilon_i - \mu < \epsilon_{\rm c}}
\left(\sum_{\epsilon_k - \mu > \epsilon_{\rm c}}
\frac {(\epsilon_i - \epsilon_k)\vert\Delta_{ik}\vert^2}
{(\epsilon_i + \epsilon_k - 2\mu)^2}\right)\,.
\label{SUM}
\end{eqnarray}
Here $\Delta_{ik}$ are the matrix elements of $\Delta$ in the basis 
of the single-particle Hamiltonian $h$ with $\epsilon_i$ being
its eigenvalues (see Appendix B, eq.~(\ref{spbas})). 

In uniform infinite matter, the pairing field 
$\Delta(1,2)$ is a function of 
$\vert {\vec r}_1 - {\vec r}_2 \vert$, all its nondiagonal matrix 
elements vanish and we see that 
$\delta_{\rm c}\left(E -\mu N(\mu)\right)=0$  
in the first order upon imposing the energy cutoff 
$\epsilon_{\rm c} > \epsilon_{\rm F}$. In finite systems, 
the change 
of $\delta_{\rm c}\left(E -\mu N(\mu)\right)$ vanishes 
in the first order in diagonal approximation. 
A nonvanishing right-hand side 
of eq.~(\ref{SUM}) arises only due to nonuniformity of $\Delta$, 
which means that in finite systems with more or less uniform density 
in the volume
(like heavy atomic nuclei), this would be a surface effect. 
Reliable estimates of the sum in~(\ref{SUM}) involving only 
nondiagonal matrix elements $\Delta_{ik,\,i \neq k}$ are not easy to 
obtain. They depend strongly on the actual distribution of the 
pairing field in real nuclei, in particular in the surface region.
In larger systems, with presence of volume pairing, the nondiagonal
matrix elements of $\Delta$  
are relatively small ($\sim A^{- 1/3}$, or less) so that
the change of $E-\mu N$ due to the cutoff 
is expected to be at least by a factor of $A^{-1/3}$ 
smaller compared to $E_{\rm pair}$, and therefore we can write
\begin{equation}
\vert\delta_{\rm c}E-\mu\delta_{\rm c} N(\mu)\vert \ll
\vert E_{\rm pair}\vert\,,
\label{EMC}
\end{equation}
where
\begin{equation}
\delta_{\rm c} N(\mu) = \int \d 1 \delta_{\rm c}\rho(1,1) =
N(\mu) - N_{\rm c}(\mu)\,,
\label{DCN}
\end{equation}
with $N_{\rm c}$ the number of particles corresponding to the 
cutoff density $\rho_{\rm c}$ of eq.~(\ref{RHOC}):
\begin{equation}
N_{\rm c} = \int \d 1 \rho_{\rm c}(1,1)\,.
\end{equation}
The pairing energy $E_{\rm pair}$ is given in Appendix B
by an exact expression (\ref{EPAIR})
and by a BCS estimate (\ref{epspair}).

The change in the system energy related with the cutoff may be
written as
\begin{equation}
\delta_{\rm c}E=E[{\widehat R}]-E[{\widehat R}_{\rm c}]
               =\mu\delta_{\rm c}N\,,
\end{equation}    
with $\delta_{\rm c}N$ defined in (\ref{DCN}).
Minimizing the energy functional with a fixed cutoff 
$\epsilon_{\rm c}$ would give an additional change
\begin{equation}
\delta E_{\rm c}=E[{\widehat R}_{\rm c}]-
E_{\rm c}[{\widehat R}_{\rm c}]=\mu_{\rm c}\delta N_{\rm c}\,,
\end{equation}    
where $\delta N_{\rm c}$ is the change in the particle
number within the cutoff energy space 
$\epsilon < \epsilon_{\rm c}$. In first order we have
\begin{equation}
E[{\widehat R}]=E_{\rm c}[{\widehat R}_{\rm c}]+\mu\delta_{\rm c}N
+ \mu_{\rm c}\delta N_{\rm c} = E_{\rm c}[{\widehat R}_{\rm c}] +
\mu(\delta_{\rm c}N + \delta N_{\rm c})\,.
\end{equation}
Of course, the constraint ${\rm Tr}\rho_{\rm c} =N$ has also to
be used for the cutoff functional
which gives $\delta_{\rm c}N + \delta N_{\rm c} = 0$. 

We thus come to the important conclusion that passing to the 
cutoff functional with $\epsilon_{\rm c} > \epsilon_{\rm F}$
leaves, to the first order in the weak pairing approximation, 
the energy, the variational functional $E-\mu N$ and the chemical 
potential $\mu$ of the system unchanged (to the extent that 
the estimate~(\ref{EMC}) for a given nucleus is correct). 

According to the Hohenberg-Kohn theorem~\cite{HKohn},   
the nuclear ground state properties can be described by a local
density functional. One may therefore assume that the anomalous
energy is a functional of the local density $\rho$ too.
The above consideration implies that this should be also valid 
for the cutoff functional. Both the cutoff anomalous density and
the effective pairing interaction ${\mathcal F}^\xi$ may then be 
regarded as local functionals of $\rho$. We introduce a local 
approximation to ${\mathcal F}^\xi$:
\begin{eqnarray}
{\mathcal F}^\xi(1,2;3,4) &=& {\mathcal F}^\xi
({\vec r}_1,\tau_1)\nonumber\\
                   &\times&
\delta({\vec r}_1 - {\vec r}_2)\delta({\vec r}_1 - {\vec r}_3)
\delta({\vec r}_2 - {\vec r}_4)\delta_{s_1 s_3}\delta_{s_2 s_4}
\delta_{\tau_1 \tau_3}\delta_{\tau_2 \tau_4}\,.
\end{eqnarray}
where
\begin{equation}
{\mathcal F}^\xi(\vec r,\tau) = 
{\mathcal F}^\xi([\rho_{\rm c}(\vec r,n),\rho_{\rm c}(\vec r,p)];\tau)
\end{equation}
is a local functional of the cutoff quasiparticle density 
\begin{equation}
\rho_{\rm c}(\vec r_1,\tau_1)=
{\rm Tr}_{s_1}\int \d 2\rho_{\rm c}(1,2)\delta(2,1)\,,
\quad \tau_1 \in \{n,p\}\,,
\label{rhoc}
\end{equation}
where $\delta(1,2)=
\delta(\vec r_1-\vec r_2)\delta_{s_1s_2}\delta_{\tau_1\tau_2}$.

In the case of a local functional it is convenient to work 
with the so-called anomalous density in the real space. For 
the time-reversed single-particle states 
$\vert\bar k\rangle = T\vert k\rangle$, 
in the coordinate space representation, we define
\begin{equation}
\phi_{\bar k}(1)=\int \d 2 {\mathcal T}(1,2)\phi^*_k(2)\,,\quad 
\phi^*_{\bar k}(1)=\int \d 2 {\phi}_k(2){\mathcal T}^\dagger(2,1)\,,
\label{trev}
\end{equation}
with the operator
\begin{equation}
{\mathcal T}(1,2) = \delta({\vec r}_1 - {\vec r}_2)
(-1)^{\frac 1 2 - s_1}\delta_{-s_1 s_2}\delta_{\tau_1\tau_2}\,.
\label{T12}
\end{equation}
This operator is antisymmetric,
${\mathcal T}(2,1)=- {\mathcal T}(1,2)\,,$ and has the properties
$\int \d 3{\mathcal T}(1,3){\mathcal T}(3,2)=-\delta(1,2)$ and
$\,\,\int\d 3{\mathcal T}(1,3){\mathcal T}^\dagger(3,2)=
\delta(1,2)$. The anomalous density is then defined by 
\begin{equation}
\nu_{\rm c}(\vec r_1,\tau_1) = \half {\rm Tr}_{s_1} 
\int\d 2 \,\nu_{\rm c}(1,2){\mathcal T}^\dagger(2,1)\,,
\label{nuc}
\end{equation}
and its complex conjugate by
\begin{equation}
\nu_{\rm c}^* (\vec r_1,\tau_1) = \half {\rm Tr}_{s_1} 
\int\d 2{\mathcal T}(1,2)\nu_{\rm c}^\dagger(2,1)\,.
\end{equation}
In the local approximation we get
\begin{equation}
\Delta(1,2) = {\mathcal T}(1,2)\Delta(\vec r_2,\tau_2)\,.
\label{T12D}
\end{equation}

With these definitions, the anomalous part of the cutoff 
functional reduces to the expression~(\ref{Eano}), and
for the local pairing field $\Delta(\vec r,\tau)$ one obtains
the gap equation in the simple multiplicative form~(\ref{gapeq}). 

One may notice that our approach does not imply a cutoff of
the basis since the general variational principle can be 
formulated with a ``cutoff'' local-density functional 
from which the ground state characteristics of a superfluid 
system may be calculated through the solutions of the Bogolyubov
equations~(\ref{SEQ}) at the stationary point. To construct 
the normal and anomalous densities, entering this local functional, 
only those solutions from the whole set are needed which correspond
to the eigenenergies $E_\alpha$ of the HFB Hamiltonian up to the 
cutoff\footnote{At first glance 
the situation looks similar to the HF case with a EDF without pairing 
where only the occupied orbitals up to $\epsilon_{\rm F}$ in the 
self-consistent mean field are needed to calculate the density 
(a ``natural'' HF cutoff). But it should be emphasized that the 
Bogolyubov solutions cannot be obtained from the HF ones by 
perturbative methods, they spread to infinity in the energy space 
at any nonvanishing gap, hence the convergence problem for the 
densities. This problem, as usually thought of, can be avoided only
if one uses a finite-range effective 
force ensuring the convergence of the anomalous density at some energy 
which depends on the range of nonlocality of the force. Then the HFB
equations become nonlocal leading to a technical problem; in 
practice these equations are solved with 
approximate methods, for example, by expanding the HFB wave functions
in a harmonic oscillator basis~\cite{DeGo80}.} 
$\epsilon_{\rm c} > \epsilon_{\rm F}$.
This means that one can implement a local pairing effective force 
in the HFB (or Gor'kov) equations and, most important, 
in the gap equation, renormalized with the same energy
cutoff $\epsilon_{\rm c}$, in which the anomalous density 
by construction does not diverge. The whole set of resulting 
equations can be solved, for spherical nuclei, exactly with a 
coordinate-space technique (see Appendix C). It should nevertheless 
be stressed that the main difficulty now is to find such a local 
effective density-dependent pairing interaction  
${\mathcal F}^\xi$ that would be an universal one and would 
allow to reproduce the pairing ground-state nuclear properties 
(at least for a large domain of heavy nuclei). 

\setcounter{equation}{0}
\setcounter{section}{0}
\renewcommand{\theequation}{B.\arabic{equation}}
\section*{Appendix B. Gor'kov equations and pairing energy}

In this appendix we derive the general expression for the pairing
energy using the Green's function formalism. We show that, in the
case of a weak local pairing field, in infinite matter
this expression reduces to the BCS formula used in eq.~(\ref{Eaexp}).

We start with Gor'kov equations~\cite{Gor} for the generalized 
Green's function ${\widehat G}(\epsilon)$ written in matrix form,
\begin{equation}
(\epsilon - \hat {\mathcal H})\widehat G(\epsilon) = \hat I\,,
\label{GGF}
\end{equation}
where $\hat {\mathcal H}$ is the effective single-quasiparticle
Hamiltonian (\ref{EQPH}) and
\begin{equation}
{\widehat G}(\epsilon) = \left(\matrix {
  G_{\rm s}(\epsilon)        &  F(\epsilon) \cr
  F^{\dagger}(\epsilon)   &  {\bar G}_{\rm s}(\epsilon)  \cr}
\right)\,
\end{equation}
with $G_{\rm s}$ the normal and $F$ the anomalous Green's function,
respectively.
For $G_{\rm s}$ and $F^{\dagger}$, eq.~(\ref{GGF}) yields the set of
two equations
\begin{equation}
G_{\rm s}(\epsilon) = G_0(\epsilon)
+ G_0(\epsilon)\Delta F^\dagger(\epsilon)\,, \quad
F^{\dagger}(\epsilon)
= - {\bar G}_0(\epsilon)\Delta^*G_{\rm s}(\epsilon)\,,
\label{FP}
\end{equation}
where $G_0$ and ${\bar G}_0$ are the single particle Green's 
functions connected with the normal parts $h-\mu$ and
$\mu-h^*$ of the effective Hamiltonian, respectively:
\begin{equation}
G_0(\epsilon) = \frac 1 {\epsilon - h   + \mu}\,, \quad
{\bar G}_0(\epsilon) = \frac 1 {\epsilon + h^* - \mu}\,.
\end{equation}
The Green's functions $G_0$ and ${\bar G}_0$ are related to each 
other by ${\bar G}_0(\epsilon) = - G^*_0(-\epsilon^*)\,.$

Similarly, for ${\bar G}_{\rm s}$ and $F$ we can write the set of 
two equations
\begin{equation}
{\bar G}_{\rm s}(\epsilon)
= {\bar G}_0(\epsilon)
- {\bar G}_0(\epsilon)\Delta^* F(\epsilon)\,, \quad
F(\epsilon)
= G_0(\epsilon)\Delta {\bar G}_{\rm s}(\epsilon)\,.
\end{equation}
From the Gor'kov equations it follows that
\begin{equation}
{\bar G}_{\rm s}(\epsilon) = - G^*_{\rm s}(-\epsilon^*)\,, \quad
F^\dagger(\epsilon) = - F^*(-\epsilon^*)\,.
\label{GFe}
\end{equation}
The ``zero'' Green's function $G_0$ is diagonal in the basis of
eigenfunctions of the Hamiltonian $h$ ($k$--representation),
\begin{equation}
\int \d 2 h(1,2)\phi_k(2) = \epsilon_k\phi_k(1)\,,
\label{spbas}
\end{equation}
and for its spectral decomposition we get
\begin{equation}
G_0(1,2;\epsilon) =\sum_k \left[\frac{n_k}{\epsilon
- \epsilon_k + \mu - \mathrm{i}\delta}
+ \frac{1-n_k}{\epsilon
- \epsilon_k + \mu  + \mathrm{i}\delta}
\right]\phi_k (1)\phi^*_k (2)\,,
\label{SG0}
\end{equation}
where $n_k=1$ if $\epsilon_k < \mu$ and $n_k=0$ if $\epsilon_k > \mu$.

For systems with even particle number, the spectral expansions
of $G_{\rm s}$, $F$ and $F^\dagger$, in coordinate space representation,
may be written in the form:
\begin{equation}
G_{\rm s}(1,2;\epsilon) = \sum_{(E_{\alpha}>0)}\left[
\frac{V^*_{\alpha}(1) V_{\alpha}(2)}
{\epsilon + E_{\alpha} -\mathrm{i}\delta}
+ \frac{U_{\alpha}(1)U^*_{\alpha}(2)}
{\epsilon - E_{\alpha} + \mathrm{i}\delta}
\right]\,,
\label{SG}
\end{equation}
\begin{equation}
F(1,2;\epsilon) =\sum_{(E_{\alpha}>0)}\left[
\frac{V^*_{\alpha}(1) U_{\alpha}(2)}
{\epsilon + E_{\alpha} - i \delta}
+ \frac{U_{\alpha}(1)V^*_{\alpha}(2)}
{\epsilon - E_{\alpha} + \mathrm{i}\delta}
\right]\,,
\label{SF}
\end{equation}
\begin{equation}
F^\dagger(1,2;\epsilon) =\sum_{(E_{\alpha}>0)}\left[
\frac{U^*_{\alpha}(1) V_{\alpha}(2)}
{\epsilon + E_{\alpha} - \mathrm{i}\delta}
+ \frac{V_{\alpha}(1)U^*_{\alpha}(2)}
{\epsilon - E_{\alpha} + \mathrm{i}\delta}
\right]\,,
\label{SFP}
\end{equation}
with $E_{\alpha}$ the exact eigenvalues and $U_{\alpha}$,
$V_{\alpha}$ the exact eigenfunctions of eqs. (\ref{SEQ}).
The corresponding expressions in the $k$--representation
are:
\begin{equation}
G_{{\rm s}\,ij}(\epsilon) =\sum_{(E_{\alpha}>0)}\left[
\frac{V_{i \alpha}^*  \,V_{j \alpha}}
{\epsilon + E_{\alpha} - \mathrm{i}\delta}
+ \frac{U_{i\alpha} U_{j\alpha}^*}
{\epsilon - E_{\alpha} + \mathrm{i}\delta}
\right]\,,
\end{equation}
\begin{equation}
F_{ij}(\epsilon) =\sum_{(E_{\alpha}>0)}\left[
\frac{V_{i \alpha}^*  \,U_{j \alpha}}
{\epsilon + E_{\alpha} - \mathrm{i}\delta}
+ \frac{U_{i\alpha}V_{j\alpha}^*}
{\epsilon - E_{\alpha} + \mathrm{i}\delta}
\right]\,,
\end{equation}
\begin{equation}
F^\dagger_{ij}(\epsilon) = \sum_{(E_{\alpha}>0)}\left[
\frac{U_{i \alpha}^*  \,V_{j \alpha}}
{\epsilon + E_{\alpha} - \mathrm{i}\delta}
+ \frac{V_{i\alpha} U_{j\alpha}^*}
{\epsilon - E_{\alpha} + \mathrm{i}\delta}
\right]\,,
\label{SFPk}
\end{equation}
with $U_{k\alpha} = \int \d 1\, \phi^*_k(1)U_\alpha(1)\,,$ and
$V_{k\alpha} = \int \d 1\, \phi_k(1)V_\alpha(1)\,.$
In diagonal approximation,
which is exact in uniform infinite matter,
the only nonzero matrix elements
of $G_{\rm s}$, $F$ and $F^{\dagger}$ are:
\begin{equation}
G_{{\rm s}\,kk}(\epsilon) =
\frac{\vert v_k\vert ^2 }{\epsilon + E_k -\mathrm{i}\delta}
+ \frac{\vert u_k\vert ^2 }{\epsilon - E_k + \mathrm{i}\delta}\,,
\end{equation}
\begin{equation}
F_{k \bar k}(\epsilon) = -F_{\bar k k}(\epsilon)
= -\frac{\Delta_k}{2E_k}\left(\frac 1 {\epsilon + E_k - 
\mathrm{i}\delta}
- \frac 1 {\epsilon - E_k + \mathrm{i}\delta}\right)\,,
\end{equation}
\begin{equation}
F^\dagger_{k \bar k}(\epsilon) = - F^\dagger_{\bar k k}(\epsilon)
=  \frac{\Delta^*_k}{2E_k}\left(\frac 1 {\epsilon + E_k 
-\mathrm{i}\delta}
- \frac 1 {\epsilon - E_k + \mathrm{i}\delta}\right)\,,
\end{equation}
with $\vert\bar k\rangle$ being the time reversed of
$\vert k\rangle$ as defined by (\ref{trev}).
Here we have used the customary conventions:
\[\Delta_k=-2E_kv^*_ku_k\,, \quad
E_k = \sqrt {(\epsilon_k - \mu)^2 + \vert \Delta_k\vert ^2}\,,  \]
\[\vert u_k \vert ^2 =
\frac 1 2 (1 + \frac {\epsilon_k - \mu}{E_k})\,, \quad
\vert v_k\vert ^2 =
\frac 1 2 (1 - \frac {\epsilon_k - \mu}{E_k})\,. \]
The corresponding spectral expansions for ${\bar G}_{\rm s}$ may be
found from those for $G_{\rm s}$ by using the relations (\ref{GFe}).
It is easily seen that the expansions written above for
the anomalous Green's functions $F$ and $F^\dagger$ may be also
obtained from each other by using (\ref{GFe}).

In what follows the Green's function method is used to extract
the pairing contribution to the total energy of
the system. To isolate the pure pairing part one has to consider
two effects: (i) the direct influence of the pairing gap
$\Delta$ at fixed single-particle Hamiltonian $h - \mu$
and (ii) the ``polarization" mechanism due to variation of $h$
and $\mu$ in the presence of the pairing field.

To obtain the first change in the density, when
$\Delta \to 0$ but $h$ and $\mu$ fixed, we introduce a
``zero'' density matrix:
\begin{equation}
\rho_0(1,2)=\int\nolimits_C \frac{\d \epsilon}{2\pi\mathrm{i}}
G_0(1,2;\epsilon)\,.
\end{equation}
For the difference $\delta \rho =\rho - \rho_0$, eq.~(\ref{FP}) 
yields
\begin{equation}
\label{delrho}
\delta\rho =\int\nolimits_C\frac{\d \epsilon}{2\pi\mathrm{i}}
(G_{\rm s}(\epsilon) - G_0(\epsilon))
=\int\nolimits_C\frac{\d \epsilon}{2\pi\mathrm{i}}G_0(\epsilon)
\Delta F^{\dagger}(\epsilon)\,.
\end{equation}
Using (\ref{SG0}) and (\ref{SFPk}) we obtain in the 
$k$--representation:
\begin{equation}
\delta\rho_{i j} = - \sum_{E_\alpha > 0}\sum_{l}
\frac{\Delta_{i l}} {E_\alpha + \vert \epsilon_i - \mu\vert}
[n_i V_{l \alpha} U_{j \alpha}^* +
(1 - n_i) U_{l\alpha}^* V_{j \alpha }]\,,
\end{equation}
or, in diagonal approximation,
\begin{equation}
\delta\rho(\vec r ,\tau) = \sum_k \sum_{s_z}(1-2n_k)
\frac{\vert \Delta_k \vert ^2}
{2E_k(E_k + \vert \epsilon_k - \mu\vert)}
\vert \phi_k(\vec r, s_z, \tau)\vert ^2\,.
\end{equation}
One should keep in mind that $\rho$ and $\rho_0$ belong
to states with different particle number:
\begin{equation}
\delta N(\mu) = N(\mu) - N_0(\mu) \ne 0\,.
\end{equation}

The second change in the density we get in the situation
without any pairing. Then we have the HF vacuum and the
corresponding HF Green's function,
\begin{equation}
G_{\rm HF}(\epsilon)
= \frac 1 {\epsilon - h_{\rm HF}
+ \mu_{\rm HF}}\,,
\end{equation}
with $h_{\rm HF}$ being the self-consistent HF single-particle
Hamiltonian. The HF density matrix is given by
\begin{equation}
\rho_{\rm HF}(1,2) = \int\nolimits_C 
\frac{\d \epsilon}{2\pi\mathrm{i}}G_{\rm HF}(1,2;\epsilon)\,
\end{equation}
with the average number of particles
$\langle {\rm HF}\vert \hat N\vert {\rm HF}\rangle =
N_{\rm HF}(\mu_{\rm HF})$, i.e. ${\rm Tr}\rho_{\rm HF} = N$.

Collecting contributions from both pairing-induced effects
we get the total variation of the density matrix:
\begin{equation}
\delta \rho_{\rm pair} = \rho - \rho_{\rm HF}
= \delta \rho + \delta \rho_0\,,
\end{equation}
with $\delta \rho_0 = \rho_0 - \rho_{\rm HF}$.
From the definition of the Green's function we find
\begin{equation}
G^{-1}_{\rm HF} - G^{-1}_{0} = 
h - \mu - h_{\rm HF} + \mu_{\rm HF}\,,\\
\end{equation}
or
\begin{equation}
G_0 - G_{\rm HF} = G_{\rm HF}(\delta U_{\rm pair} - 
\delta\mu)G_{0}\,,
\end{equation}
where $\delta U_{\rm pair} = h - h_{\rm HF}\,$ and
$\delta\mu = \mu-\mu_{\rm HF}$.

The expression for $\delta\rho_{0}$ can then be written as
\begin{eqnarray}
\delta \rho_{0}&=&\int\nolimits_C
\frac{\d \epsilon}{2\pi\mathrm{i}}
(G_0(\epsilon) -G_{\rm HF}(\epsilon)) \nonumber \\
&=&\int\nolimits_C\frac{\d \epsilon}{2\pi\mathrm{i}}
G_{\rm HF}(\epsilon)(\delta U_{\rm pair}
-\delta \mu)G_{0}(\epsilon)\,,   \label{DR0}
\end{eqnarray}
Dealing with the first-order terms of perturbation theory,
the integral over $\epsilon$ may be replaced by the static
particle--hole propagator ${\mathcal A}$ (see Ref.\cite{Migdal}):
\begin{equation}
\int\nolimits_C \frac{\d \epsilon}{2\pi\mathrm{i}}
G_{\rm HF}(\epsilon)G_{0}(\epsilon)
\approx\int\nolimits_C \frac{\d \epsilon}{2\pi\mathrm{i}}
G_{\rm HF}(\epsilon)G_{\rm HF}(\epsilon) \equiv{\mathcal A}\,.
\end{equation}
Its convolution with any perturbation operator which
is diagonal in the HF basis, i.e. commutes with $h_{\rm HF}$,
vanishes (no polarization effect). This property
will be used to obtain the energy variation connected with
$\delta \rho_{0}$ (see eqs.~(\ref{EN01}) and (\ref{EN02}) below).
The average number of particles of each kind is fixed:
$\langle {\rm HFB}\vert \hat N\vert {\rm HFB}\rangle
= \langle{\rm HF}\vert \hat N \vert {\rm HF}\rangle$.
The total density variation due to pairing, of course, does
not change the particle number, so
\begin{equation}
\delta N_{\rm pair}
=\delta N(\mu) + \delta N_{0}(\mu ,\mu_{\rm HF} ) = 0\,.
\label{MuN}
\end{equation}

By definition, the pairing energy is the total change of the system
energy due to pairing correlations:
\begin{equation}
E_{\rm pair} = E_{\rm normal}[\rho] + E_{\rm anomal}[\rho,\nu]
- E_{\rm normal}[\rho_{\rm HF}]\,.
\end{equation}
Analogously to the procedure applied above to the density matrix,
we can split the pairing energy into two parts
\begin{equation}
\label{enersum}
E_{\rm pair} = \delta E + \delta E_0 \,,
\end{equation}
where the difference
\[\delta E  =  E_{\rm normal}[\rho] + E_{\rm anomal}[\rho,\nu]
- E_{\rm normal}[\rho_0] \]
is the ``direct'' pairing contribution and
\[\delta E_0  =  E_{\rm normal}[\rho_0]
- E_{\rm normal}[\rho_{\rm HF}]  \]
is the polarization pairing energy.

In the first step we calculate $\delta E$. The anomalous part 
of the energy (\ref{EA2}) may be written in the form:
\begin{equation}
E_{\rm anomal}[\rho,\nu]
=\frac 12{\rm Tr}\int\nolimits_C\frac{\d \epsilon}{2\pi\mathrm{i}}
\Delta F^{\dagger}(\epsilon)\,.
\end{equation}
This can be expressed, by using the Gor'kov equations, as
\begin{equation}
E_{\rm anomal}=\frac 12 {\rm Tr} \int\nolimits_C 
\frac{\d \epsilon}{2\pi\mathrm{i}}
(\epsilon - h + \mu)G_{\rm s}(\epsilon)\,.
\end{equation}
On the other hand, the change of the normal part of the energy
functional, in first-order perturbation theory,
is given by
\begin{equation}
E_{\rm normal}[\rho]-E_{\rm normal}[\rho_0]={\rm Tr}(h\delta\rho)
\equiv \mu\delta N + {\rm Tr}[(h - \mu)\delta\rho]\,.
\end{equation}
This expression may be written in another form:
\[E_{\rm normal}[\rho] - E_{\rm normal}[\rho_0] = \mu \delta N +
{\rm Tr}\int\nolimits_C\frac{\d \epsilon}{2\pi\mathrm{i}}(h-\mu)
(G_{\rm s}(\epsilon) - G_0(\epsilon))\,.  \]
Adding $E_{\rm anomal}$, we find the first (``direct'') change in
the energy connected with the density variation $\delta\rho$:
\begin{eqnarray}
\delta E = \mu\delta N &+&
\frac 12{\rm Tr}\int\nolimits_C \frac{d\epsilon}{2\pi\mathrm{i}}
(\epsilon + h-\mu) G_{\rm s}(\epsilon) \nonumber \\ 
&-&{\rm Tr}\int\nolimits_C\frac{\d \epsilon}{2\pi\mathrm{i}}
(h-\mu)G_0(\epsilon)\,,
\end{eqnarray}
which can also be written, by using the Gor'kov
equations~(\ref{FP}), as
\begin{equation}
\delta E =\mu \delta N
+\frac 12{\rm Tr}\int\nolimits_C \frac{\d \epsilon}{2\pi\mathrm{i}}
(\epsilon + h - \mu)G_0(\epsilon)\Delta F^{\dagger}(\epsilon)\,.
\label{DE}
\end{equation}

For the second (``polarization'') contribution to the total pairing
energy in eq.~(\ref{enersum}), related to the density 
variation~(\ref{DR0}), we get:
\begin{eqnarray}
\delta E_0 & = & {\rm Tr}(h_{\rm HF} \delta \rho_0)
 =  \mu_{\rm HF} \delta N_0 \nonumber\\
& + & {\rm Tr}\int\nolimits_C\frac{\d \epsilon}{2\pi\mathrm{i}}
(h_{\rm HF} - \mu_{\rm HF})
G_{\rm HF}(\epsilon)(\delta U_{\rm pair} - \delta\mu)G_0(\epsilon)\,.
\label{EN01}
\end{eqnarray}
The last term vanishes in first order. Thus, for the second
change in the total energy, again in the first order, we find
\begin{equation}
\delta E_0 = \mu_{\rm HF}\delta N_0 \approx \mu \delta N_0\,.
\label{EN02}
\end{equation}

Collecting both contributions and taking into account the particle
conservation~(\ref{MuN}), which gives $\mu (\delta N + \delta N_0 )
= 0$, we arrive at the resulting formula for the pairing energy:
\begin{equation}
E_{\rm pair}
=\frac 12 {\rm Tr} \int \frac{\d \epsilon}{2\pi\mathrm{i}}
(\epsilon + h - \mu)G_0(\epsilon)\Delta F^{\dagger}(\epsilon)\,.
\end{equation}
In $k$--representation, using the spectral decompositions of 
eqs.~(\ref{SG0}) and~(\ref{SFPk}), this can be written as
\begin{eqnarray}
E_{\rm pair} & = & - \frac 14\sum_i \sum_{(E_{\alpha}>0)}
\frac{E_\alpha - \vert\epsilon_i - \mu\vert}
{E_\alpha + \vert\epsilon_i - \mu\vert}
\sum_j \Delta_{ij} \nonumber\\
&  &  \left[
(V_{j\alpha}U_{i\alpha}^* - U_{j \alpha}^*  V_{i \alpha})
-(1 - n_i)(V_{j\alpha}U_{i\alpha}^* + U_{j \alpha}^*  V_{i \alpha})
\right]\,.
\label{EPAIR}
\end{eqnarray}
In diagonal approximation we get:
\begin{equation}
\label{EPAIRDIAG}
E_{\rm pair} = -\frac 1 2 \sum_k
\frac {\vert \Delta_k\vert ^2}{2E_k}
\frac{E_k - \vert \epsilon_k - \mu \vert}
{E_k + \vert \epsilon_k - \mu \vert}\,.
\end{equation}

For infinite nuclear matter the sum over $k$ is replaced by an 
integral in the momentum space, and one gets the pairing energy 
density (for one kind of particles):
\begin{equation}
\varepsilon_{\rm pair}=-\frac 1 2 \int \frac{\d \vec p}{(2\pi)^3}
\frac{\vert \Delta(\vec p) \vert ^2} {E(\vec p)}
\frac{E(\vec p) - \vert \epsilon(\vec p) - \epsilon_{\rm F}\vert}
{E(\vec p) + \vert \epsilon(\vec p) - \epsilon_{\rm F}\vert}\,.
\end{equation}
In the simplest model we assume
\begin{equation}
\vert \Delta(\vec p)\vert = \vert \Delta(p_{\rm F})\vert
\equiv \Delta\,,\qquad
\epsilon(\vec p) = \frac{{\vec p}^2}{2m}\,,
\label{SM}
\end{equation}
which leads to the expression
\begin{equation}
\varepsilon_{\rm pair}
= - \frac 3 8 \varrho_0 \frac{\Delta^2}{\epsilon_{\rm F}}
\int_{-1}^\infty \d t\, \frac{\sqrt{1+t} }{\sqrt{\delta^2+t^2}}
\frac {\sqrt{\delta^2 + t^2} - \vert t \vert}
{\sqrt{\delta^2 + t^2} + \vert t \vert}\,,
\label{tint}
\end{equation}
with
\[t = \frac{\epsilon(\vec p)}{\epsilon_{\rm F}} - 1\,,\;\;
\delta = \frac \Delta {\epsilon_{\rm F}}\,,\;\;
\varrho_0 = \frac {p_{\rm F}^3}{3\pi^2}\,.  \]
The integral in (\ref{tint}) at $\delta \ll 1$ can be
easily estimated to yield\\
$1 - \delta^2(6\ln2 + 1 - 2\ln\delta)/32$
+ higher order corrections in $\delta$. Numerically this
integral deviates very little from unity even at unreasonably
large $\delta$ (less than by 2.5\% up to $\delta \approx 0.5$).

Thus, in the weak pairing approximation for an infinite system,
as long as the condition $\Delta\ll\epsilon_{\rm F}$ holds
(which is generally assumed for nuclear matter near the
saturation point), the pairing energy density is
\begin{equation}
\varepsilon_{\rm pair} = - \frac 3 8 \varrho_0
\frac{\Delta^2(p_{\rm F})}{\epsilon_{\rm F}} =
- \frac 1 4 \frac{m^*p_{\rm F}}{\pi^2}\Delta^2(p_{\rm F}) \,.
\label{epspair}
\end{equation}
We want to conclude this Appendix with the remark that in the case
of contact pairing force leading to a local pairing field
$\Delta(\vec r)$ there is no difficulty related with the cutoff
$\epsilon_{\rm c}$ to calculate the total energy of the system: in
the energy space as a function of $\epsilon_{\rm c}$ it converges
rapidly ($\delta_{\rm c}E_{\rm pair}\sim 1/\epsilon_{\rm c}
\sqrt{\epsilon_{\rm c}}$) although kinetic and interaction energies,
if taken separately, diverge ($\sim+\sqrt{\epsilon_{\rm c}}$ and
$\sim-\sqrt{\epsilon_{\rm c}}$, respectively).

\setcounter{equation}{0}
\setcounter{section}{0}
\renewcommand{\theequation}{C.\arabic{equation}}
\section*{Appendix C. Solving the Gor'kov equations
in the coordinate-space representation}

In this Appendix we describe the coordinate-space technique which
allows, in the case of spherical symmetry, to solve the Gor'kov
equations for given local mean-field and pairing potentials exactly.
This technique has been invented in \cite{BSTF87}, where the most
interesting effects, arising due to the proper treatment of the
coupling between bound orbitals and particle continuum, were 
carefully investigated. Among them are the lowering of the chemical 
potential $\mu$ (an increase of the binding energy), the 
term-repulsion phenomenon and the appearance of width for deep-hole
states lying below $2\mu$ from the continuum threshold. It has then
been applied, within the EDF approach with a density-independent 
pairing $\delta$-force, to some superfluid even-even 
nuclei~\cite{STF88}.

The fact that the coordinate-space HFB approach for finite systems
can properly treat the positive-energy continuum, in contrast to
BCS-like methods in which the presence of an unphysical ``particle
gas'' is almost unavoidable, was apparently first pointed out in
\cite{Bul80}. It has been shown there that the HFB approach 
naturally leads to a localized nuclear wave function and gives the 
correct asymptotics for the
normal and anomalous densities. These properties of the HFB solutions
are especially important for the nuclei close to the drip lines.
A method of solving the HFB equation directly in the coordinate
representation, but approximating the continuum by a set of the
discrete states in a spherical box, has been used in \cite{DFT84} for
the self-consistent description of a chain of tin isotopes with the
Skyrme effective interaction. The asymptotic behavior of the
quasiparticle wave functions, as well as of the density
distributions, have been studied with this method \cite{DFT84,DNW96}.

Here we give a more detailed description of the technique suggested
in \cite{BSTF87} for spherical even-even nuclei. We shall present
also, by using the so-called ``uniform filling approximation" to
preserve the spherical symmetry, an extension of this technique for
the odd systems. With this extension the blocking effect appears in
a natural way. In what follows, the isospin variables are omitted to
simplify the notation, and the derivation is the same both for
neutrons and protons.

In order to separate the spin-angular variables in the
Gor'kov matrix equation (\ref{GGF}), it is convenient to introduce
a unitary transformation for the generalized Green's function:
\begin{equation}
\hat G={\hat T}^\dagger\hat{\mathcal G}{\hat T}\,,\quad
\hat{\mathcal G}=\hat T\hat G\hat T^{\dagger}\,,
\label{trG}
\end{equation}
where $\hat T$ is given by
\begin{equation}
\hat T=\left( \begin{array}{cc}
   \delta(1,2) & 0  \\ 0  & {\mathcal T}(1,2)
	       \end{array}  \right)\,.
\end{equation}
The operator ${\mathcal T}(1,2)$ is defined by eq.~(\ref{T12}).
Applying this transformation to (\ref{GGF}),
we get the equation for $\hat{\mathcal G}$:
\begin{equation}
(\epsilon -\hat{\widetilde {\mathcal H}})\hat{\mathcal G}=\hat{I}\,,
\label{1}
\end{equation}
where the transformed Hamiltonian reads
\begin{eqnarray}
\hat{\widetilde {\mathcal H}} =\hat T \hat{\mathcal H}
{\hat T}^\dagger &=&\left(\begin{array}{cc}
 \hat h -\mu & \;{\hat \Delta}{\hat {\mathcal T}}^\dagger \\
-\hat {\mathcal T}{\hat \Delta}^*
& \; \mu - {\hat {\mathcal T}} {\hat h}^*{\hat {\mathcal T}}^\dagger
	       \end{array} \right) \nonumber \\
&=&\left( \begin{array}{cc}
 h(1,2) - \mu\delta(1,2)      & \Delta(\vec r_1)\delta(1,2) \\
\Delta^*(\vec r_1)\delta(1,2) & \mu\delta(1,2) - h(1,2)
	       \end{array} \right)\,.  \label{trH}
\end{eqnarray}
Here time-reversal invariance,
${\hat {\mathcal T}} {\hat h}^*{\hat {\mathcal T}}^\dagger = \hat h$,
is assumed (no magnetic field or rotation) and the local (real)
pairing field $\Delta(\vec{r})$,
which is diagonal in spin variables. is introduced according
to eq.~(\ref{T12D}). The transformed generalized Green's function
$\hat{\mathcal G}$ entering eq.~(\ref{1}) is given by
\begin{equation}
\hat{\mathcal G}(1,2;\epsilon)=
\left(\begin{array}{cc}
G_{\rm s}                 & \; F{\mathcal T}^\dagger \\
{\mathcal T}F^\dagger   & \; {\mathcal T}{\bar G}_{\rm s}
{\mathcal T}^\dagger \end{array} \right)\,.  \label{Gtrans}
\end{equation}
Because of the time-reversal symmetry, required for the Bogolyubov
quasiparticle vacuum, this transformed Green's function
may be written, for even-even spherical nuclei, in a form where
the spin-angular parts are the same for all its $2\times2$
components:
\begin{equation}
\hat{\mathcal G}(1,2;\epsilon)=
\frac{1}{r_1 r_2}\sum_{jlm}\hat{g}_{jl}(r_1,r_2;\epsilon)
\phi_{jlm}(\vec{n}_1,s_1) \phi^*_{jlm}(\vec{n}_2,s_2)\,,
\label{gG}
\end{equation}
with $\phi_{jlm}(\vec{n},s)=
C^{jm}_{lm-s\,\frac 12 s}\mathrm{i}^lY_{l\,m-s}(\vec{n})$
being the $s$-component of the usual spherical spinors
($s=\pm \textstyle{\frac 1 2}$). For the generalized radial Green's
function ${\hat g}_{jl}$, which has the matrix form
\begin{equation}
\hat{g}_{jl}(r,r';\epsilon) =
\left(\begin{array}{cc}
g^{11}_{jl}(r,r';\epsilon)  &\; g^{12}_{jl}(r,r';\epsilon)\\
g^{21}_{jl}(r,r';\epsilon)  &\; g^{22}_{jl}(r,r';\epsilon)
	       \end{array} \right)\,,
\label{gjl}
\end{equation}
one gets the equation
\begin{equation}
\left( \begin{array}{cc}
 \epsilon-h_{jl}+\mu & -\Delta \\
 -\Delta &  \epsilon+h_{jl}-\mu
\end{array} \right) \hat{g}_{jl}(r_1,r_2;\epsilon)=
                   \left( \begin{array}{cc}
 \delta(r_1-r_2) & 0 \\
 0 & \delta(r_1-r_2)
	       \end{array} \right) \,,
\label{RGF}
\end{equation}
where $h_{jl}=\textstyle{\frac {\hbar^2}{2m}}
(-\textstyle{\frac{d^2}{dr^2}+\frac{l(l+1)}{r^2}})+U_{jl}(r)$ is
the single-quasiparticle Hamiltonian in the $jl$ channel (see below).

The solution of the matrix equation~(\ref{RGF}) can be constructed
by using the set of the four linearly independent solutions
\[\hat{y}_{i,jl}(r)=\left( \begin{array}{c}
 u_{i,jl}(r) \\ v_{i,jl}(r)
                    \end{array} \right) \,,
\,\,\, i=\hbox{1--4} \,, \]
which satisfy the homogeneous system of equations obtained
from~(\ref{RGF}) by setting the right hand side to zero:
\begin{eqnarray}
\left(\frac{\hbar^2}{2m}\frac{d^2}{d r^2}-f_{jl}(r)+\mu+
\epsilon\right)u_{i,jl}(r)
-\Delta(r)v_{i,jl}(r)= 0\,, \nonumber \\
\Delta(r)u_{i,jl}(r)
+\left(\frac{\hbar^2}{2m}\frac{d^2}{d r^2}-f_{jl}(r)+\mu
-\epsilon\right)v_{i,jl}(r) = 0 \,.
\label{5}
\end{eqnarray}
Here
\[f_{jl}(r)=U_{\rm c}(r)+U_{sl}(r)\langle\vec{\sigma}\vec{l}
\rangle_{jl}+\frac{\hbar^2l(l+1)}{2mr^2} \,,  \]
$U_{\rm c}$ and $U_{sl}$ are the central and spin-orbit mean-field
potentials, respectively, $\langle\vec{\sigma}\vec{l}\rangle_{jl}=
j(j+1)-l(l+1)-\textstyle{\frac 3 4}\,,\,
m$ is the nucleon mass. We shall seek the radial Green's function
$\hat{g}_{jl}$ entering eq.~(\ref{gG}) in the following form (for
simplicity, the $\epsilon$-dependence and the $jl$ indices are not
shown):
\begin{eqnarray}
\hat{g}(r_1,r_2) &=&
\frac{2m}{\hbar^2}[\left(\alpha\hat{y}_1(r_1)+
\beta\hat{y}_2(r_1)\right)
\tilde{\hat{y}}_3(r_2) \nonumber \\
&+&\left(\gamma\hat{y}_1(r_1)
+\delta\hat{y}_2(r_1)\right)\tilde{\hat{y}}_4(r_2)]
\theta(r_2-r_1) \nonumber \\
&+& \frac{2m}{\hbar^2}[\hat{y}_3(r_1)
\left(\alpha\tilde{\hat{y}}_1(r_2)
+\beta\tilde{\hat{y}}_2(r_2)\right) \nonumber \\
&+& \hat{y}_4(r_1)\left(\gamma\tilde{\hat{y}}_1(r_2)+
\delta\tilde{\hat{y}}_2(r_2)
\right)]\theta(r_1-r_2)\,,  \label{6}
\end{eqnarray}
where $\tilde{\hat{y}}_i=(u_i\,,v_i)$ is the transposed $i$-th
solution, $\theta(x)$ is the step function, $\alpha, \beta, \gamma$,
and $\delta$ are the coefficients to be determined. The solutions
$\hat{y}_1$ and $\hat{y}_2$ are chosen to be regular at
$r\rightarrow 0$ and the other two, $\hat{y}_3$
and $\hat{y}_4$, to be regular at $r\rightarrow \infty$:
\begin{equation} \begin{array}{ll}
\hat{y}_1(r\!\rightarrow \!\!0)=\left(\!\! \begin{array}{c}1 \\
\zeta_0 \end{array}\!\!\right) (q_+r)^{l+1}\,, &
\hat{y}_2(r\!\rightarrow\!\!0)=\left(\!\!\begin{array}{c}\zeta_0 \\
1 \end{array}\!\!\right) (q_-r)^{l+1}\,, \\
\hat{y}_3(r\!\rightarrow \!\!\infty)=\left(\!\!\begin{array}{c}1 \\
\zeta_\infty \end{array}\!\!\right) \e^{\mathrm{i}k_+r}\,, &
\hat{y}_4(r\!\rightarrow \!\!\infty)=\left(\!\!
\begin{array}{c}\zeta_\infty \\
1 \end{array}\!\!\right) \e^{\mathrm{i}k_-r}\,.
\end{array} \label{7} \end{equation}
Here
\begin{equation} \begin{array}{l}
q_\pm=\left[\frac{2m}{\hbar^2}\left(\mu-U_0\pm\sqrt{\epsilon^2
-\Delta_0^2}\right)\right]^{1/2},\;
\zeta_0=-\Delta_0/\left(\epsilon+\sqrt{\epsilon^2-\Delta_0^2}
                     \right)\,,\\
k_\pm=\left[\frac{2m}{\hbar^2}\left(\mu \pm \sqrt{\epsilon^2
-\Delta_\infty^2}\,\right)\right]^{1/2},\;\;
\zeta_\infty=-\Delta_\infty/\left(\epsilon+\sqrt{\epsilon^2-
\Delta_\infty^2}\,\right)\,; \\
U_0=U_{\rm c}(0)+U_{sl}(0)\langle\vec{\sigma}\vec{l}\rangle\,,\;
\Delta_0=\Delta(0)\,,\; \Delta_\infty=
\Delta(r \!\rightarrow \!\infty)\,.
\end{array}          \label{8}
\end{equation}
It is important to notice that on the physical $k$--sheet the
imaginary parts of the asymptotic momenta $k_\pm$ should be chosen
positive\footnote{To be more precise, the solutions $\hat{y}_3$ and
$\hat{y}_4$ are given asymptotically at $r\rightarrow \infty$ by the
Whittaker functions $W_{-\mathrm{i}\eta_\pm,\,l+1/2}
(-2\mathrm{i}k_\pm r)$, respectively,
with $\eta_\pm = mZe^2/\hbar^2 k_\pm$ being the corresponding Coulomb
parameters ($\eta_\pm = 0$ for neutrons).}: ${\rm Im}\, k_\pm >0$.

Using the system of equations (\ref{5}), it can be easily verified
that for any two arbitrary solutions,
\[\textstyle{\hat{y}_i=
\left(\!\! \begin{array}{c}u_i \\ v_i \end{array}\!\!\right)}
\;\;\; {\rm and} \;\;\; \textstyle{\hat{y}_j=\left(\!\! 
\begin{array}{c}u_j \\ v_j \end{array}\!\!\right)} \,, \]
the combination $C_{ij}=U_{ij}-V_{ij}$ composed of the Wronskians
$U_{ij}=u_iu_j'-u_i'u_j$ and $V_{ij}=v_iv_j'-v_i'v_j$,
where the primes denote differentiation with respect to $r$,
does not depend on the coordinate $r$,
and for solutions satisfying the boundary conditions (\ref{7}) one 
gets $C_{12}=C_{34}=0$. It can be seen also that
$U_{13}U_{24}-U_{23}U_{14}=U_{12}U_{34}$ and
$V_{13}V_{24}-V_{23}V_{14}=V_{12}V_{34}$. With these relations one
easily obtains an equation for determining the four coefficients
entering eq.~(\ref{6}):
\begin{equation}
\left(\begin{array}{cc}\alpha & \beta \\ \gamma & \delta \end{array}
\right) = \left( \begin{array}{cc}C_{13}& C_{14} \\
 C_{23} & C_{24} \end{array} \right)^{-1} =
\frac{1}{D}\left( \begin{array}{cc}C_{24}& -C_{14} \\
 -C_{23} & C_{13} \end{array} \right)\,,
\label{9}\end{equation}
where $D(\epsilon)=C_{13}C_{24}-C_{23}C_{14}$ is the determinant of 
the matrix $C$.

The formulas (\ref{trG}), (\ref{gG}) and (\ref{6})--(\ref{9})
constitute an exact solution of the Gor'kov equation for the 
generalized Green's function in the coordinate representation for 
superfluid spherical nuclei with a localized pairing field 
$\Delta(\vec r)$.

\begin{figure}[ht]
\begin{center}
\leavevmode
\epsfxsize=22pc 
\epsfbox{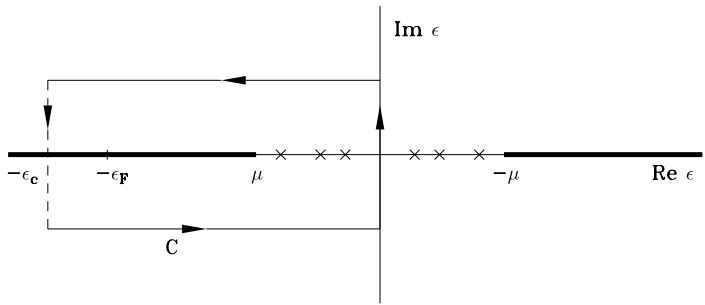} 
\end{center}
\caption{The contour $C$ in the energy plane to integrate the 
generalized Green's function for determining the generalized density
matrix (for even nuclei). The crosses mark the positions of the 
quasiparticle poles, the heavy lines represent the branch cuts on 
the ${\rm Im} \epsilon = 0$ axis where the energy spectrum is 
continuous, $\mu$ is the chemical potential, $\epsilon_{\rm F}$ is 
the Fermi energy and $\epsilon_{\rm c}$ is the energy
cutoff (see text).}
\label{f:f24}
\end{figure}

In the case of vanishing pairing one has $u_2=u_4=v_1=v_3=0$ and
$\beta=\gamma=0$. Then the radial generalized Green's 
function~(\ref{gjl}) becomes diagonal with $g^{11}$ and $g^{22}$ 
in the following form:
\begin{eqnarray}
g^{11}_{jl}(r,r';\epsilon)\vert_{\Delta = 0} &=& \frac {2m}{\hbar^2}
\frac {u_{1;jl}(r_<;\epsilon)u_{3;jl}(r_>;\epsilon)}
{U_{13}(\epsilon)}\,, \nonumber \\
g^{22}_{jl}(r,r';\epsilon)\vert_{\Delta = 0} &=& -\frac {2m}{\hbar^2}
\frac {v_{2;jl}(r_<;\epsilon)v_{4;jl}(r_>;\epsilon)}
{V_{24}(\epsilon)}\,,  \label{gjlHF}
\end{eqnarray}
where $r_<$ and $r_>$ denote the lesser and the greater of $r$ and 
$r'$, respectively. Here $g^{11}$ is interpreted as the HF Green's 
function for particles ($\epsilon > \mu$) while $g^{22}$ as that for
holes ($\epsilon < \mu$)\footnote{Eq.(\ref{gjlHF}) for the 
single-particle Green's function allows one to construct the 
particle-hole propagator in the coordinate representation and to 
solve, for closed-shell nuclei, the RPA equations exactly, including
the whole particle continuum. An earlier application of this method 
for the study of pion condensation in finite nuclear systems may be 
found in \cite{STF75}, and, for calculating the continuum--RPA 
multipole response functions, in \cite{SB75}.}. By integrating the 
latter along the contour $C$ of Fig.~\ref{f:f24} one gets the 
HF $jl$ density matrix $\rho_{jl}(r,r')\vert_{\Delta=0}$ which is 
normalized by $(2j+1)\int \d r \rho_{jl}(r,r')\vert_{\Delta=0} = 
N_{jl}$ where $N_{jl}$ is the number of particles occupying orbitals
with the given quantum numbers $jl$. This density matrix can be 
expressed by a finite sum of separable terms, i.e. by a sum of 
residues of $g^{22}\vert_{\Delta=0}$ at the poles which are zeros 
of the Wronskian $V_{24}$. The equation $V_{24}=0$ defines the 
spectrum of the HF hole states, and this spectrum is utterly 
discrete. 

The situation with pairing is completely different.
It is clear that the poles (if any) of the generalized Green's 
function are determined by the equation $D(\epsilon)=0$ which gives
the discrete spectrum of the Bogolyubov quasiparticle states. Due 
to the invariance of the equation~(\ref{5}) under simultaneous 
replacements $\epsilon\rightarrow -\epsilon$ and $(u,\,v) 
\rightarrow (v,\,-u)$ the energy spectrum is symmetrical around the
point $\epsilon=0$. For bound systems, by definition, one has 
$\mu < 0$. If the system is finite, i.e. if the density distribution
has a finite range, one expects $\Delta_\infty=0$, otherwise the 
system would be unstable with respect to two particle emission. 
From eqs.~(\ref{6})--(\ref{9}) it follows then that the 
quasiparticle spectrum is discrete within the energy region
$\vert \epsilon \vert < \vert \mu \vert$ (here $\hat{g}_{jl}$ is a 
real function at the axis ${\rm Im}\,\epsilon =0$) and continuous if
$\vert \epsilon \vert > \vert \mu \vert$ (there $\hat{g}_{jl}$ is a 
complex function). These features are reflected in Fig.~\ref{f:f24}
where the branch cuts are shown by the heavy lines extending 
symmetrically to the left and to the right from the points 
$\pm \mu$, respectively, along the ${\rm Im}\,\epsilon =0$ axis. 
The case with only the branch cuts, without poles, corresponds to a
drip-line even nucleus. Performing the integration of the radial 
Green's function $\hat{g}_{jl}$ along the contour $C$ of
Fig.~\ref{f:f24} one obtains the radial part of the generalized 
density matrix\footnote{The analytical properties of the Green's 
function allow one to consider only the upper half of the contour 
with Im$\epsilon>0$: integrate Re$G$ along the vertical sections 
and Im$G$ along the horizontal section in the upper half-plane, then
for the total result take the doubled sum of these integrals.}. To 
calculate the normal and anomalous densities we need the following 
two radial ``$jl$ density matrices'':
\begin{equation}
\rho_{jl}(r,r')=\int\nolimits_C\frac{\d \epsilon}{2\pi\mathrm{i}}
g^{22}_{jl}(r,r';\epsilon)\,,\;\;
\nu_{jl}(r,r')=\int\nolimits_C\frac{\d \epsilon}{2\pi\mathrm{i}}
g^{12}_{jl}(r,r';\epsilon)\,.  \label{jlden}
\end{equation}
Then the resulting expression for the normal density reads
\begin{equation}
\rho(\vec{r})=\frac{1}{4\pi r^2} \sum_{jl} (2j+1)\rho_{jl}(r,r)\,.
\label{normden}
\end{equation}
It is normalized by $\int\d {\vec r}\rho(\vec{r})=N$ which is 
actually the condition~(\ref{MuC}) for the chemical potential 
$\mu$. For the anomalous density from~(\ref{nuF}), (\ref{nuc}), 
(\ref{Gtrans}), (\ref{gG}) and~(\ref{jlden}) we get
\begin{equation}
\nu(\vec{r})= \frac{1}{4\pi r^2} \sum_{jl}
(j+\textstyle{\frac 1 2)}\nu_{jl}(r,r)\,.
\label{anoden}
\end{equation}
These densities are used to calculate both the normal and anomalous
parts of the interaction energy in eq.~(\ref{FC}). In addition, for
the spin-orbital term, eq.~(\ref{sosph}), we define\footnote{Remember
that we do not show here the cutoff index $c$. It should be 
emphasized that because of the continuum all the densities defined 
in this Appendix cannot be expressed by a finite sum of separable 
terms. The latter would be possible only within a certain 
approximation, namely with a discretized basis {\em and\/} with an 
energy cutoff. The coordinate-space technique described above, 
which involves integration of the exactly constructed 
generalized Green's function in the complex energy plane,
appears to be exact for the $jl$--densities: both $\rho_{jl}(r,r)$
and $\nu_{jl}(r,r)$ converge as functions of $\epsilon_{\rm c}$
($\delta_{\rm c}\rho_{jl} \sim \epsilon_{\rm c}^{-1}$
and $\delta_{\rm c}\nu_{jl} \sim \epsilon_{\rm c}^{-1/4}$, 
respectively). It is exact also for the normal 
density~ \ref{normden}) since it converges,
$\delta_{\rm c} \rho \sim \epsilon_{\rm c}^{-1/2}$. However, if one
goes beyond the cutoff $\epsilon_{\rm c}$, the anomalous 
density~(\ref{anoden}) diverges, $\nu \sim \epsilon_{\rm c}^{1/4}$.
The kinetic and anomalous energies beyond $\epsilon_{\rm c}$ diverge
too, but their sum converges very rapidly
($\delta_{\rm c} E_{\rm pair}\sim \epsilon_{\rm c}^{-3/2}$, see 
Appendix B). This fact has been used in the present paper to 
formulate the generalized variational principle with the cutoff 
local EDF. We stress once more that this formulation does not imply
a cutoff of the basis.}:
\begin{equation}
\rho_{sl}(\vec{r})=\frac{1}{4\pi r^2} \sum_{jl} (2j+1)
\langle {\vec \sigma}{\vec l}\rangle_{jl}\rho_{jl}(r,r)\,.
\end{equation}

The kinetic energy of the system is given by
\begin{equation}
E_{\rm kin} = \frac{\hbar^2}{2m}\int \d r \sum_{jl}(2j+1)
\left\{ \left[-\frac{\d^2}{\d r^2}+\frac{l(l+1)}{r^2}\right]
\rho_{jl}(r,r') \right\}_{r'=r}\,.  \label{Ekinjl}
\end{equation}
Note that to compute this energy one needs not only the normal
$jl$-densities, but also the $jl$ normal density matrices
$\rho_{jl}(r,r')$.

\begin{figure}[t]
\begin{center}
\leavevmode
\epsfxsize=22pc 
\epsfbox{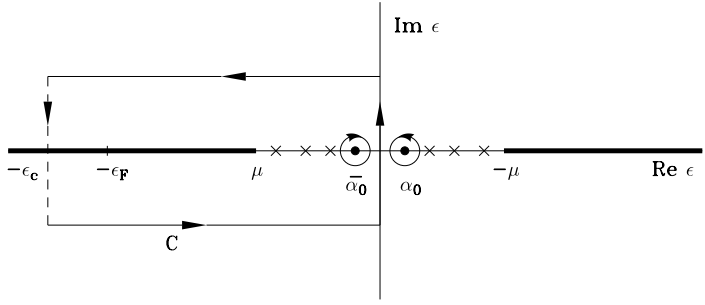} 
\end{center}
\caption{The contour $C$ for odd nuclei. The positions of the poles for 
an odd
quasiparticle with the energies $\pm E_{\alpha_0}$ are shown by heavy dots. 
Note that the ways of passing these poles go in opposite directions. 
Other notations are the same as in Fig.~\ref{f:f24}.}
\label{f:f25}
\end{figure}

For an odd system the spectral expansion of the
generalized Green's function, eqs.~(\ref{SG})--(\ref{SF}), should be
modified. Suppose that the addition of an odd particle leads to
the appearance of the quasiparticle with the energy $E_{\alpha_0}$
in the ground state of the system. Here $\alpha_0$ denotes all the
relevant quantum numbers. They can be specified, for example, by
$\alpha_0=\{njlm\}_0$ from the whole set $\{\alpha\}$ used to label
the solutions of the HFB equation~(\ref{SEQ}). Suppose further
that $E_{\alpha}$ corresponds to a certain eigenenergy of this
equation and belongs to the discrete spectrum, i.e.
$\vert \mu \vert >E_{\alpha_0}>0$. As illustrated in Fig.~\ref{f:f25},
$E_{\alpha_0}$ is located on the ${\rm Im}\,\epsilon = 0$ axis
in the vicinity of the point $\epsilon =0$. Generally, $E_{\alpha_0}$
is of the order of $\bar \Delta$, the average matrix element of the
pairing potential on the the Fermi surface. Note that the pole
nearest to $\epsilon = 0$ need not correspond to the ground
state of the system. The case of $E_{\alpha_0} \approx -\mu$
determines the position of the drip line for odd nuclei. The rules
for passing the Green's function poles for superfluid odd systems are
formulated by Migdal \cite{Migdal}. To the spectral expansions
(\ref{SG})--(\ref{SF}), which do not contain the contribution of the
odd quasiparticle, one should add the following terms:
\begin{eqnarray}
\delta G^{\rm odd}_{\rm s}(1,2;\epsilon) & = &
\frac{V^*_{\alpha_0}(1) V_{\alpha_0}(2)}{\epsilon+E_{\alpha_0}
-\mathrm{i}\delta} +
\frac{U_{\alpha_0}(1) U^*_{\alpha_0}(2)}{\epsilon-E_{\alpha_0}
-\mathrm{i}\delta}
          \nonumber \\  & + &
\frac{V^*_{\bar \alpha_0}(1) V_{\bar \alpha_0}(2)}
{\epsilon + E_{\bar \alpha_0} + \mathrm{i}\delta} +
\frac{U_{\bar \alpha_0}(1) U^*_{\bar \alpha_0}(2)}
{\epsilon - E_{\bar \alpha_0} + \mathrm{i}\delta}\,,  \label{ggodd}
\end{eqnarray}
\begin{eqnarray}
\delta F^{\rm odd}(1,2;\epsilon)   & = &
\frac{V^*_{\alpha_0}(1) U_{\alpha_0}(2)}{\epsilon+E_{\alpha_0}-
\mathrm{i}\delta} +
\frac{U_{\alpha_0}(1) V^*_{\alpha_0}(2)}{\epsilon-E_{\alpha_0}-
\mathrm{i}\delta}
       \nonumber \\     & + &
\frac{V^*_{\bar \alpha_0}(1) U_{\bar \alpha_0}(2)}
{\epsilon + E_{\bar \alpha_0} + \mathrm{i}\delta} +
\frac{U_{\bar \alpha_0}(1) V^*_{\bar \alpha_0}(2)}
{\epsilon - E_{\bar \alpha_0} + \mathrm{i}\delta}\,.  \label{ffodd}
\end{eqnarray}
The subscript $\alpha_0$ in these expressions, in Migdal's terminology,
refers to an odd quasiparticle added to a system of $N$ particles
and to an odd quasihole in a system of $N+2$ particles. Consequently,
the subscript $\bar \alpha_0$ refers to the odd quasihole in the system
of $N$ particles and to the odd quasiparticle added to a system of $N-2$
particles. Thus the four terms in~(\ref{ggodd}) and~(\ref{ffodd})
correspond to pairs of time-reversed states. But, due to
the signs of the infinitesimal imaginary parts $\pm \mathrm{i}\delta$ in the
denominators, only two, not time-reversed, terms in these expressions
should contribute to the generalized density matrix.
Closing the contour in the upper half of the energy plane, the terms
from the first lines in eqs.~(\ref{ggodd}) and~(\ref{ffodd}) will be
accounted for. By integrating the Green's functions $G_{\rm s}$ and $F$ along
the contour $C$ of Fig.~\ref{f:f24}, all the residues at the pole
$\epsilon = -E_{\alpha_0}$ $(=-E_{\bar \alpha_0})$ have been already
included. Therefore, from the so obtained density matrix we have to
subtract the contributions related to the third terms of these equations
and to add the residues from the second ones. The resulting contour for
the odd system is depicted in Fig.~\ref{f:f25} which clearly illustrates the
blocking effect: the presence of the odd particle in the level $\alpha_0$
prevents it from participating in the pairing correlations because in this
case the ``conjugate'' level $\bar \alpha_0$ should be empty. This way we get
the contributions to the normal and abnormal density matrices from the odd
qusiparticle:
\begin{equation}
\delta\rho^{\rm odd}(1,2)=-V^*_{\bar \alpha_0}(1)V_{\bar \alpha_0}(2)
                      +U_{\alpha_0}(1)U^*_{\alpha_0}(2)\,,
\label{droodd}
\end{equation}
\begin{equation}
\delta\nu^{\rm odd}(1,2)=-V^*_{\bar \alpha_0}(1)U_{\bar \alpha_0}(2)
                      +U_{\alpha_0}(1)V^*_{\alpha_0}(2)\,.
\label{dnuodd}
\end{equation}

In the following we shall consider the case when only the spherical part
of the core polarization induced by the odd quasiparticle is important,
and the deformation of the core can be neglected. Then the solutions of
the Bogolyubov equation may be chosen to have the form
\begin{eqnarray}
U_\alpha(\vec r,s)&=&\frac 1 r 
u_{njl}(r)\phi_{jlm}(\vec n,s)\,,\nonumber\\
V_\alpha(\vec r,s)&=&\frac 1 r v_{njl}(r)(-1)^{j-m+1}
\phi^*_{jl\,-m}(\vec n,s)\,,
\label{uva0}
\end{eqnarray}
where the radial wave functions $u_{njl}(r)$ and $v_{njl}(r)$ 
satisfy eq.~(\ref{5}). For the discrete states, including the odd
quasiparticle state with $\alpha=\alpha_0$, the latter obey the 
orthogonality relation $\int \d r [u^*_{n'lj}(r)u_{nlj}(r)+
v^*_{n'lj}(r)v_{nlj}(r)]=\delta_{nn'}$. We introduce further the 
uniform filling approximation. Namely, we assume that the odd 
quasiparticle occupies all the possible $m_0$ states with equal 
probability and take the averages
\begin{eqnarray}
\overline{\delta\rho^{\rm odd}}(1,2)&=&
\frac{1}{2j_0+1}\sum_{m_0}\delta\rho^{\rm odd}(1,2)\,,\nonumber\\
\overline{\delta\nu^{\rm odd}}(1,2) &=&
\frac{1}{2j_0+1}\sum_{m_0}\delta\nu^{\rm odd}(1,2)\,.
\label{avrhonu}
\end{eqnarray}
Such an averaging procedure, with the wave functions given by (\ref{uva0}),
enables us to separate the spin-angular variables
for the odd spherical system in the same way as done above for even systems.
The resulting formulas for the contribution to the radial $jl$ density
matrices (\ref{jlden}) due to the presence of the odd particle read
\begin{eqnarray}
\overline{\delta\rho^{\rm odd}_{n_0j_0l_0}}(r,r')&=&
- v^*_{n_0j_0l_0}(r) v_{n_0j_0l_0}(r')
+ u_{n_0j_0l_0}(r) u^*_{n_0j_0l_0}(r') \,,\nonumber\\
\overline{\delta\nu^{\rm odd}_{n_0j_0l_0}}(r,r')&=&
- v^*_{n_0j_0l_0}(r) u_{n_0j_0l_0}(r')
- u_{n_0j_0l_0}(r) v^*_{n_0j_0l_0}(r') \,.
\label{jlodd}
\end{eqnarray}
The contributions to the normal and anomalous densities (\ref{normden})
and (\ref{anoden}) are given by
\begin{equation}
\overline{\delta\rho^{\rm odd}_{n_0j_0l_0}}(\vec r)=\frac{1}{4\pi r^2}
\left(-\vert v_{n_0j_0l_0}(r)\vert^2+
       \vert u_{n_0j_0l_0}(r)\vert^2\right)\,.
\label{oddrho}
\end{equation}
\begin{equation}
\overline{\delta\nu^{\rm odd}_{n_0j_0l_0}}(\vec r)=\frac{1}{4\pi r^2}
u_{n_0j_0l_0}(r)v^*_{n_0j_0l_0}(r)\,.
\label{oddnu}
\end{equation}
It can be seen that $\overline{\delta\nu^{\rm odd}_{n_0j_0l_0}}(\vec r)$
has the opposite sign as $\nu(\vec r)$, eq.~(\ref{anoden}), reflecting the
fact that the blocked level cannot directly contribute to the anomalous
energy. The nucleon separation energies $S_n$ are determined by
the position of the poles close to $\epsilon = 0$. As easily
understood from Fig.~\ref{f:f25}, since these separation energies are 
measured from the continuum threshold $\epsilon = -\mu$, for an odd system 
one gets $S^{\rm odd}_n \approx -\mu - E_{\alpha_0}$, and, for an even 
system, $S^{\rm even}_n \approx -\mu + E_{\bar \alpha_0}$. Thus we have 
$S^{\rm even}_n-S^{\rm odd}_n \approx 2E_{\alpha_0} \approx
2 \bar \Delta$, i.e. the familiar odd-even effect in nuclear masses.

The rest of this Appendix is devoted to the discussion of the
asymptotic behavior of the densities $\rho(\vec r)$ and 
$\nu(\vec r)$. Consider first the asymptotic properties of the wave
functions $y_i$ used to construct the radial Green's 
functions~(\ref{6}) through which these density can be obtained. For
finite systems, the functions $v_3(r)$ and $u_4(r)$ in the solutions
$y_3$ and $y_4$ of the radial HFB equation~(\ref{5}), as given by 
the boundary conditions (\ref{7}), vanish at large distances since
$\zeta_\infty \sim \Delta_\infty = 0$. The function $u_3$ at large
$r$ behaves as $\exp\mathrm{i}k_+r$ and either decreases
$\sim \exp(-r\sqrt{2m(\vert \epsilon \vert + \mu)/\hbar^2})$ if
$\vert \epsilon \vert < - \mu$ or oscillates if $\vert \epsilon 
\vert > - \mu$, the latter corresponds to the continuum spectrum.
The function $v_4$ behaves asymptotically as $\exp\mathrm{i}k_-r$.
If the system is bound, i.e. $\mu < 0$, this function at $r \to
\infty$ always decreases exponentially 
$\sim \exp(-r\sqrt{2m(\vert \epsilon \vert - \mu)/\hbar^2})$.

To get the asymptotic laws for normal and anomalous densities,
defined by eqs.~(\ref{jlden})--(\ref{anoden}), we have to consider
now the radial Green's functions $g^{12}\propto (\gamma u_1+\delta 
u_2) v_4$ and $g^{22} \propto (\gamma v_1 + \delta v_2) v_4$. 
Asymptotically, at large $r$, the functions $u_i$ and $v_i$, 
$i=1,2$, which are regular at $r \to 0$, are given by $u_i(r) = 
c_i\exp(\mathrm{i}k_+r) + d_i\exp(-\mathrm{i}k_+r)$ and
$v_i(r) = e_i\exp(\mathrm{i}k_-r) + f_i\exp(-\mathrm{i}k_-r)$,
respectively. Using eq.~(\ref{9}) one can easily verify that, by
construction, the combination $\gamma d_1 + \delta d_2$ vanishes.
The combination $\gamma f_1 + \delta f_2$ vanishes at the poles of
the Green's function determined by the equation $D(\epsilon) = 0$.
Finally, as the result of of integration of the Green's function
along the contour $C$, we get 
\begin{eqnarray}
\rho(r \to \infty) & \sim & \frac{1}{r^2}
\exp(-2r\sqrt{2m(\vert \epsilon_0 \vert -\mu)/\hbar^2})\,,
\nonumber \\
\nu(r \to \infty) & \sim &
\frac{1}{r^2}\exp(-r[\sqrt{2m(-\vert\epsilon_0\vert-\mu)/\hbar^2}+
\sqrt{2m(\vert\epsilon_0\vert-\mu)/\hbar^2}])\,,  \label{asrhonu}
\end{eqnarray}
where $\epsilon_0$ is the position, within the contour $C$, of a
singularity point nearest to $\epsilon = 0$. This point may be a
pole of the Green's function or the beginning of the branch cut.
The latter situation corresponds to the drip-line nuclei. Generally,
$\vert \epsilon_0 \vert$ is of the order of the average matrix
element $\bar \Delta$ of the pairing potential on the Fermi surface
(typically, $\bar \Delta \approx 1$ MeV). In stable nuclei, since
the absolute value of the chemical potential is much larger, 
$\mu \approx -8$ MeV, the asymptotic behavior of both densities
(\ref{asrhonu}) differ but insignificantly. Approaching the 
drip-line, in even system, $\vert\epsilon_0\vert$, $\vert\mu\vert$ 
and $\bar\Delta$ may become of the same order:
$\vert\epsilon_0\vert\approx -\mu\approx\bar\Delta$.
In such a case one gets 
\begin{eqnarray}
\rho(r \to \infty) & \sim &
\frac{1}{r^2}\exp(-4r\sqrt{m\bar \Delta/ \hbar^2})\,,
\nonumber \\
\nu(r \to \infty)  & \sim &
\frac{1}{r^2}\exp(-2r\sqrt{m\bar \Delta/ \hbar^2})\,.
\end{eqnarray}
It follows that in this situation the anomalous density has a 
longer tail with decaying length by a factor of two greater than 
that of the normal density. Similar conclusions have been drawn 
in~\cite{DFT84,Bul80,DNW96}. On the other hand, in the genuine
drip-line even nuclei, with only one bound state -- the ground 
state, when all the HFB eigenstates are embeded in the continuum, 
there are always contributions to the densities from the 
quasiparticles with different quantum numbers $jl$ and with 
the energies close to $\mu$. The weights of these contributions
are determined by the corresponding quasiparticle level density,
but in any case the most distant tails of the densities
in drip-line nuclei are asymptotically given by the expressions
\begin{eqnarray}
\rho(r \to \infty) & \sim &
\frac{1}{r^2}\exp(-4r\sqrt{m\vert\mu\vert/\hbar^2})\,,
\nonumber \\
\nu(r \to \infty)  & \sim &
\frac{1}{r^2}\exp(-2r\sqrt{m\vert\mu\vert/\hbar^2})\,,
\end{eqnarray}
which are obtained from eq.~(\ref{asrhonu}) by taking the limit  
$\vert\epsilon_0\vert\rightarrow\vert\mu\vert$.
The technique described in this appendix permits, in principle, 
an exact calculation of the asymptotic behavior of the HFB 
densities in different situations.

\end{document}